\documentclass[12pt]{article}

\usepackage{cite}
\usepackage{axodraw}
\usepackage[dvips]{graphicx}

\def\theequation{\arabic{section}.\arabic{equation}}

\begin{document}

\begin{flushright}
%%%   MC-TH-2005-xx\\[-2pt]
{\tt hep-ph/0506107}\\
   June 2005
\end{flushright}
\bigskip

\begin{center}
{\LARGE {\bf Electroweak--Scale Resonant Leptogenesis}}\\[1.5cm] 
{\large Apostolos Pilaftsis and Thomas E.~J. Underwood}\\[0.3cm] 
{\em School of Physics and Astronomy, University of Manchester,}\\ 
{\em Manchester M13 9PL, United Kingdom}
\end{center}

\vspace{1.5cm} 

\centerline{\bf ABSTRACT} 

\noindent
{\small
We  study   minimal  scenarios  of  resonant   leptogenesis  near  the
electroweak phase transition.  These models offer a number of testable
phenomenological  signatures  for  low-energy experiments  and  future
high-energy  colliders.  Our  study extends  previous analyses  of the
relevant  network  of Boltzmann  equations,  consistently taking  into
account effects  from out of equilibrium sphalerons  and single lepton
flavours.  We show that the effects from single lepton flavours become
very  important  in  variants  of  resonant  leptogenesis,  where  the
observed   baryon   asymmetry   in   the  Universe   is   created   by
lepton-to-baryon  conversion  of  an  individual  lepton  number,  for
example that  of the $\tau$-lepton.  The predictions  of such resonant
$\tau$-leptogenesis models  for the final baryon  asymmetry are almost
independent   of  the   initial  lepton-number   and   heavy  neutrino
abundances.  These  models accommodate  the current neutrino  data and
have a number of testable phenomenological implications.  They contain
electroweak-scale heavy Majorana  neutrinos with appreciable couplings
to electrons  and muons,  which can be  probed at future  $e^+e^-$ and
$\mu^+\mu^-$   high-energy   colliders.    In   particular,   resonant
$\tau$-leptogenesis models predict sizeable $0\nu\beta\beta$ decay, as
well as  $e$- and $\mu$-number-violating  processes, such as  $\mu \to
e\gamma$ and  $\mu \to  e$ conversion in  nuclei, with rates  that are
within  reach  of  the  experiments  proposed  by  the  MEG  and  MECO
collaborations. }

\medskip
\noindent
{\small PACS numbers: 11.30.Er, 14.60.St, 98.80.Cq}

\newpage

\setcounter{equation}{0}
\section{Introduction}

The origin  of the baryon asymmetry  in our Universe  (BAU) has always
been one of  the central topics in particle  cosmology.  Recently, the
high-precision   determination   of   many  cosmological   parameters,
including  the  baryon-to-photon ratio  of  number densities,  $\eta_B
\approx 6.1  \times 10^{-10}$~\cite{WMAP}, has  given renewed momentum
for extensive  studies on this  topic~\cite{reviews}.  The established
BAU provides one  of the strongest pieces of  evidence towards physics
beyond  the  Standard  Model~(SM).   One  interesting  suggestion  for
explaining the  BAU, known as {\it  leptogenesis}~\cite{FY}, is linked
with neutrinos.   Although strictly massless in the  SM, neutrinos can
naturally acquire  their small observed  mass through the  presence of
superheavy partners and  the so-called seesaw mechanism~\cite{seesaw}.
These superheavy neutrinos  are singlets under the SM  gauge group and
may  therefore  possess  large  Majorana masses  that  violate  lepton
number~($L$)  conservation by  two units.   In an  expanding Universe,
these  heavy   Majorana  neutrinos  will  in  general   decay  out  of
equilibrium,  potentially  generating  a  net lepton  asymmetry.   The
so-produced  lepton asymmetry  will eventually  be converted  into the
observed   BAU~\cite{FY}   by    means   of   in-thermal   equilibrium
$(B+L)$-violating sphaleron interactions~\cite{KRS}.

One  difficulty faced  by  ordinary seesaw  models  embedded in  grand
unified theories (GUTs)  is associated with the natural  mass scale of
the heavy Majorana neutrinos.  This is expected to be of order the GUT
scale $M_{\rm  GUT} = 10^{16}$~GeV.   On the other  hand, inflationary
supergravity  models  generically   predict  a  reheating  temperature
$T_{\rm  reh}$ of order  $10^9$~GeV.  In  these models,  a significant
constraint  on  the upper  bound  for  $T_{\rm  reh}$ comes  from  the
requirement that  gravitinos are  underabundant in the  early Universe
and so  their late  decays do not  disrupt the nucleosynthesis  of the
light  elements~\cite{AHJMP}.  However,  the low  $T_{\rm  reh}$ gives
rise to another constraint within the context of thermal leptogenesis.
The   heavy  Majorana   neutrino,  whose   $L$-violating   decays  are
responsible for the BAU, has  to be somewhat lighter than $T_{\rm reh}
\sim 10^9$~GeV, so as to be abundantly produced in the early Universe.
Such  a mass for  the heavy  Majorana neutrino  should be  regarded as
unnaturally low for  GUT-scale thermal leptogenesis.  Finally, further
constraints             on             successful            GUT-scale
leptogenesis~\cite{DI,BBP,GCBetal,CT} may  be obtained from  solar and
atmospheric neutrino data~\cite{PDG}.

The  aforementioned problem with  a low  reheating temperature  may be
completely   avoided  in   models  that   realize   low-scale  thermal
leptogenesis~\cite{APRD,APreview,LB}.  In  particular, the  lowering of
the scale may  rely on a dynamical mechanism,  in which heavy-neutrino
self-energy effects~\cite{LiuSegre}  on the leptonic  asymmetry become
dominant~\cite{Paschos} and  get resonantly enhanced~\cite{APRD}, when
a pair of heavy Majorana neutrinos has a mass difference comparable to
the  heavy  neutrino   decay  widths.   In~\cite{PU},  this  dynamical
mechanism   was  termed  {\it   resonant  leptogenesis}~(RL).    As  a
consequence of  RL, the  heavy Majorana  mass scale can  be as  low as
$\sim$~1~TeV~\cite{APRD,APreview} in complete agreement with the solar
and atmospheric neutrino data~\cite{PU}.

A crucial model-building aspect of  RL models is that such models have
to  rely  on  a   nearly  degenerate  heavy  neutrino  mass  spectrum.
Although,  without  any  additional  lepton-flavour symmetry,  such  a
requirement would  appear very fine-tuned, there is  no theoretical or
phenomenologically  compelling reason that  would prevent  the singlet
neutrino  sector   of  the  SM   from  possessing  such   a  symmetry.
Specifically,  the  RL   model  discussed  in~\cite{APRD},  which  was
motivated  by  E$_6$  unified   theories~\cite{MV},  was  based  on  a
particular lepton symmetry in  the heavy neutrino sector.  This lepton
symmetry  was broken  very approximately  by GUT-  and/or Planck-scale
suppressed operators of dimension  5 and higher. In~\cite{PU}, another
RL  scenario  was  put  forward based  on  the  Froggatt--Nielsen~(FN)
mechanism~\cite{FN}, where two of  the heavy neutrinos naturally had a
mass difference  comparable to their decay  widths.  Recently, several
constructions  of RL  models  appeared in  the  literature within  the
context  of supersymmetric theories~\cite{HMW,softL,DLR,GJN},  or even
embedded in SO(10) unified theories~\cite{AFS,AB}.

One of the great advantages of RL models is that their predictions for
the  BAU   are  almost  independent  of   the  primordial  $L$-number,
$B$-number and heavy neutrino abundances~\cite{APreview,PU}. This fact
may be explained as follows:  in RL scenarios, the $L$-violating decay
widths  of the heavy  Majorana neutrinos  can be  significantly larger
than the Hubble expansion rate $H$ of the Universe.  As a consequence,
the heavy Majorana neutrinos can rapidly thermalize from their decays,
inverse  decays and  scatterings with  the other  SM particles  in the
plasma,  even  if there  were  no  heavy  Majorana neutrinos  at  high
temperatures.   Moreover,   in  this  high   temperature  regime,  any
pre-existing lepton asymmetry  will rapidly be driven to  zero, due to
the $L$-violating  inverse decays  and scattering processes  which are
almost  in thermal  equilibrium.  As  the Universe  cools down,  a net
lepton asymmetry can  be created at temperatures just  below the heavy
neutrino  mass as  a  consequence of  the aforementioned  CP-violating
resonant enhancement that occurs in RL models. This $L$ asymmetry will
survive   wash-out   effects   and    will   be   converted   by   the
$(B+L)$-violating sphalerons into the observed~BAU.

In this paper we provide a detailed study of a new variant of RL where
a given single lepton flavour  asymmetry is resonantly produced by the
quasi-in-equilibrium   decays  of  heavy   Majorana  neutrinos   of  a
particular  family type.   Such a  variant of  RL was  first discussed
in~\cite{APtau},  and for the  case of  the $\tau$-lepton  number this
mechanism       has        been       called       {\it       resonant
$\tau$-leptogenesis}~(R$\tau$L).   This  mechanism  makes use  of  the
property  that,  in addition  to  $B  -  L$, sphalerons  preserve  the
individual       quantum      numbers      $\frac{1}{3}       B      -
L_{e,\mu,\tau}$~\cite{KS,HT,DR,LS}.    In   a   R$\tau$L  model,   the
generated excess  in the  $L_\tau$ number will  be converted  into the
observed BAU, provided the $L_\tau$-violating reactions are not strong
enough to  wash out such  an excess.  

Although  our  focus  will  be  on  minimal  {\em  non-supersymmetric}
3-generation RL models, supersymmetry  could account for the origin of
the  electroweak-scale heavy Majorana  neutrinos.  In  particular, one
may  tie  the  singlet  Majorana  neutrino mass  scale  $m_N$  to  the
$\mu$-parameter through the vacuum expectation value (VEV) of a chiral
singlet superfield $\widehat{S}$~\cite{Borzumati}.  The proposed model
is a variant of  the so-called Next-to-Minimal Supersymmetric Standard
Model  (NMSSM)  and  is  described  by  the  following  superpotential
(summation over repeated indices implied):
\begin{equation}
W\ =\ W_{\rm MSSM} (\mu = 0)\ +\ h^{\nu_R}_{ij}\, \widehat{L}_i
\widehat{H}_2 \widehat{\nu}_{jR}\ +\ \lambda\, \widehat{S} \widehat{H}_1
\widehat{H}_2 \ +\ \frac{\rho}{2}\, \widehat{S}\,
\widehat{\nu}_{iR}\widehat{\nu}_{iR}\ +\ \frac{\kappa}{3}\,
\widehat{S}^3\ ,
\end{equation}
where $W_{\rm MSSM} (\mu = 0)$ is the superpotential of the well-known
Minimal Supersymmetric  Standard Model (MSSM)  without the $\mu$-term,
and        $\widehat{H}_{1,2}$,        $\widehat{L}_{1,2,3}$       and
$\widehat{\nu}_{1,2,3\,R}$  are the Higgs-doublet,  lepton-doublet and
right-handed  neutrino  superfields,  respectively.  Once  the  scalar
component  of  $\widehat{S}$  develops  a  VEV $v_S$,  then  both  the
would-be $\mu$-parameter, $\mu = \lambda v_S$, and the SO(3)-symmetric
singlet  scale, $m_N =  \frac{1}{2}\,\rho\, v_S$,  are expected  to be
comparable  in magnitude  (asumming  that $\lambda  \sim \rho$),  thus
providing a natural  framework for the possible existence  of 3 nearly
degenerate  electroweak-scale heavy  Majorana neutrinos~\cite{Anupam}.
In this  minimal extension  of the MSSM,  the predictions for  the BAU
will depend on  the size of the soft  SUSY-breaking mass scale $M_{\rm
SUSY}$.   However, if  $M_{\rm SUSY}$  is relatively  larger  than the
singlet  Majorana  neutrino   mass  scale  $m_N$,  e.g.~$M_{\rm  SUSY}
\stackrel{>}{{}_\sim} 2m_N$, the  dominant source of leptogenesis will
be the minimal non-supersymmetric sector that we are studying here, so
our predictions will remain almost unaffected in this case.

As mentioned above,  single lepton-flavour effects on the  net $L$ and
$B$ asymmetries play a key role in R$\tau$L models.  To properly treat
these as well  as SM chemical potential effects,  the relevant network
of the Boltzmann equations (BEs) needs to be extended consistently. In
particular, single  lepton-flavour effects can have  a dramatic impact
on the predictions  for the $B$ asymmetry.  These  predictions for the
BAU  can differ  by many  orders of  magnitude with  respect  to those
obtained in the  conventional BE formalism, which is  commonly used in
the literature.  Although  our primary interest will be  to analyze RL
models, we should stress that single lepton-flavour effects could also
significantly   affect  the   predictions  obtained   in  hierarchical
leptogenesis  scenarios.  The improved  set of  BEs derived  here will
therefore be of general use.

Another  important  question  we   wish  to  address  is  whether  the
leptogenesis scale can be lowered to energies 100--250~GeV, very close
to  the  critical  temperature  $T_c$,  where  the  electroweak  phase
transition occurs. In this temperature region, freeze-out effects from
sphaleron processes dropping out of equilibrium need to be considered,
as they  can significantly modify  the predicted values for  the final
baryon asymmetry. Our treatment  of these sphaleron freeze-out effects
will    be    approximate    and    based    on    the    calculations
of~\cite{CLMW,KS,LS}.  Our  approximate treatment is  motivated by the
fact that,  within the framework of  RL models, the creation  of a net
lepton  asymmetry  at  the  electroweak  scale does  not  require  the
electroweak phase transition to be strongly first order.

Most  importantly,  in models  where  the  BAU  is generated  from  an
individual    lepton-number   asymmetry,    a   range    of   testable
phenomenological  implications  may arise.   The  key  aspect is  that
scenarios such  as R$\tau$L can contain heavy  Majorana neutrinos with
appreciable Yukawa couplings to  electrons and muons.  The (normalized
to the SM)  $W^\pm$-boson couplings of $e$ and  $\mu$ leptons to these
heavy  Majorana  neutrinos  could  be  as  large  as  $10^{-2}$.   For
electroweak-scale heavy neutrinos,  such couplings would be sufficient
to  produce  these  particles  at  future  $e^+e^-$  and  $\mu^+\mu^-$
colliders.   Furthermore,  minimal  (non-supersymmetric)  3-generation
R$\tau$L  models  can  predict  $\mu  \to e\gamma$  and  $\mu  \to  e$
conversion in  nuclei at rates that  can be tested  by the foreseeable
experiments  MEG  at   PSI~\cite{MEG}  and  MECO  at  BNL~\cite{MECO},
respectively.  Finally, R$\tau$L  models naturally realize an inverted
hierarchy for  the light neutrino spectrum and  therefore also predict
neutrinoless  double  beta ($0\nu\beta\beta$)  decay  with a  sizeable
effective  neutrino mass  $|\langle m  \rangle |$,  as large  as $\sim
0.02$~eV.   This value  falls  within reach  of  proposals for  future
$0\nu\beta\beta$-decay experiments sensitive  to $|\langle m \rangle |
\sim  0.01$--0.05~eV~\cite{CA}, e.g.~CUORE  ($^{130}{\rm  Te}$), GERDA
($^{76}{\rm  Ge}$), EXO ($^{136}{\rm  Xe}$), MOON  ($^{100}{\rm Xe}$),
XMASS  ($^{136}{\rm   Xe}$),  CANDLES  ($^{48}{\rm   Ca}$),  SuperNEMO
($^{82}{\rm Se}$) etc.

Our paper has been organized  as follows: Section 2 presents a minimal
model for resonant $\tau$-leptogenesis. In Section 3 we derive the BEs
for  single  lepton flavours,  by  carefully  taking  into account  SM
chemical  potential  effects.   Technical  details pertinent  to  this
derivation have been relegated to  Appendix A.  In Section 4 we review
the  calculation  of  out  of  equilibrium sphaleron  effects  at  the
electroweak phase transition and apply it to leptogenesis.  In Section
5 we give several numerical  examples of R$\tau$L models, focusing our
attention  on scenarios  that can  be  tested at  future $e^+e^-$  and
$\mu^+\mu^-$ colliders and  in low-energy experiments.  In particular,
in  Section  6, we  present  predictions for  lepton-flavour-violating
(LFV) processes, such as $\mu \to e\gamma$, $\mu \to eee$ and $\mu \to
e$  conversion in  nuclei.  Finally, we  present  our conclusions  and
future prospects in Section 7.

\setcounter{equation}{0}
\section{Minimal Model for Resonant {\boldmath $\tau$}-Leptogenesis}

There   have   been   several    studies   on   RL   models   in   the
literature~\cite{APreview,LB,PU,GJN,softL,AB,HMW}.    Here,   we  will
focus our  attention on a  variant of resonant leptogenesis  where the
BAU  is   generated  by  the   production  of  an   individual  lepton
number~\cite{APtau}.    For  definiteness,   we  consider   a  minimal
(non-supersymmetric) model for R$\tau$L.

Let  us  start our discussion     by  briefly reviewing the   relevant
low-energy structure of the SM symmetrically extended with one singlet
neutrino  $\nu_{iR}$ per $i$  family  (with $i=1,2,3$).  The  leptonic
Yukawa and Majorana  sectors  of   such a  model  are  given  by the
Lagrangian
\begin{eqnarray}
  \label{Lym}
-\, {\cal L}_{\rm M,Y} &=& \frac{1}{2}\,\sum_{i,j=1}^3\,
\bigg(\, (\bar{\nu}_{iR})^C\, (M_S)_{ij}\, \nu_{jR}\ +\
\mbox{h.c.}\,\bigg)\nonumber\\
&&+\: \sum_{i=e,\mu,\tau}\, \bigg[\, \hat{h}^l_{ii}\,
\bar{L}_i\,\Phi\, l_{iR} \ +\
\bigg(\,\sum_{j=1}^3\, h^{\nu_R}_{ij}\,
\bar{L}_i\, \tilde{\Phi}\, \nu_{jR} \ +\ \mbox{h.c.}\bigg)\,\bigg]\,,
\end{eqnarray}
where    $L_i  = (\nu_{iL},  l_{iL})^T$  are    the left-handed lepton
doublets~\footnote[1]{Occasionally    we  will    also   denote the
  individual lepton numbers with $L_{e,\mu,\tau}$. This apparent
abuse of  notation  should cause no confusion  to  the reader, as  the
precise meaning  of $L_{e,\mu,\tau}$ can  be easily inferred  from the
context.},  $l_{iR}$ are the  right-handed leptons, and $\tilde{\Phi}$
is the isospin conjugate of the Higgs doublet $\Phi$.

In the  Lagrangian~(\ref{Lym}), we have defined  the individual lepton
numbers  $L_{e,\mu,\tau}$ in the  would-be charged-lepton  mass basis,
where  the charged-lepton  Yukawa matrix  $\hat{h}^l$ is  positive and
diagonal.  In fact,  without loss of generality, it  can be shown that
sphaleron transitions exhibit a U(3)  flavour symmetry and so they can
be rotated to become flavour diagonal in the same would-be mass basis.
To prove  this, one may  write the operator $O_{B+L}$  responsible for
$B+L$-violating  sphaleron  transitions  as  follows  (group-invariant
contraction   of   the   colour    and   weak   degrees   of   freedom
implied)~\cite{HT}:
\begin{equation}
  \label{OBplusL}
O_{B+L}\ =\ \prod\limits_{i=1}^3\, Q'_i\, Q'_i\, Q'_i\, L'_i\; ,
\end{equation}
where $Q'_i$ and  $L'_i$ denote the quark and  lepton doublets defined
in an arbitrary weak basis.  The operator $O_{B+L}$ contains the fully
antisymmetric     operator     combinations:    $Q'_1Q'_2Q'_3$     and
$L'_1L'_2L'_3$,    which   are    invariant    under   U(3)    flavour
rotations~\cite{GGR}.  Thus,  we can use  this U(3)-rotational freedom
to render the charged lepton and up-quark sectors flavour diagonal and
positive.

To obtain a phenomenologically relevant model, at least 3 singlet heavy
Majorana neutrinos $\nu_{1,2,3\,R}$ are needed and these have to be nearly
degenerate in mass.  To ensure the latter, we assume that to leading order,
the heavy neutrino sector is SO(3) symmetric, i.e.
\begin{equation}
  \label{MSSO3}   
M_S\ =\ m_N\, {\bf 1}_3\: +\: \Delta M_S\; , 
\end{equation}
where ${\bf 1}_3$ is the $3\times 3$ identity matrix and $\Delta M_S$ is a
general SO(3)-breaking matrix.  As we will discuss below, compatibility with
the observed light neutrino masses and mixings requires that $(\Delta
M_S)_{ij}/m_N \stackrel{<}{{}_\sim } 10^{-7}$, for electroweak-scale heavy
Majorana neutrinos, i.e.~for $m_N \approx 0.1$--1~TeV.  One could imagine that
the soft SO(3)-breaking matrix $\Delta M_S$ originates from a sort of
Froggatt--Nielsen mechanism~\cite{FN}.

In order to account for the smallness of the light neutrino masses, we
require that the neutrino  Yukawa sector possesses a leptonic U(1)$_l$
symmetry.  This  will explicitly break   the  imposed SO(3) symmetry of
the heavy  neutrino  sector to a   particular  subgroup SO(2) $\simeq$
U(1)$_l$.  For example, one  possibility relevant to R$\tau$L  is
to   couple  all  lepton  doublets  to   a particular  heavy  neutrino
combination: $\frac{1}{\sqrt{2}}\,   (\nu_{2R}  +  i  \nu_{3R})$.   In
detail, the U(1)$_l$ charges of the fields are
\begin{equation}
Q (L_{i})\    =\  Q(l_{iR}) = 1\,,\quad
Q\bigg(\frac{\nu_{2R}  +  i  \nu_{3R}}{\sqrt{2}}\bigg)\ =\ -Q\bigg(
\frac{\nu_{2R}   - i  \nu_{3R}}{\sqrt{2}}\bigg)\  =\
1\,,\quad Q(\nu_{1R}) = 0\;.
\end{equation}
As a result   of the U(1)$_l$ symmetry,   the matrix for the  neutrino
Yukawa couplings reads:
\begin{equation}
 \label{hmatrix}
h^{\nu_R} \ =\ \left(\! \begin{array}{ccc}
 0  & a\, e^{-i\pi/4}  & a\, e^{i\pi/4} \\
 0  & b\, e^{-i\pi/4}  & b\, e^{i\pi/4} \\
 0  & c\, e^{-i\pi/4}  & c\, e^{i\pi/4} \end{array} \!\right)\ +\
\delta h^{\nu_R}\; .
\end{equation}
In the  above, $a,b$ and $c$ are  arbitrary complex parameters of the model.
For  electroweak-scale heavy neutrinos, the  absolute  value of these
parameters  has   to     be   smaller   than   about~$10^{-2}$,   for
phenomenological reasons to  be discussed below and in  Section 6.  In
particular, the   requirement that an excess in  $L_\tau$ is protected from
wash-out   effects leads  to the relatively   stronger constraint $|c|
\stackrel{<}{{}_\sim} 10^{-5}$.   In addition, $\delta h^{\nu_R}$ is a
$3\times   3$ matrix that   parameterizes possible  violations of  the
U(1)$_l$ symmetry.  It should be noted that  the charged lepton  sector
and the leading SO(3)-invariant form of the heavy neutrino mass matrix
are compatible with the U(1)$_l$ symmetry.

In  this  paper  we shall  not  address  the  possible origin  of  the
smallness  of  the SO(3)-  and  U(1)$_l$-breaking parameters  $(\Delta
M_S)_{ij}$ and  $\delta h^{\nu_R}_{ij}$,  as there are  many different
possibilities  that could  be  considered for  this purpose,  e.g.~the
Froggatt--Nielsen   mechanism~\cite{PU,FN},   Planck-   or   GUT-scale
lepton-number breaking~\cite{APRD,MV}.  Instead, in our model-building
we  will  require that  the  symmetry  breaking  terms do  not  induce
radiative effects much larger than the tree-level contributions.  This
naturalness condition will be applied  to the light and heavy neutrino
mass matrices~${\bf m}^\nu$ and $M_S$, respectively.

We  start by observing  that the  U(1)$_l$  symmetry is  sufficient to
ensure the vanishing  of    the  light neutrino mass   matrix    ${\bf
m}^\nu$~\cite{BGL}.  In fact, if U(1)$_l$ is  an exact symmetry of the
theory, the light  neutrino mass matrix will vanish  to all orders  in
perturbation theory~\cite{AZPC,APmix}.    To   leading  order in   the
U(1)$_l$-breaking    parameters $\Delta  M_S$,   the  tree-level light
neutrino mass matrix ${\bf m}^\nu$ is given by
\begin{equation}
  \label{mnutree0}
{\bf m}^\nu\ =\ -\; \frac{v^2}{2}\, h^{\nu_R}\, M^{-1}_S
(h^{\nu_R})^T\ =\ \frac{v^2}{2m_N}\, \bigg(\,
\frac{h^{\nu_R}\Delta M_S \,(h^{\nu_R})^T}{m_N}\ -\ 
h^{\nu_R}\,(h^{\nu_R})^T\,\bigg)\,,
\end{equation}
where $v = 2M_W/g_w = 245$~GeV  is the vacuum expectation value of the
SM Higgs  field $\Phi$.  As a  minimal departure from  U(1)$_l$ in the
neutrino  Yukawa sector, we  consider that  this leptonic  symmetry is
broken only by $\nu_{1R}$, through
\begin{equation}
  \label{epshnu}
\delta h^{\nu_R}\ =\ \left(\! \begin{array}{ccc}
 \varepsilon_e  & 0 & 0 \\
 \varepsilon_\mu  & 0  & 0 \\
 \varepsilon_\tau  & 0  & 0 \end{array} \!\right)\; .
\end{equation}
In    this       case, the     tree-level     light   neutrino    mass
matrix~(\ref{mnutree0}) takes on the form
\begin{equation}
  \label{mnutree}
{\bf m}^\nu\ =\ \frac{v^2}{2m_N}\,\left(\! \begin{array}{ccc}
 \frac{\Delta m_N}{m_N}\,a^2 -\varepsilon^2_e  & 
\frac{\Delta m_N}{m_N}\,ab - \varepsilon_e\varepsilon_\mu & 
\frac{\Delta m_N}{m_N}\,ac -\varepsilon_e\varepsilon_\tau \\
\frac{\Delta m_N}{m_N}\,ab -\varepsilon_e\varepsilon_\mu  & 
\frac{\Delta m_N}{m_N}\,b^2-\varepsilon^2_\mu  & 
\frac{\Delta m_N}{m_N}\,bc-\varepsilon_\mu\varepsilon_\tau  \\
\frac{\Delta m_N}{m_N}\,ac -\varepsilon_e\varepsilon_\tau  & 
\frac{\Delta m_N}{m_N}\,bc -\varepsilon_\mu\varepsilon_\tau & 
\frac{\Delta m_N}{m_N}\,c^2-\varepsilon^2_\tau \end{array} \!\right)\; ,
\end{equation}
where $\Delta m_N = 2 (\Delta M_S)_{23} + i [(\Delta M_S)_{33} -
(\Delta M_S)_{22}]$. It is interesting to notice that in this type of
U(1)$_l$ breaking, the parameters $\varepsilon_{e,\mu,\tau}$ enter the
tree-level light neutrino mass matrix ${\bf m}^\nu$ quadratically.  As
a consequence, one finds that for $m_N \sim v$, these
U(1)$_l$-breaking parameters need not be much smaller than the
electron Yukawa coupling $h_e \sim 10^{-6}$.  Moreover, one should
observe that only a particular combination of soft SO(3)- and
U(1)$_l$-breaking terms $(\Delta M_S)_{ij}$ appears in ${\bf m}^\nu$
through $\Delta m_N$.  Nevertheless, for electroweak-scale heavy
neutrinos with mass differences $|\Delta m_N|/m_N
\stackrel{<}{{}_\sim} 10^{-7}$, one should have $|a|,\, |b|
\stackrel{<}{{}_\sim} 10^{-2}$ to avoid getting too large light
neutrino masses much above 0.5~eV.  As we will see more explicitly in
Section~\ref{sec:num}, for the R$\tau$L scenario under study, the
favoured solution will be an inverted hierarchical neutrino mass
spectrum with large $\nu_e\nu_\mu$ and $\nu_\mu\nu_\tau$
mixings~\cite{PDG}.

%******************************************************************
%%% Figure 1 
%******************************************************************
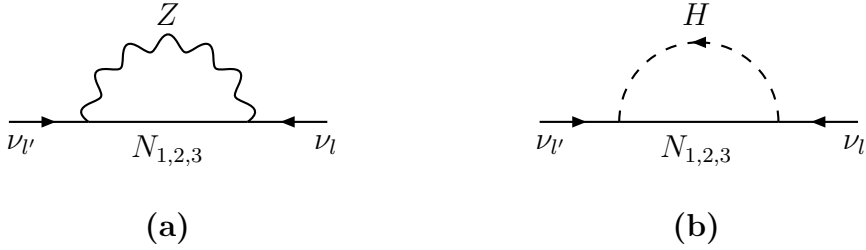
\begin{figure}[t]
\begin{center}
\begin{picture}(350,100)(0,0)
\SetWidth{0.8}

\ArrowLine(0,50)(30,50)\Line(30,50)(90,50)\ArrowLine(120,50)(90,50)
\PhotonArc(60,50)(30,0,180){3}{6.5}
\Text(5,45)[t]{$\nu_{l'}$}\Text(60,45)[t]{$N_{1,2,3}$}\Text(120,45)[t]{$\nu_l$}
\Text(60,87)[b]{$Z$}

\Text(60,10)[]{\bf (a)}

\ArrowLine(200,50)(230,50)\Line(230,50)(290,50)
\ArrowLine(320,50)(290,50)\DashArrowArc(260,50)(30,0,180){4}
\Text(205,45)[t]{$\nu_{l'}$}\Text(260,45)[t]{$N_{1,2,3}$}
\Text(320,45)[t]{$\nu_l$}
\Text(260,87)[b]{$H$}

\Text(260,10)[]{\bf (b)}

\end{picture}
\end{center}
\caption{\it Finite radiative effects contributing to the
light-neutrino mass matrix.}\label{fig:rad}
\end{figure}

In addition to the tree  level contributions given in~(\ref{mnutree}),
there   are  $Z$-   and   Higgs-boson-mediated threshold contributions
$\delta {\bf  m}^\nu$ to ${\bf m}^{\nu}$~\cite{AZPC}. The contributing
graphs are displayed  in  Fig.~\ref{fig:rad}.  In the  heavy  neutrino
mass  basis, where  $M_S \equiv {\rm  diag} (m_{N_1}\,,\, m_{N_2}\,,\,
m_{N_3})$,  with  $m_{N_1}\leq   m_{N_2}\leq m_{N_3}$, and  $h^{\nu_R}
\equiv h^\nu$, these finite radiative  corrections may conveniently be
expressed as follows~\cite{APmix}:
\begin{eqnarray}
  \label{mnurad0} 
(\delta {\bf m}^\nu)_{ll'} &=&
-\,\frac{\alpha_w}{32\pi}\, 
\sum_{\alpha = 1,2,3}\, 
\frac{h^\nu_{l\alpha}\, h^\nu_{l'\alpha}\, v^2}{m_{N_\alpha}}\; \bigg[\,
\frac{3 M^2_Z}{M^2_W}\, \bigg( B_0 (0,m^2_{N_\alpha}, M^2_Z)\: -\:
B_0(0,0,M^2_Z) \bigg)\nonumber\\ 
&&+\ \frac{m^2_{N_\alpha}}{M^2_Z}\, \bigg( B_0 (0,m^2_{N_\alpha}, M^2_H)\:
-\: B_0(0,m^2_{N_\alpha},m^2_{N_\alpha})\bigg)\, \bigg]\; ,
\end{eqnarray}
where  $\alpha_w =  g_w^2/(4\pi)$  and  $M_H$ is  the  SM Higgs  boson
mass.   In~(\ref{mnurad0}),  $B_0    (0,m^2_1,m^2_2)$  is  the   usual
Pasarino--Veltman one-loop function~\cite{PV}, i.e.
\begin{equation}
  \label{B0}
B_0 (0,m^2_1,m^2_2)\ =\ C_{\rm UV}\: +\: 1\: -\: 
\ln\bigg(\frac{m_1m_2}{\mu^2}\bigg)\ +\
\frac{m^2_1 + m^2_2}{m^2_1 - m^2_2}\, \ln\bigg(\frac{m_2}{m_1}\bigg)\, ,
\end{equation}
and   $C_{\rm    UV}$ is  a   UV   infinite   constant.  Moreover, in
writing~(\ref{mnurad0}),  we have     neglected   terms  of      order
$[(h^\nu_{l\alpha})^4\,v^3]/m^2_{N_\alpha}$,  which  are suppressed by
higher powers   of  the small  Yukawa  couplings.    It can  easily be
verified    that   the radiative lepton-number-violating  contribution
$\delta {\bf m}^\nu$ to  the light neutrino  mass matrix is  UV finite
and $\mu$-scale  independent.  For    $m^2_{N_\alpha} \gg M^2_H$   and
$(m_{N_\alpha}-m_{N_1})/m_{N_1} \ll 1$, the expression~(\ref{mnurad0})
evaluated in the original weak basis simplifies to
\begin{equation}
  \label{mnurad1} 
\delta {\bf m}^\nu \ =\
\frac{\alpha_w}{16\pi}\, \frac{M^2_H + 3 M^2_Z}{M^2_W}\;
\frac{v^2}{m_N}\;
\frac{h^{\nu_R}\Delta M_S \,(h^{\nu_R})^T}{m_N}\; .
\end{equation}
For electroweak-scale  heavy Majorana  neutrinos $m_{N_\alpha} \sim v$
and    $M_H    =  120$--200~GeV,   one     may   estimate    that  for
$(m_{N_\alpha}-m_{N_1})/m_{N_1} \stackrel{<}{{}_\sim}   10^{-7}$   and
$|a|,|b| \stackrel{<}{{}_\sim} 10^{-2}$,  the finite radiative effects
$\delta  {\bf  m}^\nu$ stay well  below 0.01~eV.   In  fact,  up to an
overall coupling-suppressed  constant, these corrections have the same
analytic form as the first term on the  RHS of~(\ref{mnutree0}).  They
can be    absorbed by  appropriately   rescaling $\Delta  m_N$ defined
after~(\ref{mnutree}).  As  a  consequence,   these  finite  radiative
effects do  not modify  the parametric  dependence of   the tree-level
light neutrino mass matrix given in~(\ref{mnutree}).

We now turn  our attention to the  heavy Majorana neutrino sector.  In
this        case,          renormalization-group   (RG)        running
effects~\cite{GJN,Antusch} become very significant. These effects   explicitly
break the SO(3)-symmetric form of the heavy neutrino mass matrix, $M_S
(M_X) = m_N\,  {\bf 1}_3$,  imposed  at some high  energy scale $M_X$,
e.g.~at the GUT  scale.  A fairly  good quantitative estimate of these
SO(3)-breaking effects can be obtained  by solving the RG equation for
the heavy neutrino mass matrix $M_S$:
\begin{equation}
  \label{RGMS}
\frac{d M_S}{dt}\ =\ -\, \frac{1}{16\pi^2}\,\bigg\{\,
\Big[\,(h^{\nu_R})^\dagger h^{\nu_R}\,\Big]\,M_S\: +\: 
M_S\, \Big[\,(h^{\nu_R})^T (h^{\nu_R})^*\Big]\,\bigg\}\; ,
\end{equation}
with $t = \ln ( M_X/\mu )$.  Considering that $h^{\nu_R}$ has
only a mild RG-scale  dependence and assuming that  $M_S (M_X) = m_N\,
{\bf 1}_3$ at some high  scale $M_X$, we may  calculate the RG effects by running
from $M_X$ to the low-energy scale $m_N \sim v$ through the relation
\begin{eqnarray}
  \label{RGestimate}
M_S(m_N) & = & M_S(M_X)\: -\: \frac{m_N}{8\pi^2}\,
 {\rm Re}\, \Big[\,(h^{\nu_R})^\dagger
h^{\nu_R}\,\Big]\, \ln\bigg(\frac{M_X}{m_N}\bigg)\nonumber\\
&&\hspace{-2cm}
=\  M_S(M_X)\: -\: \frac{|a|^2 + |b|^2}{8\pi^2}\, m_N\,
\ln\bigg(\frac{M_X}{m_N}\bigg)\, \bigg[\, {\rm diag}\, (0,1,1) \ +\ {\cal O}
\bigg(\frac{|\varepsilon_{e,\mu,\tau}|}{(|a|^2 + |b|^2)^{1/2}}
\bigg)\,\bigg]\,.\qquad\quad
\end{eqnarray}
If the scale $M_X$ of the  SO(3) symmetry imposed on $M_S (M_X)$ is to
be naturally associated with  the scale $M_{\rm GUT} \sim 10^{16}$~GeV
relevant to GUT dynamics,  it can be estimated from~(\ref{RGestimate})
that  the  mass  splittings  $|m_{N_2}-m_{N_1}|/m_N$ and  $|m_{N_3}  -
m_{N_1}|/m_N$  should be  larger than  $10^{-5}$ for  $|a|,\  |b| \sim
10^{-2}$    ($|c|,\,|\varepsilon_{e,\mu,\tau}|   \stackrel{<}{{}_\sim}
10^{-5}$).   Instead, the  mass difference  $|m_{N_3}  - m_{N_2}|/m_N$
should  be  comparatively much  smaller,  as  it  is protected  by  an
approximate U(1)$_l$ symmetry.  In  particular, we find that $|m_{N_3}
- m_{N_2}|/m_N       =      {\cal      O}(|\varepsilon_{e,\mu,\tau}a|,
|\varepsilon_{e,\mu,\tau}b|) \stackrel{<}{{}_\sim}  10^{-7}$.  At this
point we  should stress that in  the scenarios we  are considering, RG
effects  predominantly modify  the entries  $(\Delta  M_S)_{1i}$ (with
$i=1,2,3$)  in~(\ref{MSSO3})  and so  they  do  not  affect the  light
neutrino  mass  matrix~(\ref{mnutree}).   However, these  effects  may
affect the single lepton flavour asymmetries and the flavour-dependent
wash-out factors that are discussed in the next section.

In addition to RG effects, one might  worry that thermal effects could
significantly modify  the  heavy neutrino   mass  spectrum.   However,
thermal effects respect the  underlying symmetries of the theory, such
as global, chiral and  gauge symmetries~\cite{MBellac}.  Hence,  their
impact on the heavy  neutrino mass spectrum  is controlled by the size
of the SO(3)- and U(1)$_l$-breaking parameters  in the Yukawa neutrino
sector.       In        the     hard       thermal   loop        (HTL)
approximation~\cite{MBellac,Weldon},  thermal corrections give rise to
an effective heavy neutrino mass matrix  $M_S (T)$, which differs from
the one evaluated at $T=0$ by an amount~\cite{PU}
\begin{equation}
  \label{MSthermal}
M_S (T)\: -\: M_S (0)\ \approx\ \frac{1}{16}\: {\rm Re}\,
\Big[\,(h^{\nu_R})^\dagger h^{\nu_R}\,\Big]\: \frac{T^2}{m_N}\ .
\end{equation} 
By comparing~(\ref{MSthermal}) with (\ref{RGestimate}), we notice that
thermal corrections have a  parametric dependence very similar  to the
RG  effects  and  are    opposite  in sign.   Nevertheless,   if   the
SO(3)-breaking   scale  $M_X$  is identified   with $M_{\rm  GUT}$, RG
effects become larger than thermal effects by at least a factor 3, for
the temperature regime relevant to  leptogenesis $T\stackrel{<}{{}_\sim }
m_N$.

In Section~\ref{sec:num}  we will  present numerical estimates  of the
BAU  for  electroweak-scale  RL  models  that  are  motivated  by  the
naturalness of the light and  heavy neutrino sectors.  As we mentioned
above, this  condition provides a potential link  between these models
and GUT-scale physics.

\setcounter{equation}{0}
\section{Boltzmann Equations for Single Lepton Flavours}

In this section we derive a set of coupled BEs for the abundances of heavy
Majorana neutrinos and each lepton flavour. We follow a procedure analogous to
the one presented in \cite{PU}, where a number of controllable approximations
were made.  In particular, we assume Maxwell-Boltzmann statistics for the
heavy Majorana neutrinos. For the SM particles, we instead consider the proper
Bose--Einstein and Fermi--Dirac statistics, but ignore
condensate effects~\cite{KW}.  The above simplifications are expected to
introduce an error no larger than~20\%.  Furthermore, we neglect thermal
effects on the collision terms, which become less significant in the
temperature domain $T \stackrel{<}{{}_\sim} m_{N_1}$ relevant to RL.  As we
will see more explicitly in Section~\ref{sec:num}, the latter approximation
may partially be justified by the observation~\cite{PU} that the resulting BAU
predicted in RL models is independent of the initial abundances of the heavy
neutrinos and any initial baryon or lepton asymmetry.

Various definitions and notations that will be useful in deriving the BEs are
introduced in Appendix A.  Adopting the formalism of \cite{KW,MAL}, the
evolution of the number density, $n_a$, of all particle species $a$ can be
modelled by a set of BEs.  These are coupled first order differential
equations and may be generically written down as\footnote[2]{This formalism
  neglects coherent time-oscillatory terms describing particle oscillations in
  terms of number densities, as well as off-diagonal number densities
  $n_{a\bar{b}}$, for the destruction of a particle species $b$ and the
  correlated creation of a particle species $a$, where $a$ and $b$ could
  represent the 3 lepton flavours or the 3 heavy neutrinos~$N_{1,2,3}$.
  Although these effects can be modelled as well~\cite{Keldysh/Schwinger},
  their impact on the BAU is expected to be negligible.  Specifically,
  coherent time-oscillatory terms between heavy Majorana neutrinos will
  rapidly undergo strong damping, as a consequence of the quasi-in-thermal
  equilibrium dynamics governing RL models. This results from the fact
  that the decay widths $\Gamma_{N_{1,2,3}}$ of the heavy neutrinos are much
  larger than the expansion rate of the Universe.  Additionally, the
  correlated off-diagonal number densities $n_{a\bar{b}}$ will be
  Yukawa-coupling suppressed ${\cal O} ((h^\nu)^2)$ with respect to the
  diagonal ones $n_{a,b}$, if the heavy neutrinos and the charged leptons are
  defined in the diagonal mass basis.  In particular, the contribution of
  $n_{a\bar{b}}$ to $n_{a,b}$ will be further suppressed ${\cal O}
  ((h^\nu)^4)$.  We will therefore neglect the effects of the coherent
  time-oscillatory terms and the off-diagonal number densities~$n_{a\bar{b}}$
  on the~BEs.}
\begin{equation}
  \label{BEgeneric}
\frac{dn_a}{dt}\: +\: 3 H n_a\ =\ -\, \sum\limits_{aX^\prime\leftrightarrow
  Y}\,\bigg[\ \frac{n_a n_{X^\prime}}{n^{\rm eq}_a 
  n^{\rm eq}_{X^\prime}}\,\gamma (a X^\prime \to Y)\ -\ 
\frac{n_Y}{n^{\rm eq}_Y}\, \gamma (Y\to a X^\prime )\ \bigg]\; ,
\end{equation}
where all possible reactions of the form $a X^\prime \to Y$ or $Y\to a
X^\prime$, in  which $a$ can   be  created or annihilated are   summed
over. If $a$ is unstable, it could occur  as a real intermediate state
(RIS)  in a resonant  process like  $X  \to a \to  Y$.   In this case,
special treatment is required to avoid overcounting processes.

In principle, there is a  large number of   coupled BEs, one for  each
particle degree of freedom.  This number can be drastically reduced by
noting that rapidly occurring SM processes  hold most of the different
particle   degrees of   freedom and    particle   species in   thermal
equilibrium.  The non-zero chemical potentials of the particle species
other than  heavy Majorana  neutrinos  and leptons produce  effects of
$\mathcal{O}(1)$ on the  final baryon asymmetry~\cite{MPspect}.  These
effects   will be  consistently included  in    the BEs for the  heavy
Majorana neutrinos    $N_{1,2,3}$     and   the   lepton      doublets
$L_{e,\mu,\tau}$.

Although an infinite series of collision  terms could be added to each
BE,  only  a few will have   a significant contribution. Following the
procedure in  \cite{PU}, terms of order $\bar{h}^{\nu\,4}_{\pm} h_u^2$
and higher will be neglected, where $\bar{h}^{\nu}_{\pm} \sim h^{\nu}$
are the  one loop resummed   effective Yukawa couplings  introduced in
\cite{PU}.  Also neglected are terms of order $\bar{h}^{\nu\,4}_{\pm}$
for $2 \leftrightarrow 2$ scatterings with two external heavy Majorana
neutrinos. This leaves $1 \leftrightarrow 2$ decays and inverse decays
of heavy Majorana neutrinos ${\cal O}(\bar{h}^{\nu \,2}_{\pm})$ and $2
\leftrightarrow  2$  scatterings  between   heavy  Majorana neutrinos,
lepton doublets, gauge bosons, quarks  and the Higgs field, which  are
formally          of        order      $\bar{h}^{\nu\,2}_{\pm}   g^2$,
$\bar{h}_{\pm}^{\nu\,2}  g^{\prime\,2}$  and  $(\bar{h}^{\nu}_{\pm})^2
h_u^2$.

An  important  step    in the  following   derivation   is  the proper
subtraction  of    RISs.      For   example,     the  process     $L_j
\Phi~\leftrightarrow~L_k^C\Phi^\dagger$ will contain real intermediate
heavy Majorana neutrino  states.  Their  inverse decay and  subsequent
decay   have  already been   accounted  for in the   BEs   and must be
subtracted to ensure that unitarity and CPT are respected~\cite{KW}.

In analogy to  $2\leftrightarrow 2$ scatterings,  $2 \to 3$ processes,
such  as $L_j Q^C~\to~L_k^C   \Phi^\dagger u^C$, may  also contain the
heavy  neutrinos $N_\alpha$ as  RISs.    The resonant part  of  such a
process consists of the  reaction $L_j Q^C~\to~N_\alpha u^C$, followed
by the  decay $N_\alpha~\to~L_k^C \Phi^\dagger$.   As before, to avoid
double counting, we subtract the RISs  from such a  $2 \to 3$ process.
Although the off-shell $2\to 3$ process is  a higher order effect than
those we  are considering,  the  subtracted resonant  part contributes
terms of order $(\bar{h}^\nu_{\pm})^2  h_u^2$ and must be consistently
included  within the given  approximations for the BEs.  Specifically,
the following relations among the collision terms are derived:
\begin{eqnarray}
\label{CPviolating}
\gamma^{\,\prime}(L^C_k \Phi^\dagger \to L_j \Phi) -
\gamma^{\,\prime}(L_k \Phi \to L^C_j \Phi^\dagger) & = &
-\,\frac{1}{2} \sum_{\alpha=1}^3 \bigg(B_{N_\alpha}^k
\delta_{N_\alpha}^j + B_{N_\alpha}^j \delta_{N_\alpha}^k \bigg)
\sum_{l\,=\,e,\mu,\tau} \!\!
\gamma^{N_{\alpha}}_{L_l \Phi}\,,\nonumber\\
\gamma^{\,\prime}(L_k \Phi \to L_j \Phi) -
\gamma^{\,\prime}(L^C_k \Phi^\dagger \to L^C_j \Phi^\dagger) & = &
-\,\frac{1}{2} \sum_{\alpha=1}^3 \bigg(B_{N_\alpha}^k
\delta_{N_\alpha}^j - B_{N_\alpha}^j \delta_{N_\alpha}^k \bigg)
\sum_{l\,=\,e,\mu,\tau} \!\!
\gamma^{N_{\alpha}}_{L_l \Phi}\,,\nonumber\\
\gamma^{\,\prime}(Q u^C \to L_j L_k \Phi) - 
\gamma^{\,\prime}(Q^C u \to L^C_j L^C_k \Phi^\dagger) & = &\!\!
-\,S_{jk} \sum_{\alpha=1}^3 \bigg(B_{N_\alpha}^k
\delta_{N_\alpha}^j + B_{N_\alpha}^j \delta_{N_\alpha}^k \bigg)
\sum_{l\,=\,e,\mu,\tau} \!\!
\gamma^{N_{\alpha} L_l}_{Q u^C}\,,\nonumber\\
\gamma^{\,\prime}(Q u^C \to L_j L^C_k \Phi^\dagger) - 
\gamma^{\,\prime}(Q^C u \to L^C_j L_k \Phi) & = &
-\,\frac{1}{2} \sum_{\alpha=1}^3 \bigg(B_{N_\alpha}^k
\delta_{N_\alpha}^j - B_{N_\alpha}^j \delta_{N_\alpha}^k \bigg)
\sum_{l\,=\,e,\mu,\tau} \!\!
\gamma^{N_{\alpha} L_l}_{Q u^C}\,,\nonumber\\
\gamma^{\,\prime}(L_j Q^C \to u^C \Phi^\dagger L^C_k) - 
\gamma^{\,\prime}(L^C_j Q \to u \Phi L_k) & = &
\frac{1}{2} \sum_{\alpha=1}^3 \bigg(B_{N_\alpha}^k
\delta_{N_\alpha}^j + B_{N_\alpha}^j \delta_{N_\alpha}^k \bigg)
\sum_{l\,=\,e,\mu,\tau} \!\!
\gamma^{N_{\alpha} u^C}_{L_l Q^C}\,,\nonumber\\
\gamma^{\,\prime}(L_j Q^C \to u^C \Phi L_k) - 
\gamma^{\,\prime}(L^C_j Q \to u \Phi^\dagger L^C_k) & = &
\frac{1}{2} \sum_{\alpha=1}^3 \bigg(B_{N_\alpha}^k
\delta_{N_\alpha}^j - B_{N_\alpha}^j \delta_{N_\alpha}^k \bigg)
\sum_{l\,=\,e,\mu,\tau} \!\!
\gamma^{N_{\alpha} u^C}_{L_l Q^C}\,,\nonumber\\
\end{eqnarray}
where a  prime denotes   subtraction  of RISs,    the indices $j,k   =
e,\mu,\tau$ label lepton  flavour, and $S_{jk} = (1+\delta_{jk})^{-1}$
is  a statistical  factor    that  corrects for  the     production or
annihilation  of identical   lepton  flavours.  In addition,   we have
defined the individual lepton-flavour asymmetries and branching ratios
as
\begin{eqnarray}
\delta_{N_\alpha}^l & = & \frac{\Gamma (N_{\alpha} \to L_l \Phi) -
\Gamma (N_{\alpha} \to L^C_l \Phi^\dagger)} {\sum_k \Big[ \Gamma
(N_{\alpha} \to L_k \Phi) + \Gamma (N_{\alpha} \to L^C_k
\Phi^\dagger)\Big]}\ ,\nonumber\\
B_{N_\alpha}^l & = & \frac{\Gamma (N_{\alpha} \to L_l \Phi) +
\Gamma (N_{\alpha} \to L^C_l \Phi^\dagger)} {\sum_k \Big[ \Gamma
(N_{\alpha} \to L_k \Phi) + \Gamma (N_{\alpha} \to L^C_k
\Phi^\dagger)\Big]}\ .
\label{CPbranch}
\end{eqnarray}
As CP  violation  in these  processes is  predominantly  caused by the
resonant exchange  of   heavy  Majorana  neutrinos,  the  CP-violating
collision  terms have been approximated in  terms of the CP-conserving
ones as follows:
\begin{eqnarray}
  \label{simpl} 
\delta \gamma^{N_{\alpha}}_{L_j\Phi} \!&=&\!
\delta^j_{N_{\alpha}} \sum_{l\,=\,e,\mu,\tau}\!\!\!
\gamma^{N_{\alpha}}_{L_l\Phi}\,,\qquad \delta \gamma^{N_{\alpha}
u^C}_{L_j Q^C}\ =\ \delta^j_{N_{\alpha}} \sum_{l\,=\,e,\mu,\tau}\!\!\!
\gamma^{N_{\alpha} u^C}_{L_l Q^C}\,,\nonumber\\
\delta \gamma^{N_{\alpha} L_j}_{Qu^C}\!&=&\! -\, \delta^j_{N_{\alpha}}
\sum_{l\,=\,e,\mu,\tau}\!\!\! \gamma^{N_{\alpha} L_l}_{Q u^C}\qquad
{\rm etc.}
\end{eqnarray}

Unlike the $2 \to 3$ reactions, $3 \to 2$ processes are treated differently.
Although $3 \to 2$ processes could contain real intermediate $N_{\alpha}$
states, collision terms for their associated annihilation processes have not
been included before.  For example, in the process $L_j L_k \Phi \to Q u^C$, a
real intermediate $N_{\alpha}$ state could be coherently created from $L$ and
$\Phi$ states.  This coherent RIS would then interact with another $L$ state
producing $Q$ and $u^C$.  Previously, the process $N_\alpha L \to Q u^C$ has
only been considered for heavy $N_{\alpha}$ neutrinos in a thermally
incoherent state.  This implies that $3 \to 2$ processes containing $N_\alpha$
as RISs have not yet been accounted for and should not be subtracted.  With
the help of CPT and unitarity, one may therefore obtain the following
relations for the $3 \to 2$ processes:
\begin{eqnarray}
  \label{CPT2}
\gamma (L_j L_k \Phi \to Q u^C )\: -\:
\gamma (L_j^C L_k^C \Phi^\dagger \to Q^C u) & = &
\mathcal{O}(h^{\nu\,4} h_u^2)\,,\nonumber\\
\gamma (L_j L^C_k \Phi^\dagger \to Q u^C)\: -\:
\gamma (L_j^C L_k \Phi \to Q^C u ) & = &
\mathcal{O}(h^{\nu\,4} h_u^2)\,,\nonumber\\
\gamma (L_j \Phi u \to L_k^C Q)\: -\: 
\gamma (L_j^C \Phi^\dagger u^C \to L_k Q^C) & = &
\mathcal{O}(h^{\nu\,4} h_u^2)\,,\nonumber\\
\gamma (L_j \Phi u^C \to L_k Q^C)\: -\:
\gamma (L^C_j \Phi^\dagger u \to L_k^C Q) & = &
\mathcal{O}(h^{\nu\,4} h_u^2)\,.
\end{eqnarray}
As a consequence of  this, $3 \to 2$  processes will contribute  extra
CP-conserving $2 \to 2$ collision terms, through the resonant exchange
of  real  intermediate $N_\alpha$ states.    Applying the narrow width
approximation, we find
\begin{eqnarray}
\label{CPconserving}
\gamma (L_j L_k \Phi \to Q u^C )\: +\: 
\gamma (L_j^C L_k^C \Phi^\dagger \to Q^C u) & = & S_{jk}
\sum_{\alpha=1}^3 \bigg(B_{N_\alpha}^j B_{N_\alpha}^k +
\delta_{N_\alpha}^j \delta_{N_\alpha}^k \bigg)
\sum_{l\,=\,e,\mu,\tau} \!\!
\gamma^{N_{\alpha} L_l}_{Q u^C}\,,\nonumber\\
\gamma (L_j L^C_k \Phi^\dagger \to Q u^C )\: +\: 
\gamma (L_j^C L_k \Phi \to Q^C u) & = &
\frac{1}{2} \sum_{\alpha=1}^3 \bigg(B_{N_\alpha}^j B_{N_\alpha}^k -
\delta_{N_\alpha}^j \delta_{N_\alpha}^k \bigg)
\sum_{l\,=\,e,\mu,\tau} \!\!
\gamma^{N_{\alpha} L_l}_{Q u^C}\,,\nonumber\\
\gamma (L_j \Phi u \to L_k^C Q)\: +\: 
\gamma (L_j^C \Phi^\dagger u^C \to L_k Q^C) & = &
\frac{1}{2} \sum_{\alpha=1}^3 \bigg(B_{N_\alpha}^j B_{N_\alpha}^k +
\delta_{N_\alpha}^j \delta_{N_\alpha}^k \bigg)
\sum_{l\,=\,e,\mu,\tau} \!\!
\gamma^{N_{\alpha} u^C}_{L_l Q^C}\,,\nonumber\\
\gamma (L_j \Phi u^C \to L_k Q^C)\: +\:
\gamma (L^C_j \Phi^\dagger u \to L_k^C Q) & = &
\frac{1}{2} \sum_{\alpha=1}^3 \bigg(B_{N_\alpha}^j B_{N_\alpha}^k -
\delta_{N_\alpha}^j \delta_{N_\alpha}^k \bigg)
\sum_{l\,=\,e,\mu,\tau} \!\!
\gamma^{N_{\alpha} u^C}_{L_l Q^C}\,.\nonumber\\
\end{eqnarray}

We may now employ (\ref{BEgeneric}) and write down the BEs in terms of
the number densities   of heavy Majorana  neutrinos $n_{N_\alpha}$ and
the lepton-doublet asymmetries $n_{\Delta L_{e,\mu,\tau}}$,
\begin{eqnarray}
  \label{BE1} 
\frac{dn_{N_{\alpha}}}{dt}\: +\: 3 H n_{N_{\alpha}}
\!\!&=&\!\! \bigg( 1 \: -\: \frac{n_{N_{\alpha}}}{n^{\rm
eq}_{N_{\alpha}}}\,\bigg) \sum_{k\,=\,e,\mu,\tau} \bigg(
\gamma^{N_{\alpha}}_{L_k\Phi}\: +\: \gamma^{N_{\alpha} L_k}_{Q u^C}\:
+\: \gamma^{N_{\alpha} u^C}_{L_k Q^C}\: +\: \gamma^{N_{\alpha} Q}_{L_k
u}\nonumber\\ 
\!\!&&\!\!+\, \gamma^{N_{\alpha} V_\mu}_{L_k\Phi} \: +\:
\gamma^{N_{\alpha} L_k}_{\Phi^\dagger V_\mu}\: +\:
\gamma^{N_{\alpha}\Phi^\dagger }_{L_k V_\mu}\, \bigg)\nonumber\\
\!\!&&\!\!- \sum_{k\,=\,e,\mu,\tau} \frac{n_{\Delta L_k}}{2\,n^{\rm
eq}_{l_k}}\, \bigg[\, \delta\gamma^{N_{\alpha}}_{L_k\Phi}\: +\:
\delta\gamma^{N_{\alpha} u^C}_{L_k Q^C}\: +\: \delta\gamma^{N_{\alpha}
Q}_{L_k u}\: +\: \delta\gamma^{N_{\alpha} V_\mu}_{L_k \Phi}\: +\:
\delta\gamma^{N_{\alpha}\Phi^\dagger }_{L_k V_\mu}\nonumber\\
\!\!&&\!\! +\: \frac{n_{N_{\alpha}}}{n^{\rm eq}_{N_{\alpha}}}\,
\bigg(\, \delta\gamma^{N_{\alpha} L_k}_{Q u^C}\: +\:
\delta\gamma^{N_{\alpha} L_k}_{\Phi^\dagger
V_\mu}\,\bigg)\,\bigg]\;,\\[3mm] 
  \label{BE2} 
\frac{dn_{\Delta L_j}}{dt}\: +\: 3 H n_{\Delta
L_j} \!\!&=&\!\! \sum^3_{\alpha = 1} \bigg(\,
\frac{n_{N_{\alpha}}}{n^{\rm eq}_{N_{\alpha}}} \: -\: 1\,\bigg)\,
\bigg( \delta\gamma^{N_{\alpha}}_{L_j \Phi}\: -\:
\delta\gamma^{N_{\alpha} L_j}_{Q u^C}\: +\: \delta\gamma^{N_{\alpha}
u^C}_{L_j Q^C}\: +\: \delta\gamma^{N_{\alpha} Q}_{L_j u}\nonumber\\
\!\!&&\!\!+\, \delta\gamma^{N_{\alpha} V_\mu}_{L_j \Phi}\: -\:
\delta\gamma^{N_{\alpha} L_j}_{\Phi^\dagger V_\mu}\: +\:
\delta\gamma^{N_{\alpha}\Phi^\dagger }_{L_j V_\mu}\, \bigg)\nonumber\\
\!\!&&\!\!-\, \frac{n_{\Delta L_j}}{2\,n^{\rm eq}_{l_j}}\, \bigg[\,
\sum^3_{\alpha = 1} \bigg(\,\gamma^{N_{\alpha}}_{L_j \Phi}\: +\: 2
\gamma^{N_{\alpha} u^C}_{L_j Q^C}\: +\: 2 \gamma^{N_{\alpha} Q}_{L_j
u} \: +\: 2 \gamma^{N_{\alpha} V_\mu}_{L_j \Phi}\: +\: 2
\gamma^{N_{\alpha}\Phi^\dagger }_{L_j V_\mu}\nonumber\\ \!\!&&\!\!
+\:2 \gamma^{N_{\alpha} L_j}_{Q u^C}\: +\: 2 \gamma^{N_{\alpha}
L_j}_{\Phi^\dagger V_\mu} \: +\: \frac{n_{N_{\alpha}}}{n^{\rm
eq}_{N_{\alpha}}}\, \bigg(\, \gamma^{N_{\alpha} L_j}_{Q u^C}\: +\:
\gamma^{N_{\alpha} L_j}_{\Phi^\dagger
V_\mu}\,\bigg)\,\bigg)\nonumber\\ \!\!&&\!\! +\!
\sum_{k\,=\,e,\mu,\tau} \bigg(\,\gamma^{\,\prime
L_j\Phi}_{\,L_k^C\Phi^\dagger} \: +\: \gamma^{L_j
L_k}_{\Phi^\dagger\Phi^\dagger} \: +\:
\gamma^{\,\prime\,L_j\Phi}_{\,L_k\Phi} \:+ \: \gamma^{L_j L^C_k}_{\Phi
\Phi^\dagger}\,\bigg)\,\bigg]\nonumber\\ \!\!&&\!\! -\!
\sum_{k\,=\,e,\mu,\tau} \frac{n_{\Delta L_k}}{2\,n^{\rm eq}_{l_k}}\,
\bigg[\, \gamma^{\,\prime L_k\Phi}_{\,L_j^C\Phi^\dagger} \: +\:
\gamma^{L_k L_j}_{\Phi^\dagger\Phi^\dagger} \: -\:
\gamma^{\,\prime\,L_k\Phi}_{\,L_j\Phi} \:- \: \gamma^{L_k L^C_j}_{\Phi
\Phi^\dagger}\nonumber\\ \!\!&&\!\!+ \: \sum^3_{\alpha = 1}
\delta^j_{N_{\alpha}} \delta^k_{N_{\alpha}} \sum_{l\,=\,e,\mu,\tau}
\bigg(\, \gamma^{N_{\alpha} u^C}_{L_l Q^C}\: +\: \gamma^{N_{\alpha}
Q}_{L_l u} \: +\: \gamma^{N_{\alpha} V_\mu}_{L_l \Phi}\: +\:
\gamma^{N_{\alpha}\Phi^\dagger }_{L_l V_\mu}\nonumber\\ \!\!&&\!\!+\:2
\gamma^{N_{\alpha} L_l}_{Q u^C}\: +\: 2 \gamma^{N_{\alpha}
L_l}_{\Phi^\dagger V_\mu}\,\bigg)\,\bigg]\,.
\end{eqnarray}
In the above set of BEs, we  have only kept  terms to leading order in
$n_{\Delta L_j}/n^{\rm eq}_{l_j}$, and implemented the relations given
in (\ref{CPviolating})--(\ref{CPconserving}).

All SM species in the thermal  bath, including  the lepton doublets
$L_{e,\mu,\tau}$, possess  non-zero  chemical     potentials.    These
chemical potentials can be   expressed in terms of  the lepton-doublet
chemical potentials  only, under the assumption  that SM processes are
in  full thermal  equilibrium   \cite{HT}.  This  analysis  yields the
following relations:
\begin{eqnarray}
  \label{ChemPot}
\mu_{V} \!&=&\! 0\,,\qquad
\mu_{\Phi}\ =\ \frac{4}{21} \sum_{l\,=\,e,\mu,\tau}\!\!\! \mu_{L_l}\,,
\qquad
\mu_{Q}\ =\ -\frac{1}{9} \sum_{l\,=\,e,\mu,\tau}\!\!\! \mu_{L_l}\,,\qquad 
\mu_{u} \ = \ \frac{5}{63} \sum_{l\,=\,e,\mu,\tau}\!\!\!
\mu_{L_l}\,,\nonumber\\ 
\mu_{e_l} \!&=&\! \mu_{L_l} - \frac{4}{21} \sum_{l\,=\,e,\mu,\tau}\!\!\!
\mu_{L_l}\;, 
\end{eqnarray}
where  $\mu_x$ denotes the  chemical potential  of a  particle species
$x$.   The  relations~(\ref{ChemPot}) can  be  used  to implement  the
effects  of the SM  chemical potentials  in the  BEs.  They  result in
corrections to  the so-called  wash-out terms in  both the  lepton and
heavy neutrino BEs. At this point  we should also note that the BEs in
their present form  are most accurate above $T_c$.   As $T$ approaches
$T_c$,  the assumption  that the  sphaleron processes  are  in thermal
equilibrium    becomes   less    valid.    This    will    result   in
$\mathcal{O}(v/T)$  corrections to  the relations  in (\ref{ChemPot}).
The inclusion of  the bulk of these corrections  will be considered in
the next section.

To numerically solve  the  BEs, it  proves convenient to  introduce  a
number  of new  variables.  In the  radiation  dominated epoch  of the
Universe relevant to baryogenesis, the  cosmic time $t$ is related  to
the temperature $T$ through
\begin{equation}
  \label{Tt}
t \ =\ \frac{z^2}{2\, H(z=1)}\ ,
\end{equation}
where
\begin{equation}
  \label{zeta}
z\ =\ \frac{m_{N_1}}{T}\ ,\qquad H(z)\ \approx\ 17.2\, \times\, 
\frac{m^2_{N_1}}{z^2\, M_{\rm Planck}}\ ,
\end{equation}
with $M_{\rm Planck} = 1.2\times 10^{19}$~GeV.  Also, we normalize the
number density of a particle species, $n_a$, to  the number density of
photons, $n_\gamma$, thereby defining the new parameter $\eta_a$,
\begin{equation}
  \label{etas}
\eta_a (z) \ =\ \frac{n_a (z)}{n_\gamma (z)}\ , 
\end{equation}
with
\begin{equation}
  \label{ngamma}
n_\gamma (z)\ =\ \frac{2\,T^3}{\pi^2}\ =\ 
               \frac{2\, m^3_{N_1}}{\pi^2}\,\frac{1}{z^3}\ .
\end{equation}
To allow the BEs  to be written in   a slightly more compact form,  we
will use the summation conventions
\begin{equation}
  \label{sumconv}
\gamma^{N_{\alpha}\,X}_{L\,Y}\ =\ 
\sum_{l\,=\,e,\mu,\tau} \!\!\!
\gamma^{N_{\alpha}\,X}_{L_l\,Y}\;, \qquad 
\eta_{\Delta L}\ =\ 
\sum_{l\,=\,e,\mu,\tau} \!\!\! \eta_{\Delta L_l}\;,
\end{equation}
where  $X$ and $Y$  stand for any  particle state other than $L_l$ and
$N_\alpha$.

Using  (\ref{BE1})--(\ref{sumconv}) and incorporating corrections  due
to the  SM chemical potentials, the  BEs for heavy  Majorana neutrinos
and lepton doublets are written down
\begin{eqnarray}
  \label{BEN} 
\frac{d \eta_{N_{\alpha}}}{dz} &=& \frac{z}{H(z=1)}\
\Bigg[\,\Bigg( 1 \: -\: \frac{\eta_{N_{\alpha}}}{\eta^{\rm eq}_{N_{\alpha}}}\,
\Bigg)\, \sum_{k\,=\,e,\mu,\tau} \bigg(\,
\Gamma^{D\; (\alpha k)} \: +\: \Gamma^{S\; (\alpha k)}_{\rm Yukawa}\: +\:
\Gamma^{S\; (\alpha k)}_{\rm Gauge}\, \bigg) \nonumber\\ &&-\,
\frac{2}{3}\, \sum_{k\,=\,e,\mu,\tau} \eta_{\Delta L_k}\,
\delta^{\,k}_{N_{\alpha}}\, \bigg(\,
\widehat{\Gamma}^{D\; (\alpha k)} \: +\: 
\widehat{\Gamma}^{S\; (\alpha k)}_{\rm Yukawa} \: +\:
\widehat{\Gamma}^{S\; (\alpha k)}_{\rm Gauge}\, \bigg)\,\Bigg]\,,\\[15mm] 
  \label{BEL} 
\frac{d \eta_{\Delta L_j}}{dz} &=& 
\frac{z}{H(z=1)}\, \Bigg\{\, \sum\limits_{\alpha=1}^3\,
\delta^{\,j}_{N_{\alpha}}\ \Bigg(
\frac{\eta_{N_{\alpha}}}{\eta^{\rm eq}_{N_{\alpha}}} \: -\: 1\,\Bigg)\, 
\sum_{k\,=\,e,\mu,\tau} \bigg(\,
\Gamma^{D\; (\alpha k)} \: +\: \Gamma^{S\; (\alpha k)}_{\rm Yukawa}\:
+\: \Gamma^{S\; (\alpha k)}_{\rm Gauge}\, \bigg) \nonumber\\ 
&&-\,\frac{2}{3}\, \eta_{\Delta L_j}\, \Bigg[\, \sum\limits_{\alpha=1}^3\,
B_{N_{\alpha}}^{\,j}\,
\bigg(\, \widetilde{\Gamma}^{D\; (\alpha j)} \: +\: 
\widetilde{\Gamma}^{S\;(\alpha j)}_{\rm Yukawa}\: +\: 
\widetilde{\Gamma}^{S\; (\alpha j)}_{\rm Gauge}\: +\: 
\Gamma^{W\; (\alpha j)}_{\rm Yukawa} + \Gamma^{W\;(\alpha j)}_{\rm Gauge}
\,\bigg)\nonumber\\
&& \qquad\qquad\,\,\;
\: +\: \sum_{k\,=\,e,\mu,\tau}\,\bigg(\,\Gamma^{\Delta L =2\:(j
k)}_{\rm Yukawa} \: +\:\Gamma^{\Delta L =0\:(j k)}_{\rm
Yukawa}\,\bigg) \Bigg] \nonumber\\
&&-\, \frac{2}{3}\,\sum_{k\,=\,e,\mu,\tau} \eta_{\Delta L_k}\,
\Bigg[\,\sum\limits_{\alpha=1}^3\,
\delta^{\,j}_{N_{\alpha}}\,\delta^{\,k}_{N_{\alpha}}
\bigg(\,\Gamma^{W\;(\alpha k)}_{\rm Yukawa}\:
+\: \Gamma^{W\; (\alpha k)}_{\rm Gauge}\,\bigg)\nonumber\\ 
&& \qquad\qquad\qquad\qquad\qquad\quad\,\, \: +\: \Gamma^{\,\Delta L =2\:
(k j)}_{\rm Yukawa} \: -\: \Gamma^{\,\Delta L =0\: (k j)}_{\rm Yukawa}
\Bigg]\,\Bigg\}\,,
\end{eqnarray}
where 
\begin{eqnarray}
  \label{GD}
\Gamma^{D\; (\alpha l)} & = & \frac{1}{n_\gamma}\
\gamma^{N_{\alpha}}_{L_l\Phi}\;, \\[3mm]
  \label{GDT}
\widehat{\Gamma}^{D\; (\alpha l)} & = &
\widetilde{\Gamma}^{D\; (\alpha l)} \,=\, 
\frac{1}{n_\gamma}\,
\Bigg(1+\frac{4}{21}\,\frac{\eta_{\Delta L}}{\eta_{\Delta L_l}} \Bigg)\,
\gamma^{N_{\alpha}}_{L\Phi}\;, \\[3mm]
  \label{GSY}
\Gamma^{S\; (\alpha l)}_{\rm Yukawa} & = & \frac{1}{n_\gamma}\
\bigg(\, \gamma^{N_{\alpha} L_l}_{Q u^C}\: +\:  \gamma^{N_{\alpha}
u^C}_{L_l Q^C}\:  +\: \gamma^{N_{\alpha} Q}_{L_l u}\, \bigg)\; ,\\[3mm]
  \label{GSYhat}
\widehat{\Gamma}^{S\;(\alpha l)}_{\rm Yukawa} &=& \frac{1}{n_\gamma}\
\Bigg[\,\Bigg(-\frac{\eta_{N_{\alpha}}}{\eta^{\rm eq}_{N_{\alpha}}}
+\frac{4}{21}\,\frac{\eta_{\Delta L}}{\eta_{\Delta L_l}}\Bigg)\,
\gamma^{N_{\alpha} L}_{Q u^C}\: +\:
\Bigg(1+\frac{1}{9}\,\frac{\eta_{\Delta L}}{\eta_{\Delta L_l}}
-\frac{5}{63}\,\frac{\eta_{N_{\alpha}}}{\eta^{\rm eq}_{N_{\alpha}}}\,
\frac{\eta_{\Delta L}}{\eta_{\Delta L_l}}\Bigg)\,
\gamma^{N_{\alpha} u^C}_{L Q^C}\nonumber\\[1mm]
& & +\:\Bigg(1+\frac{5}{63}\,\frac{\eta_{\Delta L}}{\eta_{\Delta L_l}}
-\frac{1}{9}\,\frac{\eta_{N_{\alpha}}}{\eta^{\rm eq}_{N_{\alpha}}}\,
\frac{\eta_{\Delta L}}{\eta_{\Delta L_l}}\Bigg)\, 
\gamma^{N_{\alpha} Q}_{L u}\, \Bigg]\;,\\[3mm]
  \label{GSYtilde}
\widetilde{\Gamma}^{S\;(\alpha l)}_{\rm Yukawa} &=& \frac{1}{n_\gamma}\
\Bigg[\,\Bigg(\frac{\eta_{N_{\alpha}}}{\eta^{\rm eq}_{N_{\alpha}}}
+\frac{4}{21}\,\frac{\eta_{\Delta L}}{\eta_{\Delta L_l}}\Bigg)\, 
\gamma^{N_{\alpha} L}_{Q u^C}\: +\:
\Bigg(1+\frac{1}{9}\,\frac{\eta_{\Delta L}}{\eta_{\Delta L_l}}
+\frac{5}{63}\,\frac{\eta_{N_{\alpha}}}{\eta^{\rm eq}_{N_{\alpha}}}\,
\frac{\eta_{\Delta L}}{\eta_{\Delta L_l}}\Bigg)\,
\gamma^{N_{\alpha} u^C}_{L Q^C}\nonumber\\[1mm]
& & +\: \Bigg(1+\frac{5}{63}\,\frac{\eta_{\Delta L}}{\eta_{\Delta L_l}}
+\frac{1}{9}\,\frac{\eta_{N_{\alpha}}}{\eta^{\rm eq}_{N_{\alpha}}}\,
\frac{\eta_{\Delta L}}{\eta_{\Delta L_l}}\Bigg)\,
\gamma^{N_{\alpha} Q}_{L u}\, \Bigg]\;,\\[3mm]
  \label{GSG}
\Gamma^{S\; (\alpha l)}_{\rm Gauge} & = & \frac{1}{n_\gamma}\ 
\bigg(\, \gamma^{N_{\alpha} L_l}_{\Phi^\dagger V_\mu}\: +\:
\gamma^{N_{\alpha} V_\mu}_{L_l\Phi}\: +\:
\gamma^{N_{\alpha}\Phi^\dagger }_{L_l V_\mu}\, \bigg)\;,\\[3mm]
 \label{GSGhat}
\widehat{\Gamma}^{S\; (\alpha l)}_{\rm Gauge} &=& \frac{1}{n_\gamma}\ 
\Bigg[\,\Bigg(-\frac{\eta_{N_{\alpha}}}{\eta^{\rm eq}_{N_{\alpha}}}
+\frac{4}{21}\,\frac{\eta_{\Delta L}}{\eta_{\Delta L_l}}\Bigg)\,
\gamma^{N_{\alpha} L}_{\Phi^\dagger V_\mu}
\: +\:\Bigg(1+\frac{4}{21}\,\frac{\eta_{\Delta L}}{\eta_{\Delta L_l}}\Bigg)\,
\gamma^{N_{\alpha} V_\mu}_{L \Phi}\nonumber\\[1mm]
& & +\: \Bigg(1-\frac{4}{21}\,\frac{\eta_{N_{\alpha}}}{\eta^{\rm
eq}_{N_{\alpha}}}\, 
\frac{\eta_{\Delta L}}{\eta_{\Delta L_l}}\Bigg)\,
\gamma^{N_{\alpha}\Phi^\dagger }_{L V_\mu}\, \Bigg]\; ,\\[3mm]
\widetilde{\Gamma}^{S\; (\alpha l)}_{\rm Gauge} &=& \frac{1}{n_\gamma}\ 
\Bigg[\,\Bigg(\frac{\eta_{N_{\alpha}}}{\eta^{\rm eq}_{N_{\alpha}}}
+\frac{4}{21}\,\frac{\eta_{\Delta L}}{\eta_{\Delta L_l}}\Bigg)\, 
\gamma^{N_{\alpha} L}_{\Phi^\dagger V_\mu}
\: +\:\Bigg(1+\frac{4}{21}\,\frac{\eta_{\Delta L}}{\eta_{\Delta L_l}}\Bigg)\,
\gamma^{N_{\alpha} V_\mu}_{L \Phi}\nonumber\\[1mm]
& & +\:\Bigg(1+\frac{4}{21}\,
              \frac{\eta_{N_{\alpha}}}{\eta^{\rm eq}_{N_{\alpha}}}\, 
\frac{\eta_{\Delta L}}{\eta_{\Delta L_l}}\Bigg)\,
\gamma^{N_{\alpha}\Phi^\dagger }_{L V_\mu}\, \Bigg]\; ,\\[3mm]
  \label{GWY}
\Gamma^{W\; (\alpha l)}_{\rm Yukawa} & = & \frac{1}{n_\gamma}\
\Bigg[\,\Bigg(2+\frac{4}{21}\,\frac{\eta_{\Delta L}}{\eta_{\Delta L_l}}\Bigg)\,
\gamma^{N_{\alpha} L}_{Q u^C}\: +\:
\Bigg(1+\frac{17}{63}\,\frac{\eta_{\Delta L}}{\eta_{\Delta L_l}}\Bigg)\,
\gamma^{N_{\alpha} u^C}_{L Q^C}\nonumber\\[1mm]
& & +\:\Bigg(1+\frac{19}{63}\,\frac{\eta_{\Delta L}}{\eta_{\Delta L_l}}\Bigg)\,
\gamma^{N_{\alpha} Q}_{L u}\, \Bigg]\; ,\\[3mm]
  \label{GWG}
\Gamma^{W\; (\alpha l)}_{\rm Gauge} & = & \frac{1}{n_\gamma}\ 
\Bigg[\,\Bigg(2+\frac{4}{21}\,\frac{\eta_{\Delta L}}{\eta_{\Delta L_l}}\Bigg)\,
\gamma^{N_{\alpha} L}_{\Phi^\dagger V_\mu}\: +\:
\Bigg(1+\frac{4}{21}\,\frac{\eta_{\Delta L}}{\eta_{\Delta L_l}}\Bigg)\,
\gamma^{N_{\alpha} V_\mu}_{L \Phi}\nonumber\\[1mm]
& & +\: \Bigg(1+\frac{8}{21}\,\frac{\eta_{\Delta L}}{\eta_{\Delta L_l}}\Bigg)\,
\gamma^{N_{\alpha} \Phi^\dagger }_{L V_\mu} \, \Bigg]\;,\\[3mm]
  \label{GDL2}
\Gamma^{\,\Delta L =2\:(j k)}_{\rm Yukawa} &=& \frac{1}{n_\gamma}\ 
\Bigg[\,\Bigg(1+\frac{4}{21}\,\frac{\eta_{\Delta L}}{\eta_{\Delta L_j}}\Bigg)\,
\bigg(\gamma^{\,\prime\,L_j\Phi}_{\,L_k^C\Phi^\dagger} \: + \:
\gamma^{L_j L_k}_{\Phi^\dagger\Phi^\dagger}\,\bigg)\,\Bigg]\; ,\\[3mm]
  \label{GDL0}
\Gamma^{\,\Delta L =0\:(j k)}_{\rm Yukawa} &=& \frac{1}{n_\gamma}\ 
\Bigg[\,\Bigg(1+\frac{4}{21}\,\frac{\eta_{\Delta L}}{\eta_{\Delta L_j}}\Bigg)\,
\gamma^{\,\prime\,L_j\Phi}_{\,L_k\Phi} \: + \:
\gamma^{L_j\Phi^\dagger}_{L_k\Phi^\dagger} \: + \:
\gamma^{L_j L^C_k}_{\Phi \Phi^\dagger}
\,\Bigg]\; .
\end{eqnarray}
Notice that  the would-be singularities  in  the limit of a  vanishing
$\eta_{\Delta       L_l}$  in~(\ref{GDT})--(\ref{GSYtilde})        and
(\ref{GSGhat})--(\ref{GDL0}) are  exactly cancelled  by  corresponding
factors $\eta_{\Delta  L_l}$ that multiply  the collision terms in the
BEs~(\ref{BEN}) and     (\ref{BEL}).  We should   also  note  that the
flavour-diagonal $\Delta L=0$ processes, with $k=j$, do not contribute
to the BEs,  as it can be  explicitly checked in (\ref{BEL}). Finally,
it is  worth commenting on  the earlier  form of  the BEs, obtained in
\cite{PU}. This can be recovered from (\ref{BEN})--(\ref{GDL0}), after
summing over  the three lepton-doublet BEs,  with  the assumption that
$n_{\Delta   L_i} = \frac13   n_{\Delta L}$,  and after neglecting  SM
chemical potential corrections.

\setcounter{equation}{0}
\section{Out of Equilibrium Sphaleron Effects}

In the SM, the combination of the baryon and lepton numbers, $B+L$, is
anomalous~\cite{thooft}.    Although  at   low  energies   this  $B+L$
violation is  unobservably small, at  temperatures close to  and above
the  electroweak phase  transition, e.g.~for  $T \stackrel{>}{{}_\sim}
150$~GeV, thermal fluctuations more efficiently overcome the so-called
sphaleron barrier allowing rapid $B+L$ violation in the SM~\cite{KRS}.

The temperature dependence  of   the rate of   $B+L$  violation is  of
particular interest in   models of low-scale leptogenesis. Any  lepton
asymmetry produced after  the $(B+L)$-violating interactions  drop out
of thermal equilibrium will not  be converted into a baryon asymmetry.
Therefore,   in   electroweak-scale  leptogenesis  scenarios,  it   is
important to consider the rate of $B+L$ violation in the BEs, in order
to offer a more reliable estimate of the final baryon asymmetry.

The  rate of $(B+L)$-violating transitions has  been studied in detail
in \cite{AM,CLMW} for temperatures satisfying the double inequality
\begin{equation}
  \label{BLcond}
M_W(T)\ \ll\ T\ \ll\ \frac{M_W(T)}{\alpha_w}\;,
\end{equation}
where $\alpha_w = g^2/4\pi$ is the  SU(2)$_L$ fine structure constant,
$M_W(T) = g\,v(T)/2$ is the $T$-dependent $W$-boson mass and
\begin{equation}
  \label{vT}
v(T)\ =\ v(0) \left(\,1-\frac{T^2}{T_c^2}\,\right)^{\frac{1}{2}}
\end{equation}
is the $T$-dependent VEV of the Higgs field.  The critical temperature
of the electroweak phase transition, $T_c$, is given by \cite{CKO}
\begin{equation}
T_c\ =\ v(0) \left(\,\frac{1}{2}+\frac{3\,g^2}{16\,\lambda}+
\frac{g^{\prime\,2}}{16\,\lambda}+
\frac{h_{t}}{4\,\lambda}\,
\right)^{-\frac{1}{2}},
\end{equation}
where $\lambda$ is the quartic  Higgs self-coupling, $g^\prime$ is the
U(1)$_Y$ gauge coupling and $h_t$ is the top-quark Yukawa coupling.

The rate of $B+L$ violation per unit volume is \cite{AM}
\begin{equation}
  \label{BLrate}
\gamma_{\Delta (B+L)}\ \equiv\ \frac{\Gamma}{V}\ =\
\frac{\omega_-}{2\,\pi}\,{\cal N}_{\rm tr} ({\cal N}V)_{\rm rot}
\left(\frac{\alpha_W\,T}{4\,\pi} \right)^3 \alpha_3^{-6}\,e^{-E_{\rm
sp} / T} \kappa\; .
\end{equation}
According to (\ref{BLcond}), this expression is valid for temperatures
$T\stackrel{<}{{}_\sim}    T_c$.       The       various    quantities
in~(\ref{BLrate}) are  related to   the   sphaleron dynamics and   are
discussed in \cite{AM,CLMW}.  Following the notation of~\cite{AM}, the
parameters $\omega_-$, ${\cal   N}_{\rm tr}$ and ${\cal  N}_{\rm rot}$
are functions of $\lambda / g^2$, $V_{\rm rot} = 8\pi^2$ and $\alpha_3
= g_3^2 / 4\pi$, where
\begin{equation}
g^2_3\ =\ \frac{g^2\,T}{2\,M_W(T)}\ .
\end{equation}
$E_{\rm sp}$ is the energy of the sphaleron and is given by
\begin{equation}
E_{\rm sp}\ =\ A\,\frac{2\,M_W(T)}{\alpha_W}\ ,
\end{equation}
where $A$ is a function  of $\lambda / g^2$ and  is of order 1 for all
phenomenologically relevant values of $\lambda / g^2$.  The dependence
of  the parameter $\kappa$ on    $\lambda  /g^2$ has been   calculated
in~\cite{AM,CLMW},  using  various techniques.  The   results of those
studies are  summarized  in Table~\ref{BLparams}, where the  values of
$\kappa$ and the  other sphaleron-related parameters are exhibited for
$\lambda /g^2 =  0.278$, which  corresponds  to a SM  Higgs-boson mass
$M_H$ of 120~GeV.

\begin{table}[t]
\begin{center}
\begin{tabular}{|c||c|c|c|c|c|}
\hline &&&&&\\[-11pt] 
$\lambda / g^2$ & $\omega_-$ & ${\cal N}_{\rm
rot}$ & ${\cal N}_{\rm tr}$ & $\kappa$ & $A$\\[2pt] \hline \hline
0.278 & 0.806 $(g v)$ & 11.2 & 7.6 & 0.135 -- 1.65 & 1\\ \hline
\end{tabular}
\end{center}
\caption{\sl Values  of the various  parameters  in (\ref{BLrate}) for
$\lambda/g^2  =  0.278$, corresponding  to  a  SM Higgs-boson  mass of
120~GeV.}\label{BLparams}
\end{table}

Given the present experimental limits on the SM Higgs-boson mass, $M_H
\stackrel{>}{{}_\sim} 115$~GeV, it can be shown that the
electroweak phase transition  in the SM  will either be a weakly first
order  one,  or even a  second or higher order   one, without bubble
nucleation   and the  formation of   large  spatial inhomogeneities in
particle densities.  Therefore, the use of  a formalism describing the
$(B+L)$-violating sphaleron dynamics in terms of spatially independent
$B$- and $L$-number densities $n_{B}$ and  $n_{L_j}$ may be justified.
Further refinements to this approach will be presented elsewhere.

We should bear in mind  that heavy Majorana neutrino decays, sphaleron
effects and  other processes  considered  in the  BEs  (\ref{BEN}) and
(\ref{BEL}) act  directly on the number  densities of SU(2)$_L$ lepton
doublets, $n_{\Delta  L_{e,\mu,\tau}}$.  However, the quantity usually
referred to as    lepton number, $L$,  has  a  contribution   from the
right-handed charged  leptons $l_{e,\mu,\tau R}$  as well.  In thermal
equilibrium, one   may  relate the asymmetry  in  right-handed charged
leptons  to the asymmetry   in SU(2)$_L$  lepton  doublets by   virtue
of~(\ref{ChemPot}), leading to the result
\begin{equation}
\eta_{\Delta l_{iR}} = \frac{1}{2}\,\eta_{\Delta L_i} 
- \frac{2}{21}\,\eta_{\Delta L}\,.
\end{equation}
The  change  in  lepton  flavour  can  be thought  of  as  having  two
components,    one    component    termed    leptogenesis    due    to
lepton-number-violating processes considered in Section 3, and another
due to the $(B+L)$-violating sphalerons:
\begin{equation}
\frac{d\eta_{L_i}}{dz} \ =\ \frac{d\eta_{L_i}}{dz}\Bigg|_{\rm leptogenesis} +\
\frac{d\eta_{L_i}}{dz}\Bigg|_{\rm sphaleron}\,,
\label{components}
\end{equation}
where, up to $\mathcal{O}(v/T)$ corrections,
\begin{equation}
\frac{d \eta_{L_i}}{dz}\Bigg|_{\rm leptogenesis} =\ 
\frac{3}{2}\,\frac{d \eta_{\Delta L_i}}{dz}\ 
-\ \frac{2}{21}\,\frac{d \eta_{\Delta L}}{dz}\; .
\label{LRconvert}
\end{equation}
The BEs determining  the leptogenesis  component of~(\ref{components})
have been discussed   in Section~3.  We    shall now discuss  the  BEs
determining  the sphaleron   component  of~(\ref{components}), and the
generation of a net $B$-number asymmetry.

Within the context of the above formalism, the sphaleron components of
the BEs for $n_{B}$ and $n_{L_j}$ are given by
\begin{eqnarray}
  \label{BLexp}
\frac{d n_{B}}{dt}\: +\: 3 H\,n_{B} & = & 
n_G\, \Big( e^{\beta(\mu_B Q_B+\mu_L Q_L)}\: -\: 
e^{\beta(\mu_B Q^\prime_B+\mu_L Q^\prime_L)}\: -\: 
e^{-\beta(\mu_B Q_{\bar{B}}\: +\: \mu_L Q_{\bar{L}})}\nonumber\\
&&+\, e^{-\beta(\mu_B Q^\prime_{\bar{B}}+\mu_L Q^\prime_{\bar{L}})}\Big)\,
\gamma_{\Delta(B+L)}\,,\nonumber\\
\frac{dn_{L_j}}{dt}\: +\: 3 H n_{L_j} & = & 
\Big(\, e^{\beta(\mu_B Q_B+\mu_{L} Q_{L})}\: -\: 
e^{\beta(\mu_B Q^\prime_B+\mu_{L} Q^\prime_{L})}\nonumber\\ 
&&-\, e^{-\beta(\mu_B Q_{\bar{B}}+\mu_{L} Q_{\bar{L}})}\:
+\: e^{-\beta(\mu_B Q^\prime_{\bar{B}}+\mu_{L} Q^\prime_{\bar{L}})}\,
\Big)\,\gamma_{\Delta(B+L)}\;,
\end{eqnarray}
where  $n_G$ is   the   number    of   generations and   $\beta      =
1/T$. Furthermore, $Q_{B(L)}$ is the  baryonic (or leptonic) charge of
the  system   before  the  $(B+L)$-violating sphaleron  transition and
$Q_{B(L)}^\prime$  is the charge  after the transition. Klinkhamer and
Manton showed \cite{KM} that a sphaleron carries a baryon (and lepton)
number  of  $n_G/2$, therefore  $Q_{B}  -  Q_{B}^\prime  = n_G/2$  and
$Q_{L_i} -  Q_{L_i}^\prime = 1/2$.  Finally,  assuming that the baryon
and  lepton  chemical  potentials  are   small  with  respect  to  the
temperature, the BEs~(\ref{BLexp}) may be approximated by
\begin{eqnarray}
  \label{BLlinear}
\frac{d n_{B}}{dt}\: +\: 3 H\,n_{B} \!& = &\!
-n_G\,\beta\,\bigg(\,n_G\,\mu_B + \sum_i \mu_{L_i}\,\bigg)\, 
\gamma_{\Delta(B+L)}\,,\nonumber\\
\frac{dn_{L_j}}{dt}\: +\: 3 H n_{L_j} \!& = &\!
\beta\,
\bigg(\, n_G\,\mu_B + \sum_i \mu_{L_i}\,\bigg)\, \gamma_{\Delta(B+L)}\;.
\end{eqnarray}
Notice   that the  BEs~(\ref{BLlinear})   are  linear in the  chemical
potentials, which is  a  very useful approximation for  our  numerical
estimates.

We now need  to determine the relation  between the baryon and  lepton
chemical  potentials and  their  respective  number  densities.  These
relations can be found by considering the effective potential, $V$, of
the Higgs and the SU(2)$_L$ and U(1)$_Y$ gauge fields.  They have been
computed   in~\cite{LS} at   finite temperatures,  for  small chemical
potentials, $\mu_B,\,\mu_L \ll T$ and when $v(T) \stackrel{<}{{}_\sim}
(\mbox{a  few})\times T$. In  this  framework,  the neutrality of  the
system  with respect  to   gauge charges   can be  accounted   for  by
minimizing  the potential with respect  to  the temporal components of
the SU(2)$_L$  and   U(1)$_Y$ gauge fields,   $W^a_0$  ($a=1,2,3$) and
$B_0$, respectively. The baryon and  lepton number densities are  then
given by
\begin{equation}
n_{B}\ =\ -\frac{\partial V}{\partial \mu_B}\ ,\qquad
n_{L_i}\ =\ -\frac{\partial V}{\partial \mu_{L_i}}\ .
\end{equation}
For the SM with 3 generations and 1 Higgs doublet, we obtain
\begin{eqnarray}
  \label{chemical}
\mu_B &=& 3\,n_{B}\,\frac{77\,T^2 + 27\,v^2(T)}{132\,T^4 +
51\,T^2\,v^2(T)}\ -\ 2\,\frac{22\,T^2 + 3\,v^2(T)}{132\,T^4 +
51\,T^2\,v^2(T)}\sum_{j\,=e,\mu,\tau}\!\!\!n_{L_j}\;,\nonumber\\[12pt] 
\mu_{L_i} &=& \frac{2}{51\,T^2}\,\bigg(
51\,n_{L_i}\: -\: 3\,n_{B}\: +\:
4\!\!\sum_{j\,=e,\mu,\tau}\!\!\!n_{L_j}\,\bigg)\nonumber\\ 
&&-\,\frac{484}{153\,\Big(44\,T^2+17\,v^2(T)\Big)}\ \bigg(\,
3\,n_{B}\: -\: 4\!\!\sum_{j\,=e,\mu,\tau}\!\!\!n_{L_j}\,\bigg)\;.
\end{eqnarray}

Employing the relations~(\ref{chemical}), we may now extend the system
of  BEs~(\ref{BEN})   and   (\ref{BEL}),   by   explicitly taking  the
$(B+L)$-violating sphaleron transitions into account,
\begin{eqnarray}
  \label{Bsph}
\frac{d \eta_{B}}{dz} & = & -\, \frac{z}{H(z=1)}\\
&&\times\,\bigg[\, \eta_{B}\: +\: 
\frac{28}{51}\, \sum_{j\,=e,\mu,\tau}\!\!\!\eta_{L_j}\: 
+\: \frac{225}{561}\,\frac{v^2(T)}{T^2}\,
\bigg(\, \eta_{B}\: +\:\frac{108}{225}\, \sum_{j\,=e,\mu,\tau}\!\!\!
\eta_{L_j}\,\bigg)\,\bigg]\ \Gamma_{\Delta (B+L)}\;,\nonumber\\[12pt]
  \label{Lsph}
\frac{d \eta_{L_i}}{dz} & = &  
\frac{d\eta_{L_i}}{dz}\Bigg|_{\rm leptogenesis}
\: +\: \frac13\,
\frac{d \eta_{B}}{dz}\ ,
\end{eqnarray}
with 
\begin{equation}
\Gamma_{\Delta (B+L)}\ =\ \frac{(3366/\pi^2 )\, T^2}{
132\, T^2\: +\: 51\, v^2(T) }\
\frac{\gamma_{\Delta (B+L)}}{n_{\gamma}}\ .
\end{equation}
The leptogenesis  component  of (\ref{Lsph})  may be determined  using
relation  (\ref{LRconvert}),  along  with   the   BEs (\ref{BEN})  and
(\ref{BEL}).

Observe that  in  the limit   of infinite sphaleron   transition rate,
$\Gamma_{\Delta (B+L)}/H(z = 1)\to  \infty$, and at high  temperatures
$T\gg v(T)$, the   conversion of lepton-to-baryon  number densities is
given by the known relation:
\begin{equation}
\eta_{B}\ =\ -\,\frac{28}{51}\:
\sum_{j\,=e,\mu,\tau}\!\!\!\eta_{L_j} \; .
\end{equation}
To account for the  $T$-dependent $(B+L)$-violating sphaleron effects,
our numerical estimates given in the next section will be based on the
BEs~(\ref{BEN}),  (\ref{BEL}),  (\ref{LRconvert}),  (\ref{Bsph})   and
(\ref{Lsph}).

\setcounter{equation}{0}
\section{Numerical Examples}\label{sec:num}

We shall now analyze R$\tau$L  models that comply with the constraints
obtained from the  existing low-energy neutrino data~\cite{PDG,JVdata}
and provide  successful baryogenesis.  As was  discussed in Section~2,
our  specific choice  of model  parameters  will be  motivated by  the
naturalness of the light and heavy neutrino sectors.

Phenomenologically relevant R$\tau$L models  can be constructed for an
SO(3) invariant  heavy neutrino  mass of the  size of  the electroweak
scale, e.g.~$m_N = 250$~GeV  [cf.~(\ref{MSSO3})], if $|a| \sim |b| \gg
|c|$ and $|a|, |b| \sim 10^{-2}$.  To protect the $\tau$-lepton number
from   wash-out   effects,   we    also   require   that   the   small
U(1)$_l$-breaking parameters $|\varepsilon_{e,\mu,\tau}|$ be no larger
than  about $10^{-6}$  and $|c|  \stackrel{<}{{}_\sim}  10^{-5}$.  For
definiteness, the model parameters determining the light neutrino mass
spectrum are chosen to be (in arbitrary complex units)
\begin{equation}
  \label{nupars1}
\frac{\Delta m_N}{m_N}\,a^2 \ =\  4\,,\qquad 
\varepsilon_e \ =\ 2\: +\: \frac{21}{250}\; ,\qquad
\varepsilon_\mu \ =\ \frac{13}{50} \; ,\qquad 
\varepsilon_\tau \ =\ -\, \frac{49}{128}\ , \quad
\end{equation}
where the ratio $b/a$ is kept fixed:
\begin{equation}
  \label{nupars2}
\frac{b}{a}\ =\ \frac{19}{50}\ .
\end{equation}
The actual  values selected for the relevant  parameters $a$, $(\Delta
M_S)_{22}$, $(\Delta M_S)_{33}$ and  $(\Delta M_S)_{23}$ vary with the
SO(3)  invariant mass  $m_N$. As  we will  see in  more  detail below,
Table~\ref{mnu-params}   illustrates  choices   of   these  parameters
consistent  with the  light neutrino  data.   For $m_N$  in the  range
100--1000~GeV,   the  chosen  parameters   are  consistent   with  the
naturalness condition  mentioned in  Section~2, whilst giving  rise to
phenomenologically rich models.

In  our numerical analysis, we  will focus on  4 examples, with $m_N =
100$,  250, 500, and~1000~GeV.  Clearly, the model parameters selected
in~(\ref{nupars1}),  (\ref{nupars2}) and Table~\ref{mnu-params}  imply
that  all the scenarios have  the same tree-level  light neutrino mass
matrix:
\begin{equation}
{\bf m}^\nu\ \approx\ 
-\left(\! \begin{array}{ccc}
-1.27 &  3.63  &   2.96\\
3.63  &  1.89  &  0.370\\
2.96  & 0.370  & -0.544\end{array} \!\right) 
\times 10^{-2}\ \mathrm{eV}.
\end{equation}
This leads to an  inverted hierarchy of   light neutrino masses,  with
mass differences and   mixings compatible with the   current $3\sigma$
bounds~\cite{JVdata}.  Adopting  the convention $m_{\nu_3} < m_{\nu_1}
< m_{\nu_2}$, we find the mass squared differences and mixing angles
\begin{eqnarray}
m_{\nu_2}^2 - m_{\nu_1}^2 \!& = &\! 7.54 \times
10^{-5}\,\,\mathrm{eV}^2\,,\qquad m_{\nu_1}^2 - m_{\nu_3}^2 \ =\ 
2.45 \times 10^{-3}\,\,\mathrm{eV}^2\,,\nonumber\\ 
\sin^2 \theta_{12} \! & = &\! 0.362 \,,\qquad
\sin^2 \theta_{23} \ = \ 0.341 \,,\qquad
\sin^2 \theta_{13} \ = \ 0.047 \;.
\end{eqnarray}
Since the mass  matrix (\ref{mnutree}) is rank  2, one  light neutrino
will be massless at the tree level ($m_{\nu_3}  = 0$), thus fixing the
absolute scale of the light neutrino hierarchy.

\begin{table}
\begin{center}
\begin{tabular}{|c|c|c|c|}
\hline &&&\\[-15pt] 
$a/(m_N)^{\frac{1}{2}}$ (GeV)$^{-\frac{1}{2}}$ & $(\Delta
M_S)_{22}/m_N$ & $(\Delta M_S)_{33}/m_N$ & $(\Delta
M_S)_{23}/m_N$\\[2pt] \hline \hline 
$6.0 \times 10^{-4}$ & $4.0 \times 10^{-9}$ & $5.2 \times 10^{-9}$ &
$(6.8 - 0.6\,i) \times 10^{-9}$\\ \hline 
\end{tabular}
\end{center}
\caption{\sl  Choices   of the  parameters  $a$, $(\Delta  M_S)_{22}$,
$(\Delta M_S)_{33}$ and   $(\Delta M_S)_{23}$, consistent with   light
neutrino data.}
\label{mnu-params}
\end{table}

The  remaining soft  SO(3)-breaking  parameters, $(\Delta  M_S)_{11}$,
$(\Delta  M_S)_{12}$, $(\Delta  M_S)_{13}$,  do not  affect the  light
neutrino mass spectrum.  These together  with the parameter $c$ play a
key  role  in   obtaining  the  correct  BAU  and   are  exhibited  in
Table~\ref{baryon-param}.   We   choose  $(\Delta  M_S)_{11}$   to  be
relatively  large, $(\Delta  M_S)_{11}  \sim 10^{-5}\,m_N$,  providing
large  mass  differences $|m_{N_2}  -  m_{N_1}|/m_N$  and $|m_{N_3}  -
m_{N_1}|/m_N \sim 10^{-5}$.  Such  a choice is consistent with thermal
and RG  effects running from the  GUT scale $\sim  10^{16}$~GeV to the
electroweak scale $\sim  m_N$ (see also the discussion in  Section 2). The
other  two  soft SO(3)-breaking  parameters,  $(\Delta M_S)_{12}$  and
$(\Delta M_S)_{13}$, are selected so  as to give the observed BAU.  

To  assess  the  degree  of  cancellation between  tree-level  and  RG
contributions  to  $\Delta  M_S$,  we introduce  the  parameter  $r_C$
defined as
\begin{equation}
  \label{rC}
r_C\ \equiv\ \prod\limits_{(i,j)}\ \frac{|(\Delta M^{\rm RG}_S)_{ij}|}{
|(\Delta M_S)_{ij}|}\ .
\end{equation}
In~(\ref{rC}), the  product $(i,j)$ is taken  over contributions where
$|(\Delta M^{\rm  RG}_S)_{ij}| > |(\Delta  M_S)_{ij}|$.  The~parameter
$r_C$  is always  greater than  1 and  represents that  the  degree of
cancellation is 1 part in $r_C$.  {}From the values of $r_C$ displayed
in Table~\ref{baryon-param},  we observe that  electroweak-scale heavy
Majorana neutrinos are favoured by naturalness.

The baryon  asymmetry predicted  for  each model  can be determined by
solving   the    BEs~(\ref{BEN}),    (\ref{BEL}),   (\ref{LRconvert}),
(\ref{Bsph})  and (\ref{Lsph}), and  using the collision terms derived
in Appendix A and \cite{PU}.  These  collision terms are calculated in
the basis where   the charged-lepton and heavy-Majorana  mass matrices
are positive and diagonal.  They have  been appropriately expressed in
terms of the one-loop  resummed  effective  Yukawa couplings  derived   in
\cite{PU}.  It is worth noting  that all SM reactions, including those
involving the $e$-Yukawa coupling, are in full thermal equilibrium for
the temperatures relevant  to our scenarios, $T  \stackrel{<}{{}_\sim}
10$~TeV~\cite{CKO,BCST}.   Moreover,  since heavy  Majorana   neutrino
decays are thermally blocked  at temperatures $T \stackrel{>}{{}_\sim}
3 m_{N_\alpha}$~\cite{GNRRS}, we will only display numerical estimates
of the evolution of lepton and baryon asymmetries,  for $z = m_{N_1}/T
\stackrel{>}{{}_\sim}  0.1$. Nevertheless, as   we will see below, the
predictions for  the final  BAU are  relatively  robust in RL  models,
because of the near or complete independence  on the primordial baryon
and lepton number abundances.

Some  of the Yukawa  and   gauge-mediated collision terms  contain  IR
divergences, which  are usually regulated  in thermal field  theory by
considering    the     thermal       masses    of     the    exchanged
particles~\cite{MBellac}.  To assess the theoretical errors introduced
by the choice of a universal thermal mass regulator (see the discussion in
Appendix A),   we  have estimated the  response   of the final  baryon
asymmetry under variations of the IR mass  regulator $m_{\rm IR}$.  We
find that the predicted BAU only varies by $\pm 7~\%$, for a variation
of $m_{\rm IR}$ by $\pm 25~\%$.

\begin{table}
\begin{center}
\begin{tabular}{|c||c|c|c|c||c|}
\hline &&&&&\\[-15pt] 
$m_N$ (GeV) & $(\Delta M_S)_{11}/m_N$ & $(\Delta M_S)_{12}/m_N$ &
$(\Delta M_S)_{13}/m_N$ & $c$ & $r_C$\\[2pt] \hline \hline
$100$& $1.0 \times 10^{-5}$ & $-1.00 \times 10^{-9}$ & $-5.5 \times
10^{-10}$ & $1.0 \times 10^{-7}$ & $5$\\ 
$250$& $1.0 \times 10^{-5}$ & $-1.36 \times 10^{-9}$ & $-8.0 \times
10^{-10}$ & $1.5 \times 10^{-7}$ & $39$\\ 
$500$& $1.0 \times 10^{-5}$ & $-1.36 \times 10^{-9}$ & $-8.8 \times
10^{-10}$ & $ 2.0 \times 10^{-7}$ & $264$\\ 
$1000$& $1.0 \times 10^{-5}$ & $-1.9 \times 10^{-9}$ & $-1.0 \times
10^{-9}$ & $ 2.5 \times 10^{-7}$ & $1240$\\ \hline 
\end{tabular}
\end{center}
\caption{\sl Choices of $(\Delta M_S)_{11}$, $(\Delta M_S)_{12}$,
$(\Delta M_S)_{13}$ and $c$, which, in conjunction with those in Table
\ref{mnu-params}, lead to successful baryogenesis.  $r_C$
parameterizes the degree of cancellation between radiatively
induced and tree-level contributions to $\Delta M_S$.}
\label{baryon-param}
\end{table}

\begin{figure}
\begin{center}
\includegraphics[scale=0.36]{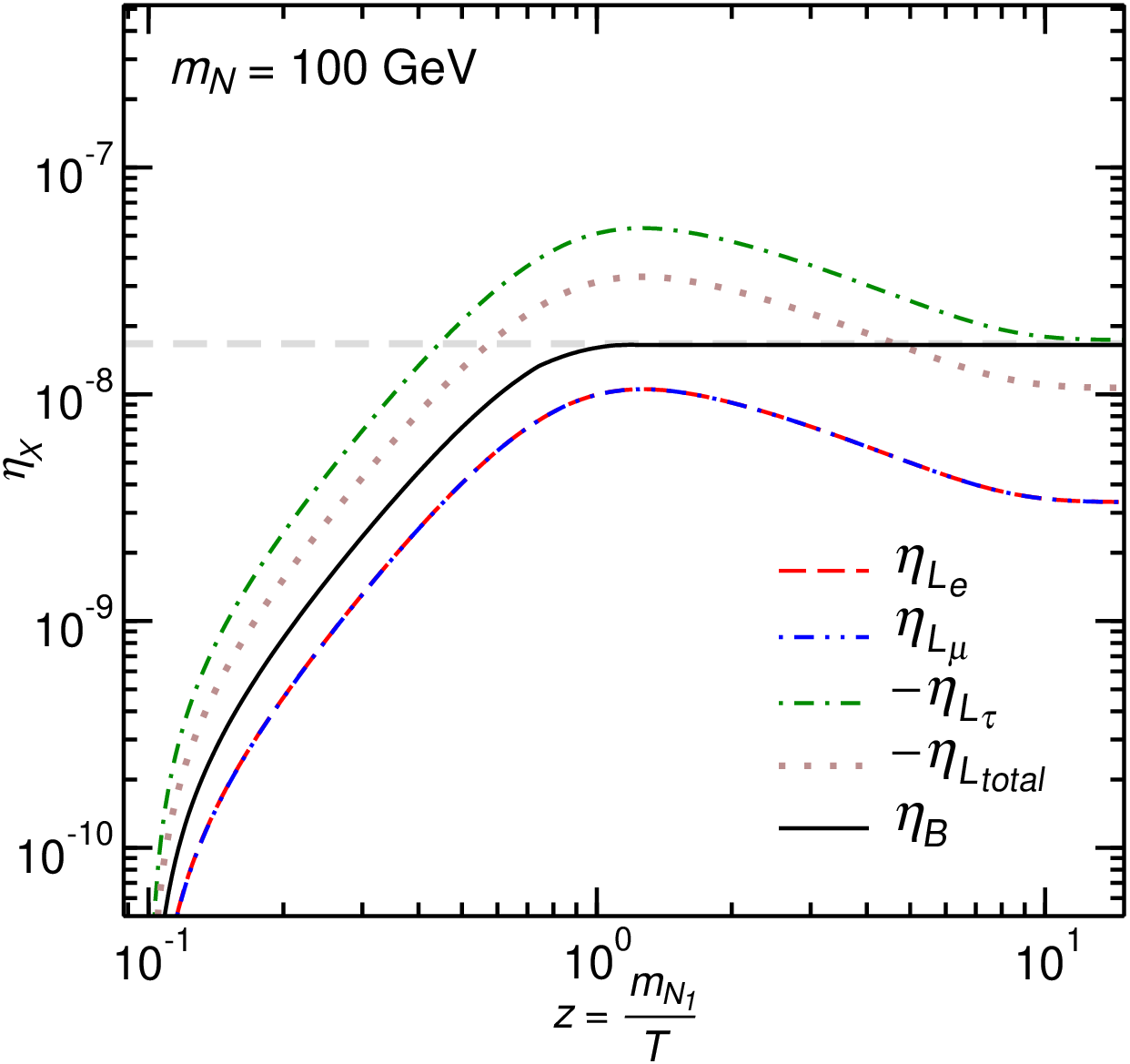}\hspace{6mm}
\includegraphics[scale=0.36]{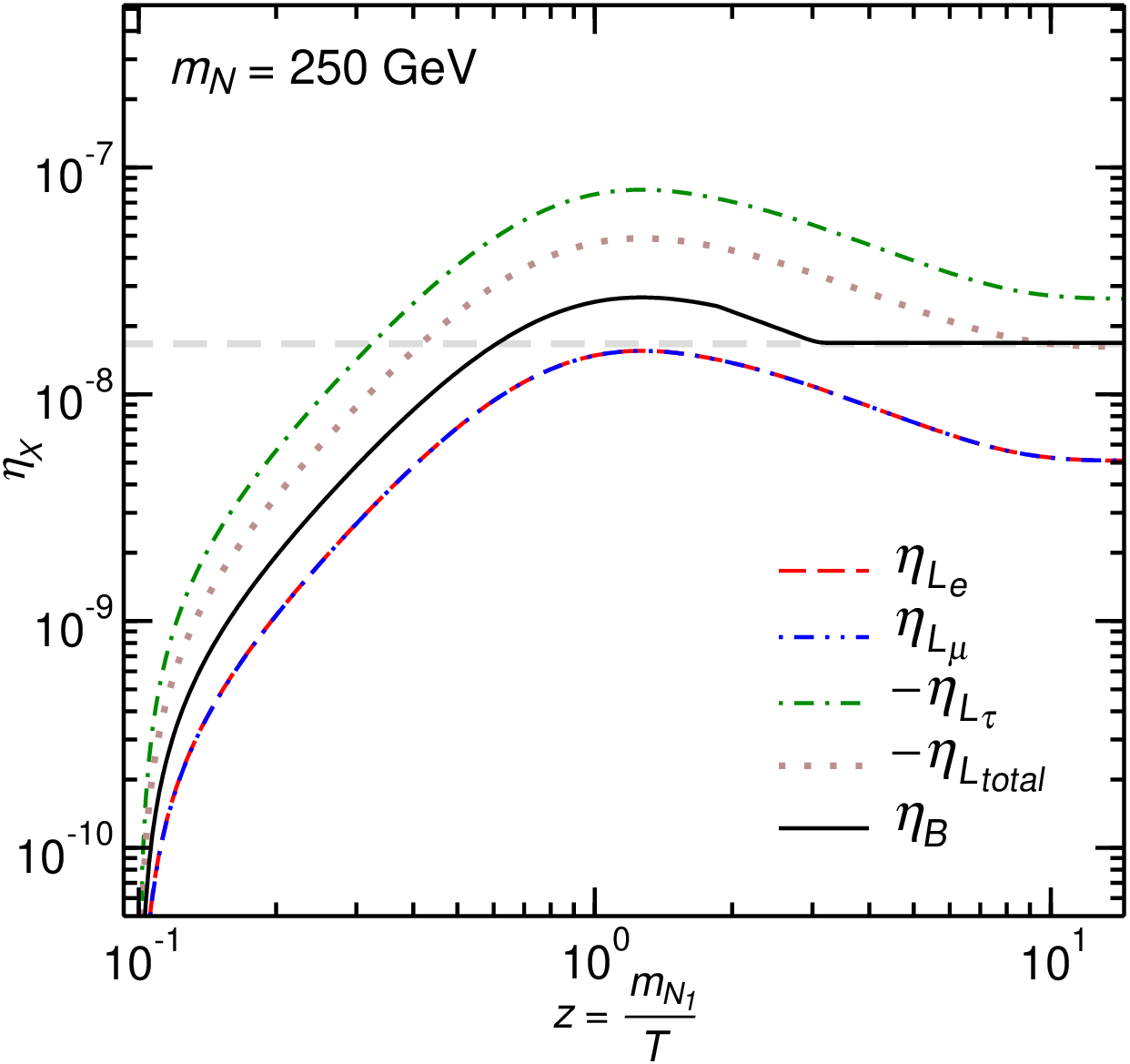}\\[6mm]
\includegraphics[scale=0.36]{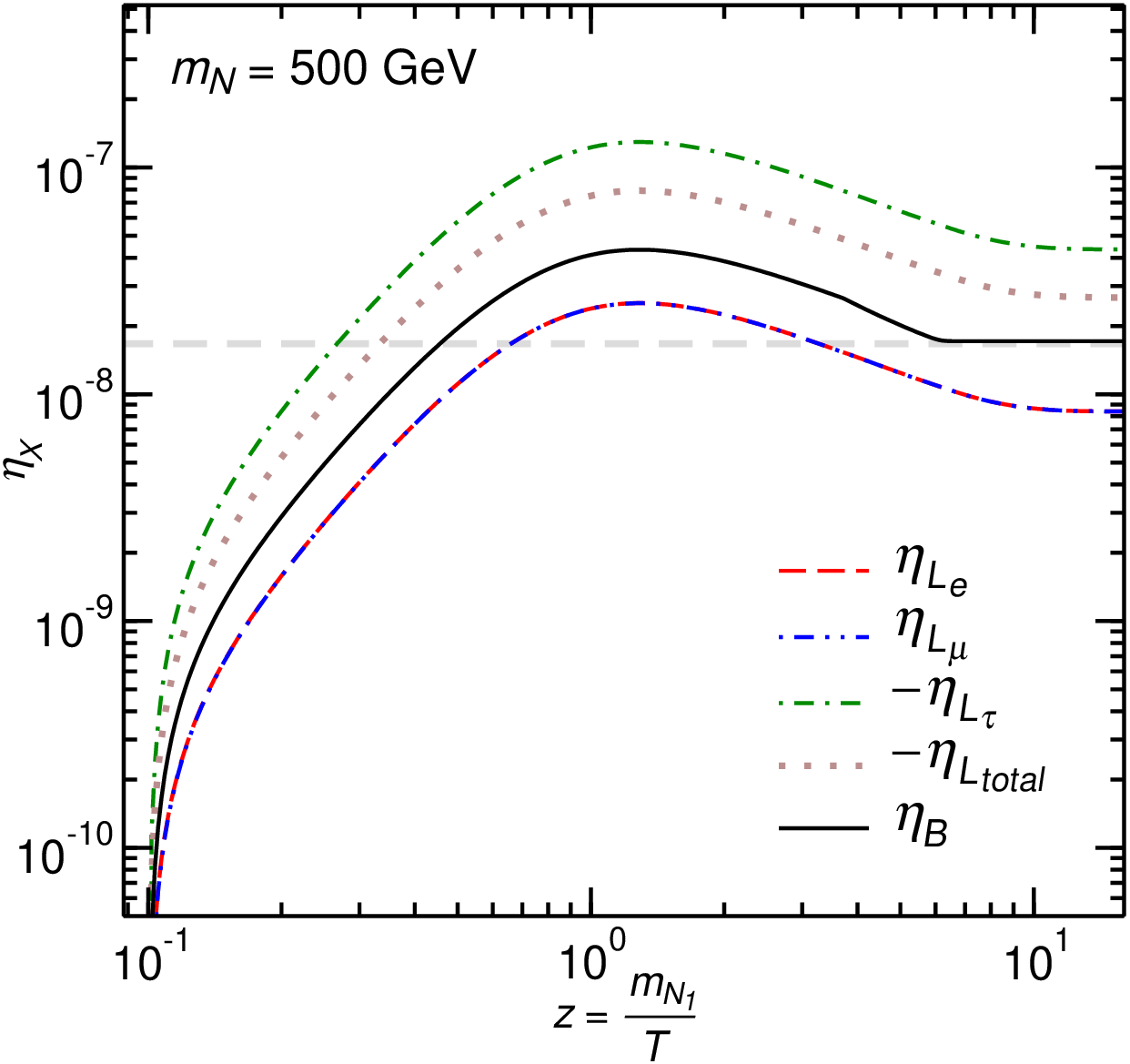}\hspace{6mm}
\includegraphics[scale=0.36]{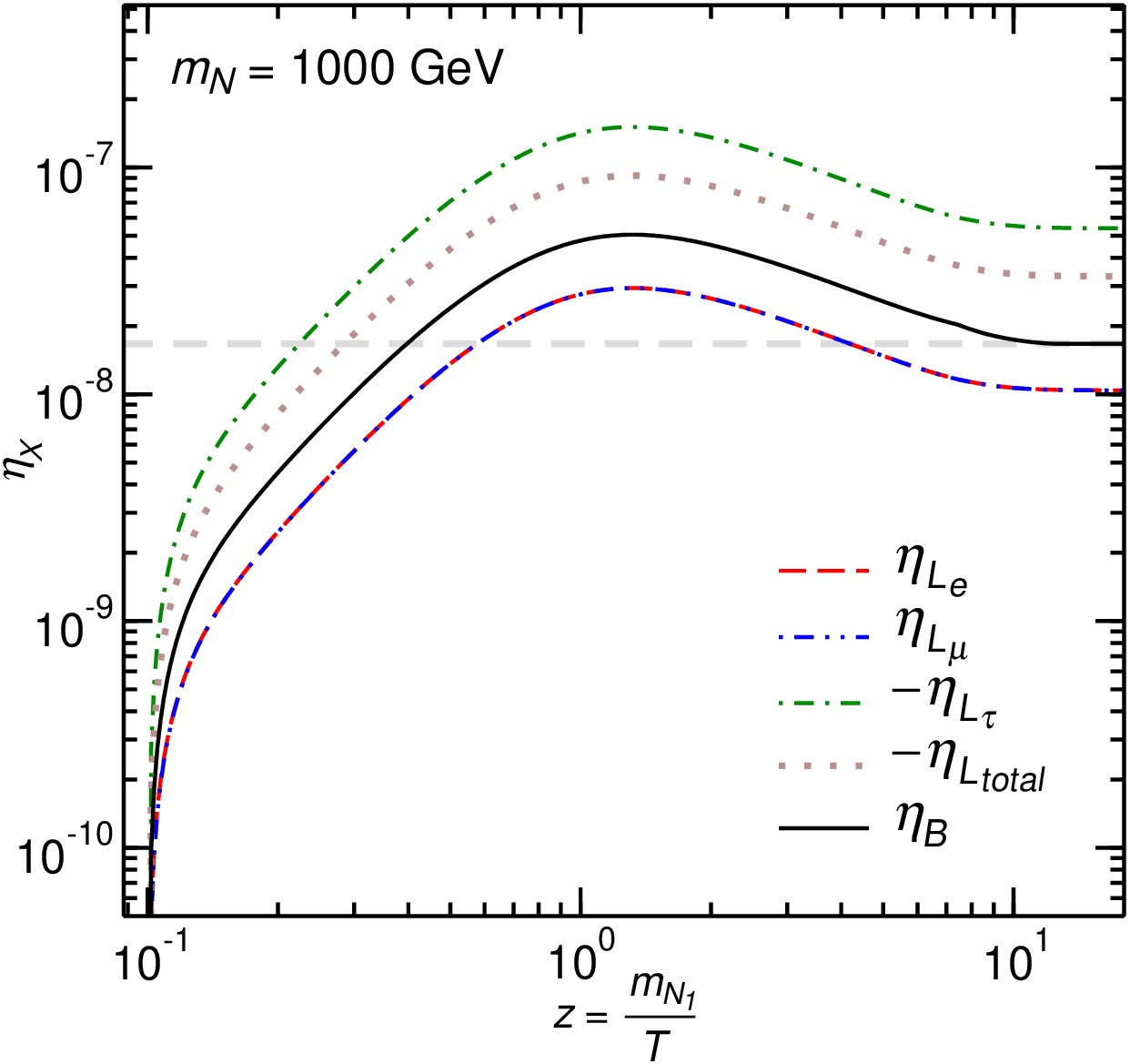}
\end{center}
\caption{\sl The predicted evolution of $\eta_{L_l}$ and $\eta_B$, for
models with $m_N =$~100, 250, 500 and 1000~GeV, and
$\eta_{N_{\alpha}}^{\mathrm{in}} = 1$.  The model parameters are given
in (\ref{nupars1}), (\ref{nupars2}), and Tables~\ref{mnu-params}
and~\ref{baryon-param}.  The horizontal grey dashed line shows the
baryon asymmetry needed to agree with observational data.}
\label{evolution-all}
\end{figure}

The  BEs   are  solved  numerically,  using  the   Fortran  code  {\tt
LeptoGen}\footnote[2]{{\tt  LeptoGen}   may  be  obtained   from  {\tt
http://hep.man.ac.uk/u/thomasu/leptogen}}.     Fig.~\ref{evolution-all}
shows  the predicted  evolution of  the baryon  and  individual lepton
asymmetries,   $\eta_B$  and   $\eta_{L_l}$,  as   functions   of  the
$T$-related parameter  $z =  m_{N_1}/T$, for each  of the  4 examples,
with  $m_N  =$~100,  250,   500  and  1000~GeV.   The  specific  model
parameters   are  given   in  (\ref{nupars1}),   (\ref{nupars2}),  and
Tables~\ref{mnu-params} and~\ref{baryon-param}.   Each scenario had an
initially thermal  heavy Majorana neutrino abundance  and zero initial
baryon and lepton asymmetries, i.e.~$\eta_{N_{\alpha}}^{\mathrm{in}} =
1$ and $\eta_{B}^{\mathrm{in}} = \eta_{L_l}^{\mathrm{in}} = 0$.  The 4
panels show that the large $L_{\tau}$ asymmetry is slightly reduced by
less significant,  but opposite sign $L_e$  and $L_{\mu}$ asymmetries.
Clearly  visible  in  each  scenario  is the  effect  of  the  rapidly
decreasing rate of $B+L$  violation; the lepton and baryon asymmetries
quickly  decouple at $T  \sim T_c$.   This decoupling  is particularly
pronounced in the $m_N =  100$~GeV scenario where the baryon asymmetry
freezes  out  exactly  when  the  lepton  asymmetry  is  maximal.   In
particular,  the rapid  decoupling  of $\eta_B$  from $\eta_{L_l}$  at
temperatures~$T$  close   to  $T_c$   has  the  virtue   that,  unlike
$\eta_{L_l}$, $\eta_B$ remains almost unaffected from ordinary SM mass
effects due  to a non-zero VEV  $v(T)$ [cf.\ (\ref{vT})],  since it is
$v(T\sim T_c) \ll v (T=0)$.

\begin{figure}
\begin{center}
\begin{picture}(450,540)
\put(15,425){\bf (a)}
\put(15,140){\bf (b)}
\put(50,275){\includegraphics[scale=0.5]{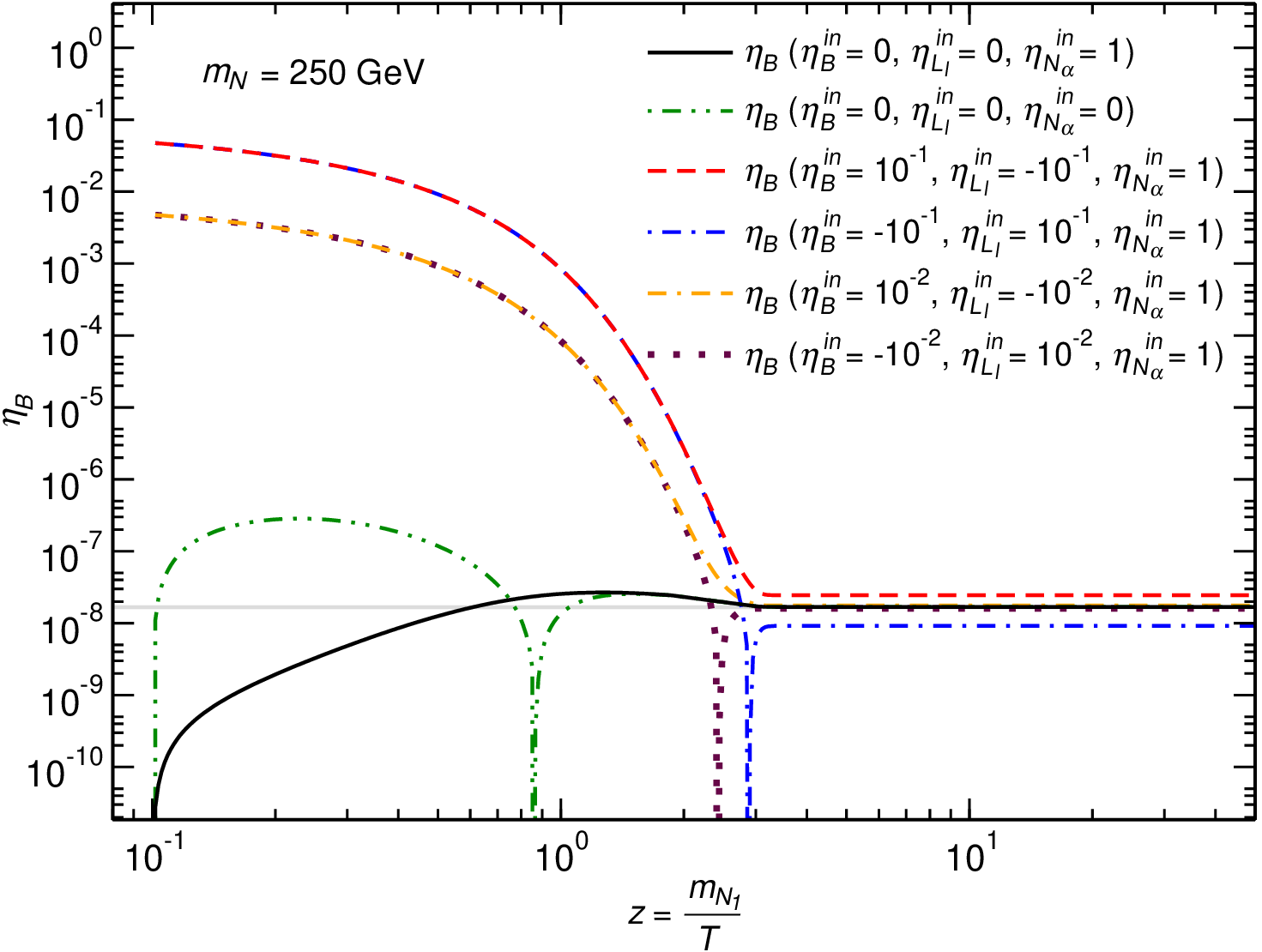}}
\put(50,-10){\includegraphics[scale=0.5]{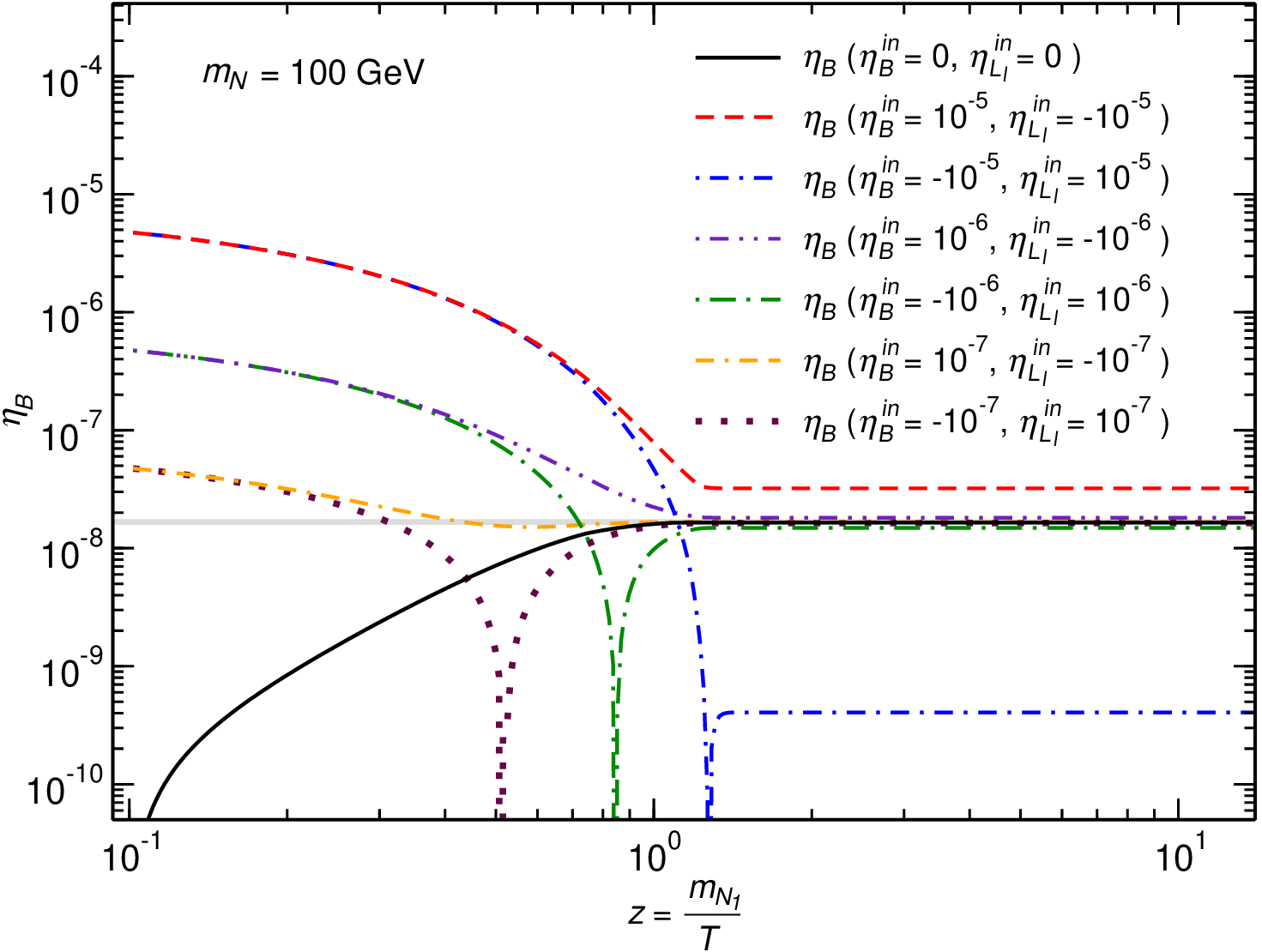}}
\end{picture}
\end{center}
\caption{\sl The  (in)dependence of the  final baryon asymmetry on the
initial lepton, baryon and  heavy Majorana neutrino abundances for (a)
$m_N =$~250~GeV and (b) $m_N =$~100~GeV.  The model parameters and the
meaning of   the   horizontal    grey  line  are   the  same    as  in
Fig.~\ref{evolution-all}.}
\label{inicondsa}
\end{figure}

Fig.~\ref{inicondsa} shows the evolution  of the baryon  asymmetry for
varying initial lepton, baryon  and heavy neutrino abundances. For the
250~GeV scenario,   Fig.~\ref{inicondsa}(a)    illustrates   the  near
independence of the   resultant   baryon  asymmetry on the     initial
conditions.   Even  for  the     most   extreme   initial   conditions
$\eta^{\mathrm{in}}_{L_l} =  \mp\,0.1$  and $\eta^{\mathrm{in}}_{B}  =
\pm\,0.1$,  the variation   in the final   baryon  asymmetry is   only
$\pm\,38\%$.  

For  heavy neutrino  masses $m_N  \stackrel{<}{{}_\sim}  250$~GeV, the
dependence  on initial  conditions becomes  stronger.  In  the  $m_N =
100$~GeV scenario, Fig.~\ref{inicondsa}(b) shows the dependence of the
final BAU on  the initial lepton and baryon  asymmetries in a R$\tau$L
scenario with $m_N = 100$~GeV.   It is interesting to observe that the
final  $B$  asymmetry  will  remain  almost unaffected,  even  if  the
primordial    baryon   asymmetry   $\eta^{\mathrm{in}}_{B}$    at   $T
\stackrel{>}{{}_\sim} 10\,  m_N$ is as large as  $10^{-6}$, namely two
orders  of  magnitude larger  than  the  one  required to  agree  with
observational data.
 
\begin{figure}[t]
\includegraphics[scale=0.73]{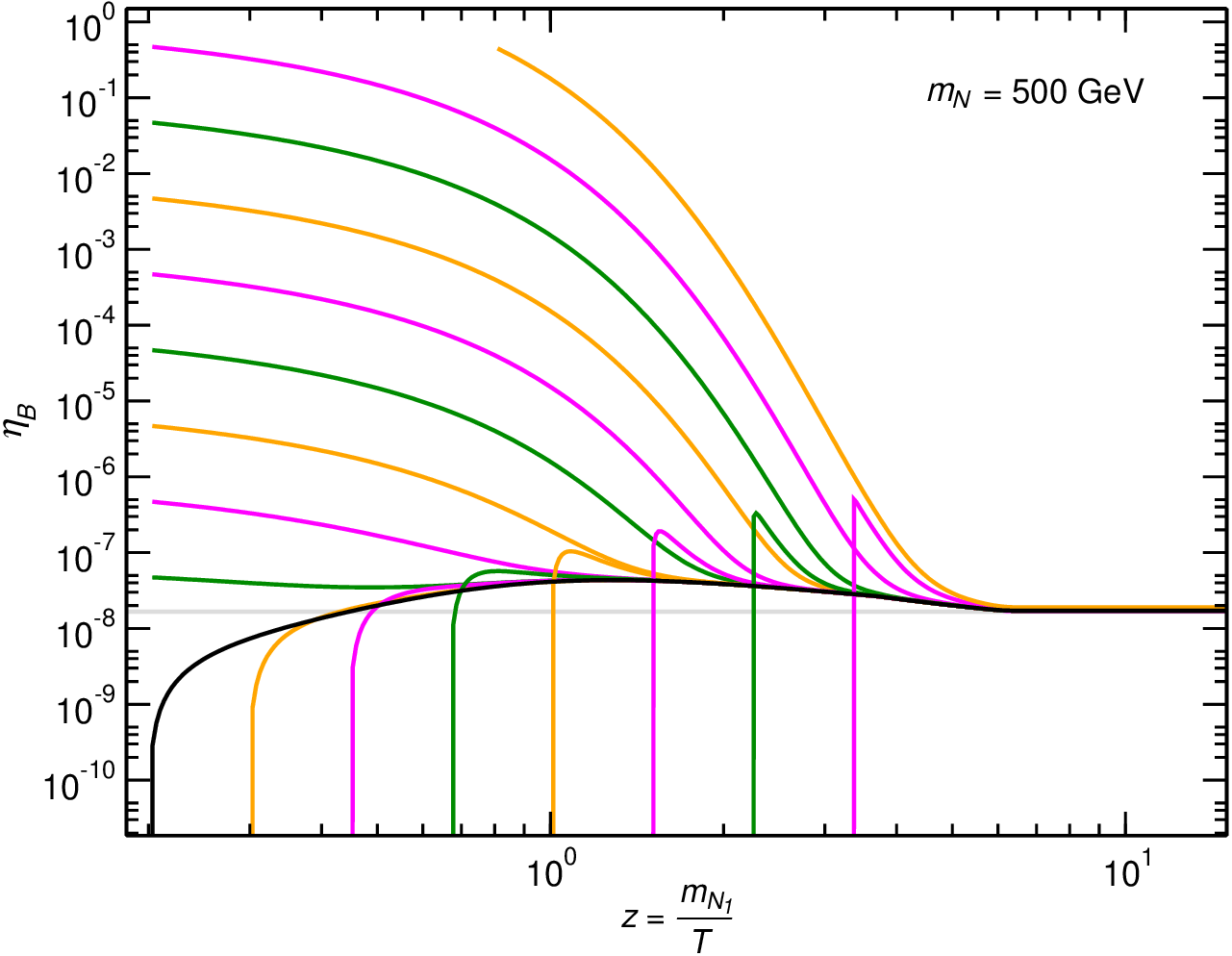}
\caption{\sl The complete independence of the final BAU on the initial
  lepton and baryon abundances for the $m_N = 500$~GeV scenario, with
model parameters the same as in Fig.~\ref{evolution-all}.}
\label{inicondsb}
\end{figure}

In R$\tau$L scenarios with $m_N > 250$~GeV, the final baryon asymmetry
is    completely  independent of  the  initial    conditions.  This is
illustrated in  Fig.~\ref{inicondsb} for  the  R$\tau$L scenario  with
$m_N=$~500~GeV.   In this numerical example,  it  is most striking  to
notice that the prediction for  the final BAU remains unchanged,  even
if  the initial  conditions are  set at  temperatures {\em below}  the
heavy neutrino mass scale $m_N$, e.g.~at $T \sim 0.5\, m_N$.

\begin{figure}[t]
\includegraphics[scale=0.73]{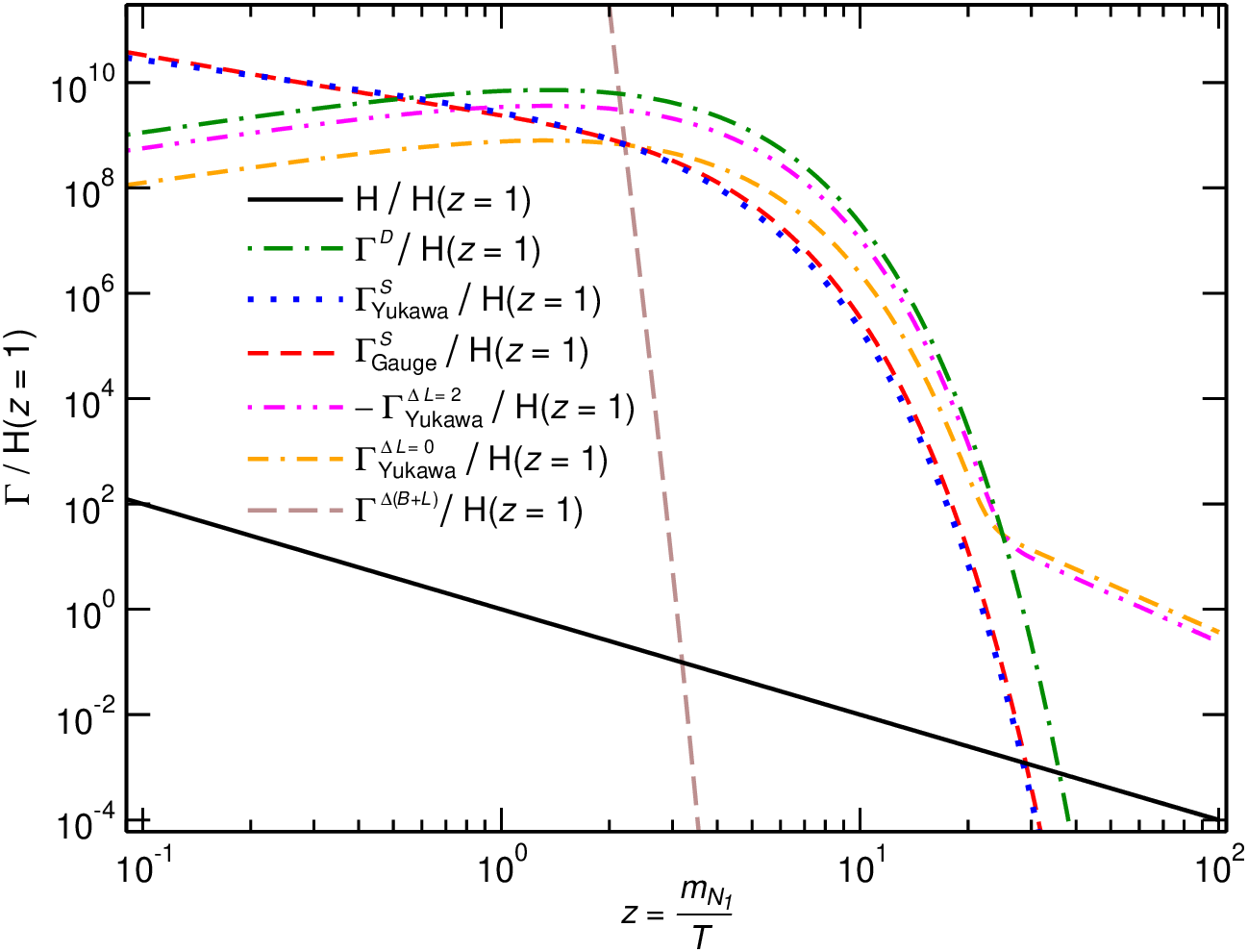}
\caption{\sl The ratio of various collision terms to the Hubble
parameter plotted for the $m_N = 250$~GeV scenario presented in
Fig.~\ref{evolution-all}.}
\label{collision-plot}
\end{figure}

Some  insight into the  independence on initial conditions is provided
by Fig.~\ref{collision-plot}. The ratios of various collision terms to
the Hubble  parameter  are plotted  for the  $m_N =$~250~GeV scenario.
These ratios show that RL can take  place almost completely in thermal
equilibrium;  in certain cases, the reaction  rates are many orders of
magnitude above the  Hubble parameter  $H  (z=1)$.  In  spite of  this
fact, RL  (R$\tau$L) can successfully  generate the required excess in
$L$ ($L_\tau$), because of the resonantly enhanced CP asymmetry.

To   allow   for a  simple   quantitative  understanding of the baryon
asymmetry in R$\tau$L (and  similar)  scenarios, we need to  introduce
the individual lepton flavour $K$-factors
\begin{equation}
K^{l}_{N_\alpha}\ =\ \frac{\Gamma (N_\alpha \to L_l \Phi)\: +\:
\Gamma(N_\alpha \to L^C_l \Phi^\dagger)}{H (z=1)}\ .
\end{equation}
Note that  the decay widths  are calculated in   terms of the one-loop
resummed effective Yukawa couplings~\cite{PU}.

\begin{table}
\begin{center}
\begin{tabular}{|cc||c|c|c|}
\hline
 $K_{N_{\alpha}}^{l}$ & & & $\alpha$ & $\qquad\qquad$ \\
 & & 1 & 2 & 3 \\
\hline \hline
 & $e$ & $1.0 \times 10^{10}$ & $1.0 \times 10^{10}$ & $25$\\ \hline
 $l$ & $\mu$ & $1.4 \times 10^{9}$ & $1.4 \times 10^{9}$ &  $20$\\ \hline
 & $\tau$ & $2.5$ & $2.5$ & $5.0$\\
\hline
\end{tabular}
\end{center}
\caption{\sl Individual lepton flavour $K$-factors for the $m_N =
250$~GeV scenario.} 
\label{RTLKfactors}
\end{table}

Table~\ref{RTLKfactors}    shows    the    various    components    of
$K_{N_{\alpha}}^{l}$ for the $m_N =$~250~GeV scenario. This explicitly
demonstrates   how  the  texture   provided  by   (\ref{hmatrix})  and
(\ref{epshnu})  allows   for  a  heavy  Majorana   neutrino  to  decay
relatively  out of equilibrium,  whilst simultaneously  protecting the
$\tau$-lepton number from being washed-out, even though large $e$- and
$\mu$-Yukawa couplings  to $N_{1,2}$ exist.  Bear in mind that  we use
the convention  $m_{N_1} < m_{N_2} < m_{N_3}$  upon diagonalization of
the heavy  Majorana neutrino mass matrix  $M_S$.  As can  be seen from
Table~\ref{RTLKfactors},    $K$-factors   $K^{e,\mu,\tau}_{N_3}   \sim
10$--100 and a  CP-asymmetry $\delta_{N_{3}}^{\tau} \sim -10^{-6}$ are
sufficient to generate a  large $\tau$-lepton asymmetry.  Although the
$K$-factors  $K^{e,\mu}_{N_{1,2}}$ associated  with $N_{1,2}$  and the
$e$ and  $\mu$ leptons are enormous of  order $10^9$--$10^{10}$, these
turn out to be harmless  to the $\tau$-lepton asymmetry, as the latter
is protected by  the low $\tau$-lepton $K$-factors $K^\tau_{N_{1,2,3}}
\sim 10$.

An  order  of magnitude    estimate  of the  final baryon   asymmetry,
including single lepton flavour effects, may be obtained using
\begin{equation}
  \label{Bestimate}
\eta_B\ \sim\ -\, 10^{-2}\,\times\, \sum_{l=1}^3\, \sum_{N_{\alpha}}\:
e^{-(m_{N_{\alpha}} - m_{N_1})/m_{N_1}}\, 
\delta^l_{N_{\alpha}}\: \frac{K^{l}_{N_{\alpha}}}{K_l\,K_{N_{\alpha}}}\ .
\end{equation}
The above estimate for $\eta_B$ is also consistent with the one stated
earlier  in~\cite{APtau}.  In~(\ref{Bestimate}),  the $K$-factors  are
summed in the following way:
\begin{eqnarray}
  \label{Kfactors}
K_{N_{\alpha}} \!&=&\!  \sum_{l = 1}^3\ K^{l}_{N_{\alpha}}\;,\qquad 
K_l \ =\ \sum_{N_{\alpha}}\, e^{-(m_{N_{\alpha}} - m_{N_1})/m_{N_1}}\,
K^{l}_{N_{\alpha}}\; . 
\end{eqnarray}
Notice that  all $K$-factors are  evaluated at $T = m_{N_1}$ (i.e.~$z=
m_{N_1}/T = 1$), where $m_{N_1}$ is the lightest of the heavy Majorana
neutrinos. The intuitive  estimate~(\ref{Bestimate}) is applicable for
all leptogenesis scenarios satisfying the approximate inequality
\begin{equation}
  \label{Kineq}
K_{lN_\alpha}\  \stackrel{>}{{}_\sim}\ 1\,,
\end{equation}
for each of the lepton flavours $l$ and the heavy Majorana neutrinos
$N_{1,2,3}$.  The inequality~(\ref{Kineq}) ensures that the energy scale
$m_{N_1}$ can be identified as the true scale of leptogenesis.

In RL scenarios, such as R$\tau$L, the importance of taking individual
lepton flavour  effects into account  can be demonstrated by comparing
(\ref{Bestimate})  with the  naive  estimate, in which  lepton flavour
effects are treated indiscriminately in a {\em universal} manner,
\begin{equation}
  \label{Bnoflavour}
\eta^{\,\mathrm{univ.}}_B\ \sim\ -\, 10^{-2}\,\times\, \sum_{N_{\alpha}}\:
e^{-(m_{N_{\alpha}} - m_{N_1})/m_{N_1}}\,\frac{\delta_{N_{\alpha}}}{K}\,,
\end{equation}
where  $K = \sum_{l=e,\mu\,\tau} K_l$.   In the R$\tau$L scenario with
$m_N = 250$~GeV, the dominant  contribution to this estimate will come
from  $N_3$,   with  a total  CP asymmetry  $\delta_{N_3}\sim10^{-3}$.
Taking the ratio of the two estimates yields
\begin{equation}
\frac{\eta^{\,\mathrm{univ.}}_B}{\eta_B}\ \sim\
\frac{\delta_{N_3}}{\delta^{\tau}_{N_{3}}}\,
\frac{K_{N_3}\,K_{\tau}}{K^{\tau}_{N_3}\,K} \ \approx\ 
\frac{\delta_{N_3}}{\delta^{\tau}_{N_{3}}}\,
\frac{|c|^2}{|a|^2+|b|^2} \ \approx\ -\,10^{-6}\,.
\end{equation}
Thus,    without considering single  lepton  flavour   effects in this
particular R$\tau$L model, one obtains an erroneous prediction for the
BAU, which is  suppressed by 6 orders  of magnitude and has the  wrong
sign.   These   estimates are confirmed  by  solving  the total lepton
number BEs presented in~\cite{PU}.

\begin{figure}[t]
\begin{center}
\includegraphics[scale=0.6]{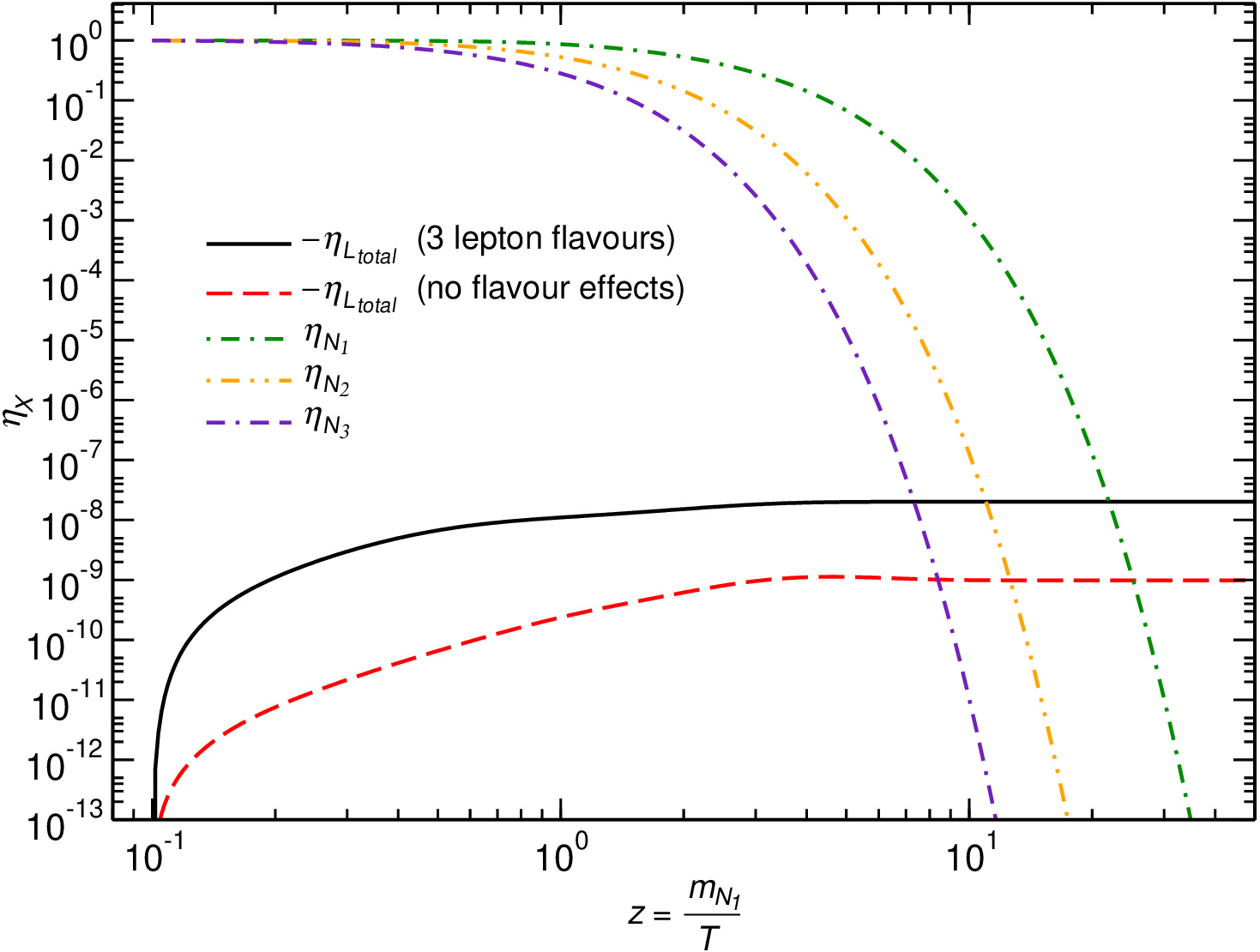}
\end{center}
\caption{\sl The predicted evolution of $\eta_{L_{\rm total}}$ and
  $\eta_{N_{\alpha}}$ for a model with a heavy neutrino spectrum:
  $m_{N_3} = 3\,m_{N_1}$, $m_{N_2} = 2\,m_{N_1}$ and $m_{N_1} =
  10^{10}$~GeV.}
\label{hierarchy-plot}
\end{figure}

In   a  hierarchical scenario,  the number    densities of the heavier
neutrinos $N_{2,3}$ at $T = m_{N_1}$ will be Boltzmann suppressed.  To
account for this phenomenon,   we have included the  Boltzmann factors
$e^{-(m_{N_\alpha}      -          m_{N_1})/m_{N_1}}$         in   the
estimates~(\ref{Bestimate}), (\ref{Bnoflavour})  and in the definition
of $K_l$.  Clearly,  in RL  models   with each  heavy neutrino  nearly
degenerate in mass, this last factor can be set to 1.

Flavour  effects   can  also  play   a  significant  role   in  mildly
hierarchical   scenarios.    Figure~\ref{hierarchy-plot}   shows   the
predicted  evolution  of the  lepton  asymmetry  in  a scenario  where
$m_{N_3}  =  3\,m_{N_1}$,  $m_{N_2}   =  2\,m_{N_1}$  and  $m_{N_1}  =
10^{10}$~GeV.  The Yukawa  texture was  chosen to  be  consistent with
light neutrino data and a normal hierarchical light neutrino spectrum.
In  this   example,  neglecting  individual   lepton  flavour  effects
introduces  an ${\cal  O}(10)$  suppression of  the  final lepton  and
baryon asymmetry.

In fully  hierarchical  scenarios satisfying (\ref{Kineq}),  it can be
seen that the estimates  (\ref{Bestimate}) and (\ref{Bnoflavour})  are
completely    equivalent.  A large   hierarchy   in the heavy neutrino
spectrum, combined with  the condition (\ref{Kineq}), implies that the
final  lepton  asymmetry is determined   entirely by the decay  of the
lightest heavy Majorana neutrino  $N_1$. This fact makes it impossible
for a single lepton flavour to be protected  from wash-out, whilst the
neutrino decays out of equilibrium. 

Likewise,  in     flavour     universal    scenarios,   where
$\eta_{L_{e,\mu,\tau}}     = \frac{1}{3}     \eta_L$,  the   estimates
(\ref{Bestimate}) and (\ref{Bnoflavour}) are completely equivalent for
both nearly degenerate and hierarchical leptogenesis scenarios.

Our  numerical  analysis  presented  in this  section  has  explicitly
demonstrated that models of  R$\tau$L can provide a viable explanation
for the  observed BAU, in  accordance with the current  light neutrino
data.  In the  next section, we will see  how the scenarios considered
here  have far reaching  phenomenological implications  for low-energy
observables   of   lepton   flavour/number  violation   and   collider
experiments.

\setcounter{equation}{0}
\section{Phenomenological Implications}

RL models, and  especially R$\tau$L models, can give  rise to a number
of  phenomenologically testable  signatures.  In  particular,  we will
analyze   the  generic   predictions  of   R$\tau$L  models   for  the
$0\nu\beta\beta$ decay, and for  the LFV processes: $\mu \to e\gamma$,
$\mu \to eee$ and $\mu \to  e$ conversion in nuclei.  Finally, we will
present simple  and realistic numerical estimates  of production cross
sections  of   heavy  Majorana   neutrinos  at  future   $e^+e^-$  and
$\mu^+\mu^-$  colliders,  and  apply  these results  to  the  R$\tau$L
models.

\subsection{{\boldmath $0\nu\beta\beta$} Decay}

Neutrinoless  double beta  decay  ($0\nu\beta\beta$) corresponds  to a
process in which  two single  $\beta$ decays~\cite{doi85,Klapdor,HKK}
occur simultaneously  in  one nucleus.  As  a consequence  of this,  a
nucleus $(Z,A)$ gets converted into a nucleus $(Z+2,A)$, i.e.\
\begin{displaymath}        
^{A}_{Z}\,X\ \to\ ^A_{Z+2}\,X\: +\: 2 e^-\; . 
\end{displaymath}
Evidently, this  process violates $L$-number  by two units and can
naturally take place in minimal RL models, in which the observed light
neutrinos are Majorana particles.   The observation of such a  process
would provide  further   information on the   structure   of the light
neutrino mass matrix ${\bf m}^\nu$. 

To  a very  good approximation, the  half life  for a $0\nu\beta\beta$
decay mediated by light Majorana neutrinos is given by
\begin{equation}
  \label{t1/2}
[T^{0\nu\beta\beta}_{1/2}]^{-1}\ =\ \frac{|\langle m
\rangle |^2}{m^2_e}\ |{\cal M}_{0\nu\beta\beta}|^2\, G_{01}\; ,
\end{equation} 
where $\langle  m  \rangle$  denotes the  effective Majorana neutrino
mass, $m_e$ is the  electron mass and ${\cal  M}_{0\nu\beta\beta}$ and
$G_{01}$  denote the appropriate nuclear  matrix element and the phase
space factor, respectively.  More details regarding the calculation of
$T^{0\nu\beta\beta}_{1/2}$ may be found in~\cite{doi85,Klapdor,HKK}.

In models of interest to us, the effective   neutrino mass can  be
related to  the entry  $\{ 11  \}\ (\equiv \{   ee \})$ of   the light
neutrino mass matrix ${\bf m}^\nu$ in (\ref{mnutree}), i.e.
\begin{equation}
|\langle m \rangle|\ =\  |({\bf m}^\nu)_{ee}|\ =\ 
\frac{v^2}{2m_N}\: \bigg|\,\frac{\Delta m_N}{m_N}\,a^2\ -\
\varepsilon^2_e\,\bigg|\; . 
\end{equation}
As has been explicitly  demonstrated in the previous section, R$\tau$L
models  realize   a  light   neutrino  mass  spectrum   with an inverted
hierarchy~\cite{KKS},  thus  giving   rise  to  a  sizeable  effective
neutrino  mass.  The  prediction for  $|\langle m  \rangle|$  in these
models is
\begin{equation}
|\langle m \rangle|\ =\  |({\bf m}^\nu)_{ee}|\ 
\approx\ 0.013~{\rm eV}\; . 
\end{equation}
Such  a prediction  lies  at the  very low  end  of the  value of  the
effective   Majorana   neutrino  mass,   reported   recently  by   the
Heidelberg--Moscow  collaboration~\cite{HMexp}.   There are  proposals
for future  $0\nu\beta\beta$-decay experiments that  will be sensitive
to  values  of $|\langle  m  \rangle|$  of order  $10^{-2}$~\cite{CA},
significantly increasing the constraints on this parameter.

\subsection{\boldmath $\mu \to e\gamma$}

\begin{figure}

\begin{center}
\begin{picture}(360,400)(0,0)
\SetWidth{0.8}

\ArrowLine(0,360)(20,360)\ArrowLine(60,360)(80,360)
\GCirc(40,360){20}{0.7}\Photon(40,340)(40,320){3}{2}
\Text(0,365)[b]{$\mu$}\Text(80,365)[b]{$e$}
\Text(42,320)[l]{$Z,\gamma$}

\Text(100,360)[]{$=$}

\ArrowLine(120,360)(140,360)\ArrowLine(180,360)(200,360)
\ArrowLine(140,360)(180,360)\Text(160,367)[b]{$N_\alpha$}
\DashArrowArc(160,360)(20,180,270){3}\PhotonArc(160,360)(20,270,360){2}{3}
\Text(145,340)[r]{$G^-$}\Text(175,340)[l]{$W^-$}
\Photon(160,340)(160,320){3}{2}
\Text(120,365)[b]{$\mu$}\Text(200,365)[b]{$e$}
\Text(162,320)[l]{$Z,\gamma$}
\Text(160,300)[]{\bf (a)}

\ArrowLine(240,360)(260,360)\ArrowLine(300,360)(320,360)
\ArrowLine(260,360)(300,360)\Text(280,367)[b]{$N_\alpha$}
\DashArrowArc(280,360)(20,270,360){3}\PhotonArc(280,360)(20,180,270){2}{3}
\Text(265,340)[r]{$W^-$}\Text(295,340)[l]{$G^-$}
\Photon(280,340)(280,320){3}{2}
\Text(240,365)[b]{$\mu$}\Text(320,365)[b]{$e$}
\Text(282,320)[l]{$Z,\gamma$}
\Text(280,300)[]{\bf (b)}

\ArrowLine(0,260)(20,260)\ArrowLine(60,260)(80,260)
\ArrowLine(20,260)(60,260)\Text(40,267)[b]{$N_\alpha$}
\PhotonArc(40,260)(20,180,270){2}{3}\PhotonArc(40,260)(20,270,360){2}{3}
\Text(25,240)[r]{$W^-$}\Text(55,240)[l]{$W^-$}
\Photon(40,240)(40,220){3}{2}
\Text(0,265)[b]{$\mu$}\Text(80,265)[b]{$e$}
\Text(42,220)[l]{$Z,\gamma$}
\Text(40,200)[]{\bf (c)}

\ArrowLine(120,260)(140,260)\ArrowLine(180,260)(200,260)
\ArrowLine(140,260)(180,260)\Text(160,267)[b]{$N_\alpha$}
\DashArrowArc(160,260)(20,180,270){3}\DashArrowArc(160,260)(20,270,360){2}
\Text(145,240)[r]{$G^-$}\Text(175,240)[l]{$G^-$}
\Photon(160,240)(160,220){3}{2}
\Text(120,265)[b]{$\mu$}\Text(200,265)[b]{$e$}
\Text(162,220)[l]{$Z,\gamma$}
\Text(160,200)[]{\bf (d)}

\ArrowLine(240,260)(260,260)\ArrowLine(300,260)(320,260)
\Photon(260,260)(300,260){2}{4}\Text(280,267)[b]{$W^-$}
\ArrowLine(260,260)(280,240)\ArrowLine(280,240)(300,260)
\Text(270,240)[r]{$N_\alpha$}\Text(295,240)[l]{$N_\beta$}
\Photon(280,240)(280,220){3}{2}
\Text(240,265)[b]{$\mu$}\Text(320,265)[b]{$e$}
\Text(282,220)[l]{$Z$}
\Text(280,200)[]{\bf (e)}

\ArrowLine(0,160)(20,160)\ArrowLine(60,160)(80,160)
\DashArrowLine(20,160)(60,160){3}\Text(40,167)[b]{$G^-$}
\ArrowLine(20,160)(40,140)\ArrowLine(40,140)(60,160)
\Text(30,140)[r]{$N_\alpha$}\Text(55,140)[l]{$N_\beta$}
\Photon(40,140)(40,120){3}{2}
\Text(0,165)[b]{$\mu$}\Text(80,165)[b]{$e$}
\Text(42,120)[l]{$Z$}
\Text(40,100)[]{\bf (f)}

\ArrowLine(120,160)(135,160)\ArrowLine(135,160)(150,160)
\ArrowLine(150,160)(180,160)\ArrowLine(180,160)(200,160)
\Text(120,165)[b]{$\mu$}\Text(142,165)[b]{$\mu$}
\Text(165,165)[b]{$N_\alpha$}\Text(200,165)[b]{$e$}
\Photon(135,160)(135,120){3}{4}
\Text(137,120)[l]{$Z,\gamma$}
\PhotonArc(165,160)(15,180,360){2}{5}\Text(165,135)[]{$W^-$}
\Text(160,100)[]{\bf (g)}

\ArrowLine(240,160)(255,160)\ArrowLine(255,160)(270,160)
\ArrowLine(270,160)(300,160)\ArrowLine(300,160)(320,160)
\Text(240,165)[b]{$\mu$}\Text(262,165)[b]{$\mu$}
\Text(285,165)[b]{$N_\alpha$}\Text(320,165)[b]{$e$}
\Photon(255,160)(255,120){3}{4}
\Text(257,120)[l]{$Z,\gamma$}
\DashArrowArc(285,160)(15,180,360){3}\Text(285,135)[]{$G^-$}
\Text(280,100)[]{\bf (h)}

\ArrowLine(0,60)(20,60)\ArrowLine(20,60)(50,60)
\ArrowLine(50,60)(65,60)\ArrowLine(65,60)(80,60)
\Text(0,65)[b]{$\mu$}\Text(57,65)[b]{$e$}
\Text(35,65)[b]{$N_\alpha$}\Text(80,65)[b]{$e$}
\Photon(65,60)(65,20){3}{4}
\Text(67,20)[l]{$Z,\gamma$}
\PhotonArc(35,60)(15,180,360){2}{5}\Text(35,35)[]{$W^-$}
\Text(40,0)[]{\bf (i)}

\ArrowLine(120,60)(140,60)\ArrowLine(140,60)(170,60)
\ArrowLine(170,60)(185,60)\ArrowLine(185,60)(200,60)
\Text(120,65)[b]{$\mu$}\Text(177,65)[b]{$e$}
\Text(155,65)[b]{$N_\alpha$}\Text(200,65)[b]{$e$}
\Photon(185,60)(185,20){3}{4}
\Text(187,20)[l]{$Z,\gamma$}
\DashArrowArc(155,60)(15,180,360){3}\Text(155,35)[]{$G^-$}
\Text(160,0)[]{\bf (j)}

\end{picture}\\[0.7cm]
\end{center}

\caption{\sl Feynman graphs pertaining to the effective $\gamma e\mu$ and
$Ze\mu$ couplings.}\label{fig:meg}

\end{figure}
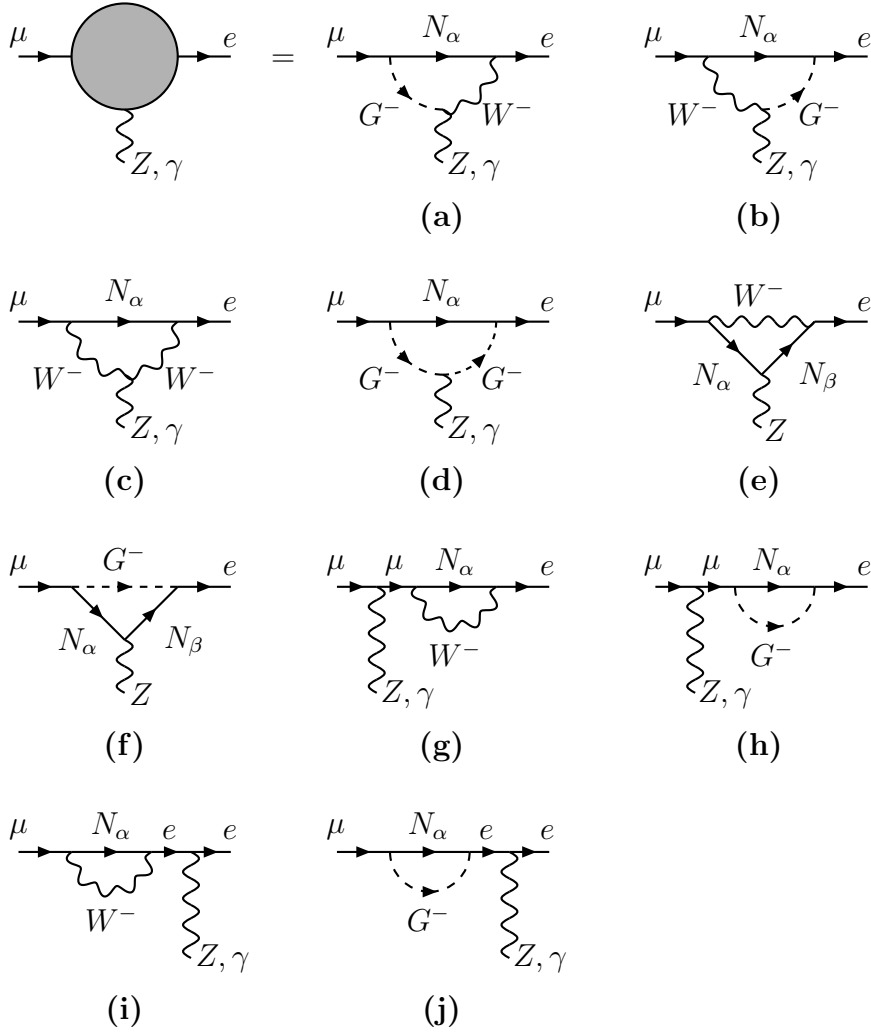

As shown  in Fig.~\ref{fig:meg},  heavy Majorana neutrinos  may induce
LFV  couplings to the  photon ($\gamma$)  and the  $Z$ boson  via loop
effects.  These  couplings give rise to  LFV decays, such  as $\mu \to
e\gamma$~\cite{CL}  and $\mu \to  eee$~\cite{IP}.  Our  discussion and
notation closely follows the extensive studies~\cite{IP,Ara}.  Related
phenomenological  analyses  of LFV  effects  in  the  SM with  singlet
neutrinos may be found in~\cite{KPS,BKPS,MPLAP,LFVrev}.

%******************************************************************
%   Figure on  \gamma - e - mu   and   Z - e - mu   couplings
%******************************************************************

To properly describe LFV in low-energy observables, we first introduce
the so-called Langacker--London (LL) parameters~\cite{LL}:
\begin{equation}
  \label{LL}
(s^{\nu_l}_L)^2\ =\  1\, -\, 
\sum_{l' = e,\mu,\tau}\, |B_{l\nu_{l'}}|^2\ \approx\ 
\Big( m^*_D\, M^{*\,-1}_S\, M^{-1}_S\, m^T_D\Big)_{ll}\ ,
\end{equation}
where $m_D =  \frac{1}{\sqrt{2}}\,h^{\nu_R}\, v$ and $B_{l\nu_l'}$ are
mixing-matrix factors close to 1  that  multiply the SM tree-level  $W
l\nu_{l'}$ vertices~\cite{AZPC}.   The LL parameters $(s^{\nu_l}_L)^2$
quantify  the deviation of the  actual squared $W l\nu_{l'}$ couplings
(summed over   all light  neutrinos)   from the corresponding   sum of
squared   couplings       in     the     SM.       The      parameters
$(s^{\nu_{e,\mu,\tau}}_L)^2$   are  constrained by  LEP and low-energy
electroweak    observables~\cite{LL,LLfit}. Independent constraints on
these    parameters     typically  give:   $(s^{\nu_{e,\mu,\tau}}_L)^2
\stackrel{<}{{}_\sim} 10^{-2}$.  As we  will see in a moment, however,
LFV observables impose much more severe constraints on products of the
LL parameters, and especially on~$s^{\nu_e}_L\, s^{\nu_\mu}_L$.

As  a first LFV  observable, we consider  the decay $\mu \to e\gamma$.
The branching fraction for the decay $\mu \to e\gamma$ is given by
\begin{equation}
  \label{Bllgamma}
B(\mu \to e \gamma )\ =\ \frac{\alpha^3_w s^2_w}{256\pi^2}\, 
\frac{m^4_\mu}{M^4_W}\, \frac{m_\mu}{\Gamma_\mu}\, |G_\gamma^{\mu e}|^2\
\approx\ \frac{\alpha^3_w s^2_w}{1024\pi^2}\, 
\frac{m^4_\mu}{M^4_W}\, \frac{m_\mu}{\Gamma_\mu}\; (s_L^{\nu_\mu})^2 
(s_L^{\nu_e})^2\, ,
\end{equation}
where   $\Gamma_\mu  =  2.997\times  10^{-19}$~GeV~\cite{PDG}   is the
experimentally measured muon decay  width, and $G_\gamma^{\mu e}$ is a
composite  form-factor defined in~\cite{IP}.   In arriving at the last
equality  in~(\ref{Bllgamma}), we have   assumed that  $m^2_N\gg
M^2_W$,  for   a  model with two     nearly degenerate heavy  Majorana
neutrinos.     In this case,  one finds    that $G_\gamma^{\mu e}  \to
\frac{e^{i\phi}}{2}\, s_L^{\nu_\mu} s_L^{\nu_e}$,  where  $\phi$ is an
unobservable  model-dependent      phase.  Confronting the theoretical
prediction~(\ref{Bllgamma})         with    the   experimental   upper
limit~\cite{PDG}
\begin{equation}
  \label{Bexpllgamma}
B_{\rm exp} (\mu \to e \gamma)\ <\ 1.2\, \times 10^{-11}\, ,\quad
\end{equation}
we obtain the following constraint:
\begin{equation}
  \label{meglimit}
s_L^{\nu_e} s_L^{\nu_\mu}\  <\ 1.2\times 10^{-4}\, .
\end{equation}
This last constraint  is  stronger by one  to two  orders of magnitude
with respect    to   those     derived  on    $(s_L^{\nu_e})^2$    and
$(s_L^{\nu_\mu})^2$ individually.

In R$\tau$L models, only two of the right-handed neutrinos, $\nu_{2R}$
and    $\nu_{3R}$,  which have    appreciable   $e$-  and $\mu$-Yukawa
couplings, will  be relevant  to  LFV effects.  In  this case, the  LL
parameters  $(s_L^{\nu_e})^2$ and $(s_L^{\nu_\mu})^2$   are, to a very
good approximation, given by
\begin{equation}
(s_L^{\nu_e})^2\ =\ \frac{|a|^2\, v^2}{m^2_N}\; ,\qquad  
(s_L^{\nu_\mu})^2\ =\ \frac{|b|^2\, v^2}{m^2_N}\; .
\end{equation}
Then, the following theoretical prediction is obtained:
\begin{equation}
  \label{RtLmeg}
B(\mu \to e \gamma )\ =\ 9 \cdot 10^{-4}\, \times\,
\frac{|a|^2\,|b|^2\, v^4}{m^4_N}\ .
\end{equation}
For the  particular scenarios  considered in Section~\ref{sec:num}, we
find  $B(\mu \to e  \gamma )  \sim   10^{-12}$. These values are  well
within reach of  the MEG collaboration,  which will be sensitive to $B
(\mu \to e\gamma ) \sim 10^{-14}$~\cite{MEG}.

\subsection{\boldmath $\mu \to eee$}

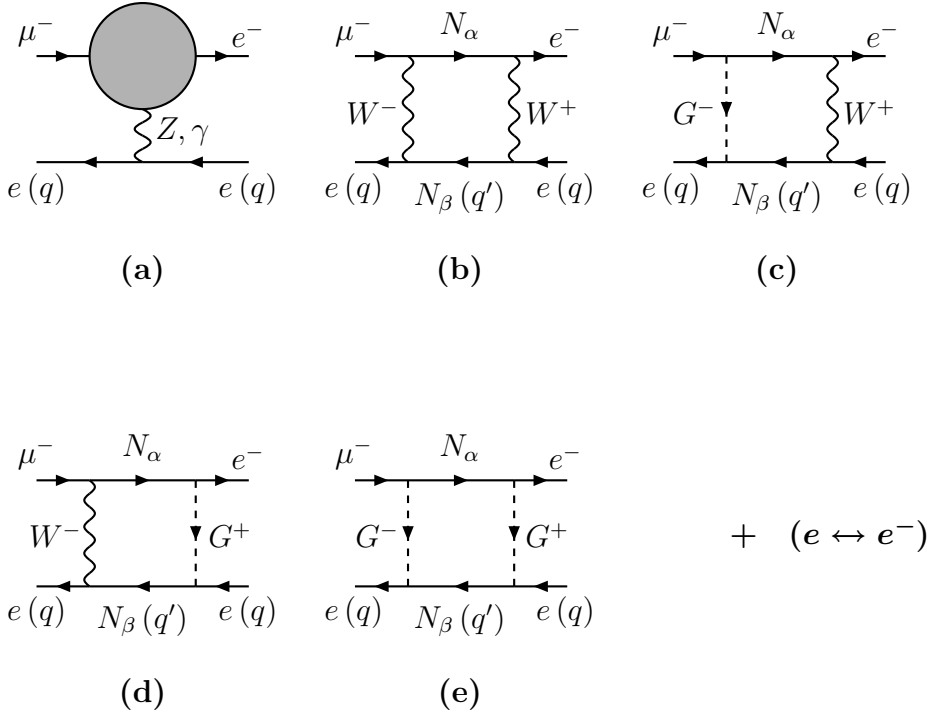
\begin{figure}

\begin{center}
\begin{picture}(360,200)(0,190)
\SetWidth{0.8}

\ArrowLine(0,360)(20,360)\ArrowLine(60,360)(80,360)
\GCirc(40,360){20}{0.7}\Photon(40,340)(40,320){3}{2}
\Text(0,365)[b]{$\mu^-$}\Text(80,365)[b]{$e^-$}
\Text(45,333)[l]{$Z,\gamma$}
\ArrowLine(40,320)(0,320)\ArrowLine(80,320)(40,320)
\Text(0,317)[t]{$e\, (q)$}\Text(80,317)[t]{$e\, (q)$}
\Text(40,280)[]{\bf (a)}

\ArrowLine(120,360)(140,360)\ArrowLine(180,360)(200,360)
\ArrowLine(140,360)(180,360)\Text(160,368)[b]{$N_\alpha$}
\ArrowLine(140,320)(120,320)\ArrowLine(200,320)(180,320)
\ArrowLine(180,320)(140,320)\Text(160,314)[t]{$N_\beta\, (q')$}
\Photon(140,360)(140,320){2}{4}\Photon(180,360)(180,320){2}{4}
\Text(120,365)[b]{$\mu^-$}\Text(200,365)[b]{$e^-$}
\Text(120,317)[t]{$e\, (q)$}\Text(200,317)[t]{$e\, (q)$}
\Text(137,340)[r]{$W^-$}\Text(185,340)[l]{$W^+$}
\Text(160,280)[]{\bf (b)}

\ArrowLine(240,360)(260,360)\ArrowLine(300,360)(320,360)
\ArrowLine(260,360)(300,360)\Text(280,368)[b]{$N_\alpha$}
\ArrowLine(260,320)(240,320)\ArrowLine(320,320)(300,320)
\ArrowLine(300,320)(260,320)\Text(280,314)[t]{$N_\beta\, (q')$}
\DashArrowLine(260,360)(260,320){3}\Photon(300,360)(300,320){2}{4}
\Text(240,365)[b]{$\mu^-$}\Text(320,365)[b]{$e^-$}
\Text(240,317)[t]{$e\, (q)$}\Text(320,317)[t]{$e\, (q)$}
\Text(257,340)[r]{$G^-$}\Text(305,340)[l]{$W^+$}
\Text(280,280)[]{\bf (c)}

\ArrowLine(0,200)(20,200)\ArrowLine(60,200)(80,200)
\ArrowLine(20,200)(60,200)\Text(40,208)[b]{$N_\alpha$}
\ArrowLine(20,160)(0,160)\ArrowLine(80,160)(60,160)
\ArrowLine(60,160)(20,160)\Text(40,154)[t]{$N_\beta\, (q')$}
\Photon(20,200)(20,160){2}{4}\DashArrowLine(60,200)(60,160){3}
\Text(0,205)[b]{$\mu^-$}\Text(80,205)[b]{$e^-$}
\Text(0,157)[t]{$e\, (q)$}\Text(80,157)[t]{$e\, (q)$}
\Text(17,180)[r]{$W^-$}\Text(65,180)[l]{$G^+$}
\Text(40,120)[]{\bf (d)}

\ArrowLine(120,200)(140,200)\ArrowLine(180,200)(200,200)
\ArrowLine(140,200)(180,200)\Text(160,208)[b]{$N_\alpha$}
\ArrowLine(140,160)(120,160)\ArrowLine(200,160)(180,160)
\ArrowLine(180,160)(140,160)\Text(160,154)[t]{$N_\beta\, (q')$}
\DashArrowLine(140,200)(140,160){3}\DashArrowLine(180,200)(180,160){3}
\Text(120,205)[b]{$\mu^-$}\Text(200,205)[b]{$e^-$}
\Text(120,157)[t]{$e\, (q)$}\Text(200,157)[t]{$e\, (q)$}
\Text(137,180)[r]{$G^-$}\Text(185,180)[l]{$G^+$}
\Text(160,120)[]{\bf (e)}

\Text(300,180)[]{\boldmath $+\quad ( e \leftrightarrow e^-)$}

\end{picture}\\
\end{center}
\vspace{2.5cm}
\caption{\sl Feynman graphs responsible for  $\mu^- \to e^- e^+ e^-$ 
($\mu \to e$ conversion in nuclei).}\label{fig:mue}

\end{figure}

As  illustrated  in Fig.~\ref{fig:mue},  quantum effects   mediated by
heavy  Majorana neutrinos may also give  rise  to the 3-body LFV decay
mode $\mu^-  \to e^-e^+e^-$.  The  branching ratio  for this LFV decay
may conveniently be expressed as
\begin{eqnarray}
  \label{Bmueee}
B(\mu \to eee ) &=& 
     \frac{\alpha_w^4}{24576\pi^3}\ \frac{m_\mu^4}{M_W^4}\
     \frac{m_\mu}{\Gamma_\mu}\bigg\{\, 2 |{\textstyle \frac{1}{2}}
        F_{\rm box}^{\mu eee}+F_Z^{\mu e}
            -2s_w^2(F_Z^{\mu e}-F_\gamma^{\mu e})|^2\nonumber\\
&&  +\, 4s_w^4 |F_Z^{\mu e}-F_\gamma^{\mu e}|^2\, +\, 
16s_w^2\: {\rm Re} \Big[(F_Z^{\mu e}+{\textstyle \frac{1}{2}}
     F_{\rm box}^{\mu eee})
      G_\gamma^{\mu e\:\ast}\Big]\nonumber\\
&&-\, 48s_w^4\: {\rm Re} \Big[\,(F_Z^{\mu e}-F_\gamma^{\mu e})
G_\gamma^{\mu e\:\ast}\,\Big]\, +\, 32s_w^4 |G_\gamma^{\mu e} |^2\: 
     \bigg(\ln\frac{m_\mu^2}{m_e^2}-\frac{11}{4}\bigg)\, \bigg\}\; .\qquad
\end{eqnarray}
The   expressions  $F_\gamma^{\mu  e}$,   $F_Z^{\mu   e}$ and  $F_{\rm
box}^{\mu   eee}$ are composite    form-factors, defined  and computed
in~\cite{IP}.  In the  limit $m^2_N \gg   M^2_W$ and up to an  overall
physically   irrelevant   phase  factor   $e^{i\phi}$, these composite
form-factors simplify to~\cite{IP}
\begin{eqnarray}
  \label{Fgamma1}
F_\gamma^{\mu e} &\approx & -\,\frac{7}{12}\, s_L^{\nu_\mu} s_L^{\nu_{e}}\,
 -\, \frac{1}{6}\, s_L^{\nu_\mu} s_L^{\nu_{e}} 
\ln\bigg(\frac{m^2_N}{M^2_W}\bigg)\, , \\  
  \label{FZ1}
F_Z^{\mu e} &\approx& 
\bigg[\, \frac{5}{2}\, -\, \frac{3}{2}\,
\ln\bigg(\frac{m^2_N}{M^2_W}\bigg)\,  \bigg]\,
s_L^{\nu_\mu}s_L^{\nu_e}\
-\ \frac{1}{2} s_L^{\nu_\mu}s_L^{\nu_e}\sum_{k=e,\mu,\tau}\ 
(s_L^{\nu_k})^2\:\frac{m^2_N}{M^2_W}\, ,\\ 
  \label{Fbox1}
F_{\rm box}^{\mu eee} &\approx & 
-2\, s_L^{\nu_\mu}s_L^{\nu_{e}}\  +\
\frac{1}{2}\; s_L^{\nu_\mu}s_L^{\nu_e} (s_L^{\nu_e})^2\: 
\frac{m^2_N}{M^2_W}\; . 
\end{eqnarray}
Correspondingly, the analytic  result~(\ref{Bmueee}) in the same limit
may be cast into the form:
\begin{eqnarray}
  \label{Amueee}
B (\mu\to e e e) \!& \simeq &\! 
\frac{\alpha_w^4}{294912\,\pi^3}\
\frac{m_\mu^4}{M_W^4}\ \frac{m_\mu}{\Gamma_\mu}\ (s_L^{\nu_\mu})^2
(s_L^{\nu_e})^2\: \Bigg\{\, 54-300s_w^2+217s_w^4
+96\,s_w^4\ln\bigg(\frac{m_{\mu}^2}{m_e^2}\bigg)\nonumber\\
&&\hspace{-0.7cm} 
-\: \Big(108-492s_w^2+800s_w^4\Big)\ln\bigg(\frac{m^2_N}{M^2_W}\bigg)
+\ \Big(54-192s_w^2+256s_w^4\Big)\ln^2\bigg(\frac{m^2_N}{M^2_W}\bigg)
\nonumber\\
&&\hspace{-0.7cm}+\ \frac{m^2_N}{M^2_W}\ \Bigg[\,
\bigg(18- 50s_w^2 - \Big(18-32s_w^2\Big)
\ln\bigg(\frac{m^2_N}{M^2_W}\bigg)\bigg)\ (s_L^{\nu_e})^2\nonumber\\[2mm]
&&\hspace{-0.7cm}-\: 
\bigg(36-172s_w^2+300s_w^4 - \Big(36-136s_w^2+192s_w^4\Big)
\ln\bigg(\frac{m^2_N}{M^2_W}\bigg)\bigg)
\sum_{l=e,\mu,\tau} (s_L^{\nu_l})^2
\Bigg]\nonumber\\[2mm]
&&\hspace{-0.7cm}+\: \frac{m^4_N}{M^4_W}\ \Bigg[\,
\frac{3}{2}(s_L^{\nu_e})^4\ -\ 6\,(1-2s^2_w)(s_L^{\nu_e})^2
\sum_{l=e,\mu,\tau} (s_L^{\nu_l})^2\nonumber\\
&&\hspace{-0.7cm}+\: 6\,\Big(1-4s_w^2+6s^4_w\Big) 
\Big(\sum_{l =e,\mu,\tau} (s_L^{\nu_l})^2\Big)^2\,
\Bigg]\, \Bigg\}\; .  
\end{eqnarray} 
It can be  seen from~(\ref{Amueee}) that  the so-called non-decoupling
terms proportional to  $m^4_N/M^4_W$ are always multiplied with higher
powers of the LL parameters.  In general,  these terms do not decouple
and become very significant~\cite{IP}, for large heavy neutrino masses
$m_N$ and fixed values  of  $s^{\nu_l}_L$, which amounts to  scenarios
with     large    neutrino     Yukawa    couplings    $|h^{\nu_R}_{ij}|
\stackrel{>}{{}_\sim} 0.5$~\cite{AZPC}.  However, these non-decoupling
terms are  negligible, as long as $s^{\nu_l}_L\,  m_N/M_W \ll 1$. This
is    actually   the case  for  the    R$\tau$L   models discussed  in
Section~\ref{sec:num}.  Neglecting terms proportional to $m^2_N/M^2_W$
and $m^4_N/M^4_W$,  we may  relate $B  (\mu\to e e   e)$ to $B(\mu \to
e\gamma)$ through:
\begin{equation}
  \label{Rmueee}
B (\mu\to e e e)\ \simeq\  
8.2 \cdot 10^{-3}\, \times
\Bigg[\,1\ -\ 0.8\ln\bigg(\frac{m^2_N}{M^2_W}\bigg)
\ +\ 0.5 \ln^2\bigg(\frac{m^2_N}{M^2_W}\bigg)
\, \Bigg]\: B(\mu\to e\gamma)\; .
\end{equation}
For  example,    for   an R$\tau$L   model    with  $m_N  =  250$~GeV,
(\ref{Rmueee}) implies
\begin{equation}
B (\mu\to e e e)\ \simeq\ 1.4\cdot 10^{-2}\,\times B(\mu \to e\gamma)\
\simeq\ 1.4\cdot 10^{-14}
\end{equation}
This value is a factor $\sim  70$ below the present experimental
bound~\cite{PDG}:  $B_{\rm exp} (\mu \to eee)  < 1.0 \times 10^{-12}$.
In this respect,  it would be  very encouraging, if higher sensitivity
experiments could be designed to probe this observable.

%******************************************************************
%   Figure \mu -> e e e  and \mu - e conversion
%******************************************************************

\subsection{Coherent {\boldmath $\mu \to e$} Conversion in Nuclei}

One of the most sensitive experiments to  LFV is the coherent conversion of 
$\mu \to e$ in nuclei, e.g.\ $\mu^-\  {}^{48}_{22}{\rm Ti} \to e^-\
{}^{48}_{22}{\rm Ti}$~\cite{FW,JDV}.    The Feynman graphs responsible
for such a process are displayed in Fig.~\ref{fig:mue}.

Our calculation of $\mu \to e$ conversion in nuclei closely
follows~\cite{FW,JDV,Ara}. We consider the kinematic approximations: $q^2
\approx -m^2_\mu$ and $p^0_e \approx |\vec{p}_e| \approx m_\mu$, which are
valid for $\mu \to e$ conversion.  Given the above approximation, the $\mu \to
e$ conversion rate in a nucleus with nucleon numbers $(N,Z)$, is given by
\begin{equation}
  \label{Bmueconv}
B_{\mu e} (N,Z)\ \equiv\ \frac{\Gamma [\mu\, (N,Z)\to e\, (N,Z) ] }{
\Gamma [ \mu\, (N,Z) \to {\rm capture}]}\
\approx\ \frac{ \alpha^3_{\rm em} \alpha^4_w m^5_\mu}{32\pi^2
M^4_W \Gamma_{\rm capt.} }\, \frac{Z^4_{\rm eff}}{Z}\,
|F(-m^2_\mu)|^2\, |Q_W|^2\, ,
\end{equation}
where $\alpha_{\rm em} = 1/137$ is  the electromagnetic fine structure
constant, $Z_{\rm eff}$ is  the  effective atomic number of  coherence
and $\Gamma_{\rm   capt.}$ is  the  muon  nuclear  capture rate.   For
${}^{48}_{22}{\rm Ti}$,  experimental measurements give  $Z_{\rm  eff}
\approx  17.6$  for ${}^{48}_{22}{\rm Ti}$~\cite{Zeff}   and $\Gamma [
\mu\  {}^{48}_{22}{\rm Ti}  \to   {\rm capture}]  \approx  1.705\times
10^{-18}$~GeV~\cite{SMR}.   Moreover,  $|F(-m^2_\mu)| \approx 0.54$ is
the nuclear form factor \cite{FPap}. Finally, $Q_W = V_u (2Z +N) + V_d
(Z+2N)$  is the coherent  charge  of the nucleus,  which is associated
with the vector current. Its explicit form is given by
\begin{eqnarray}
  \label{Vu}
V_u &=& \frac{2}{3}\, s^2_w\, \Big(\, F^{\mu e}_\gamma\, -\, 
G^{\mu e}_\gamma\, -\, F^{\mu e}_Z\, \Big)\, +\, \frac{1}{4}\, 
\Big(\, F^{\mu e}_Z\, -\, F^{\mu e uu}_{\rm box}\, \Big)\, ,\\
  \label{Vd}
V_d &=& -\, \frac{1}{3}\, s^2_w\, 
\Big(\, F^{\mu e}_\gamma\, -\, G^{\mu e}_\gamma
\, -\, F^{\mu e}_Z\, \Big)\, -\, \frac{1}{4}\, \Big(\, F^{\mu e}_Z\, +\,
F^{\mu e dd}_{\rm box}\, \Big)\, .
\end{eqnarray}
The composite form-factors  $F^{\mu  e  uu}_{\rm box}$ and  $F^{\mu  e
dd}_{\rm box}$ are  defined in~\cite{Ara}. In  the SM  with two nearly
degenerate heavy Majorana neutrinos and  in the limit $m^2_N/M^2_W \gg
1$, these form-factors can be written down in the simplified forms:
\begin{equation}
  \label{Fboxud}
F_{\rm box}^{\mu euu}\ \approx\ F_{\rm box}^{\mu edd}\ \approx\
-\, s_L^{\nu_\mu}s_L^{\nu_{e}}\; . 
\end{equation}
In the same limit $m^2_N/M^2_W \gg 1$, $B_{\mu e} (N,Z)$ is given by
\begin{eqnarray}
  \label{Amutoe}
B_{\mu e} (N,Z) & \simeq & \frac{ \alpha^3_{\rm em} 
\alpha^4_w m^5_\mu}{18432\,\pi^2
M^4_W \Gamma_{\rm capt.} }\, \frac{Z^4_{\rm eff}}{Z}\,
|F(-m^2_\mu)|^2\, (s_L^{\nu_\mu})^2 (s_L^{\nu_e})^2\nonumber\\
& &\times\, \Bigg\{\,\bigg[\,3N+(33-86s_w^2)Z+
\Big(9N-(9-32s_w^2)Z\Big)\ln\bigg(\frac{m^2_N}{M^2_W}\bigg)\bigg]^2\nonumber\\
& & +\ 6\ \frac{m^2_N}{M^2_W}\bigg[\,3N+(33-86s_w^2)Z+
\Big(9N-(9-32s_w^2)Z\Big)\ln\bigg(\frac{m^2_N}{M^2_W}\bigg)\bigg]\nonumber\\
& & \times\ \Big(N-(1-4s_w^2)Z\Big)\ 
\sum_{l =e,\mu,\tau} (s_L^{\nu_l})^2\nonumber\\
& & +\ 9\ \frac{m^4_N}{M^4_W}\ \Big(N-(1-4s_w^2)Z\Big)^2\ 
\Big(\sum_{l =e,\mu,\tau} (s_L^{\nu_l})^2\Big)^2
\,\Bigg\}\; .
\end{eqnarray}
For the ${}^{48}_{22}{\rm Ti}$ case, $B_{\mu e} (26,22)$ is related to
$B(\mu \to e\gamma)$ through
\begin{equation}
  \label{Rmue}
B_{\mu e} (26,22)\ \simeq\  
0.1\,\times \Bigg[\,1\ +\
0.5\ \ln\bigg(\frac{m^2_N}{M^2_W}\bigg)\Bigg]^2\:
B(\mu\to e\gamma)\; .
\end{equation}

On the   experimental side, the   strongest upper bound on  $B_{\mu e}
(N,Z)$ is obtained from experimental data on $\mu \to e$ conversion in
${}^{48}_{22}{\rm Ti}$ \cite{SINDRUM}:
\begin{equation}
  \label{mueconvexp}
B^{\rm exp}_{\mu  e} (26,22)\ <\ 4.3\,\times 10^{-12}\, , 
\end{equation}
at  the 90\%    CL.   However, the  proposed experiment    by the MECO
collaboration~\cite{MECO}  will  be  sensitive to conversion  rates of
order $10^{-16}$.

In the R$\tau$L model with $m_N =  250$~GeV, one obtains, on the basis
of~(\ref{Rmue}),    the  prediction for   $\mu  \to   e$ conversion in
${}^{48}_{22}{\rm Ti}$:
\begin{equation}
B_{\mu e} (26,22)\  \simeq\    0.46 \times B(\mu \to   e\gamma)\ \sim\
4.5 \times 10^{-13}\;.
\end{equation}
The  above  prediction falls well    within reach of  the  sensitivity
proposed by the MECO collaboration.

\begin{table}
\begin{center}
\begin{tabular}{|c||c|c|c|}
\hline
$m_N$~(GeV) & $B(\mu \to e\gamma)$ & $B(\mu \to eee)$ & $B_{\mu e}(26,22)$\\
\hline \hline
$100$  & $6.2 \times 10^{-12}$ & $3.8 \times 10^{-14}$ & $9.2 \times
10^{-13}$\\ 
$250$  & $9.9 \times 10^{-13}$ & $1.4 \times 10^{-14}$ & $4.5 \times
10^{-13}$\\ 
$500$  & $2.5 \times 10^{-13}$ & $9.7 \times 10^{-15}$ & $2.0 \times
10^{-13}$\\ 
$1000$ & $6.2 \times 10^{-14}$ & $4.9 \times 10^{-15}$ & $7.7 \times
10^{-14}$\\ \hline
\end{tabular}
\end{center}
\caption{\sl  Branching fractions  for the  3 LFV  processes  $\mu \to
e\gamma$,  $\mu  \to eee$  and  coherent  $\mu  \to e$  conversion  in
${}^{48}_{22}{\rm Ti}$ nuclei.}
\label{LFVrates}
\end{table}

In Table~\ref{LFVrates},  we summarize  our results for  the branching
ratios of  the 3 LFV processes:  $\mu \to e\gamma$, $\mu  \to eee$ and
coherent $\mu \to e$  conversion in ${}^{48}_{22}{\rm Ti}$ nuclei, for
each R$\tau$L model considered in Section~\ref{sec:num}.

As a final general remark, we should mention that R$\tau$L models, and
leptogenesis   models  in general,   do   not  suffer from too   large
contributions     to    the    electron    electric    dipole   moment
(EDM)~\cite{APRD,NN}, which first  arises at two  loops.  The reason is
that EDM  effects are suppressed   either by  higher  powers of  small
Yukawa   couplings of  order $10^{-4}$ and    less, or  by very  small
factors,  such as $(m_{N_1}   -  m_{N_{2,3}})/m_N \sim 10^{-7}$.   The
latter is the case  in R$\tau$L  models,  which leads  to unobservably
small  EDM effects  of  order $10^{-37}~e$~cm,   namely 10 orders   of
magnitude smaller than the present experimental limits~\cite{PDG}.

\subsection{Collider Heavy Majorana Neutrino Production}

\begin{figure}
\begin{center}
\begin{picture}(400,100)(0,0)
\SetWidth{0.8}

\ArrowLine(10,70)(50,70)\Line(50,70)(90,70)\Photon(50,70)(50,30){4}{4}
\ArrowLine(50,30)(10,30)\Line(50,30)(90,30)
\Text(0,75)[b]{$e^-\, (\mu^- )$}\Text(0,25)[t]{$e^+\, (\mu^+ )$}
\Text(95,75)[b]{$N_{2,3}$}\Text(95,25)[t]{$\nu$}
\Text(57,50)[l]{$W^-$}

\Text(50,0)[]{\bf (a)}

\ArrowLine(160,70)(200,70)\Line(200,70)(240,70)\Photon(200,70)(200,30){4}{4}
\ArrowLine(200,30)(160,30)\Line(200,30)(240,30)
\Text(150,75)[b]{$e^-\, (\mu^- )$}\Text(150,25)[t]{$e^+\, (\mu^+ )$}
\Text(245,75)[b]{$\nu$}\Text(245,25)[t]{$N_{2,3}$}
\Text(207,50)[l]{$W^-$}

\Text(200,0)[]{\bf (b)}

\ArrowLine(300,70)(320,50)\ArrowLine(320,50)(300,30)
\Photon(320,50)(360,50){4}{4}
\Line(360,50)(380,70)\Line(360,50)(380,30)
\Text(290,75)[b]{$e^-\, (\mu^- )$}\Text(290,25)[t]{$e^+\, (\mu^+ )$}
\Text(385,75)[b]{$\nu$}\Text(385,25)[t]{$N_{2,3}$}
\Text(340,43)[t]{$Z$}

\Text(340,0)[]{\bf (c)}

\end{picture}
\end{center}
\caption{\sl Diagrams related to the production of the heavy Majorana
neutrinos $N_{2,3}$ at future $e^+ e^- (\mu^+ \mu^-)$ high-energy
colliders.}
\label{fig:prod}
\end{figure}
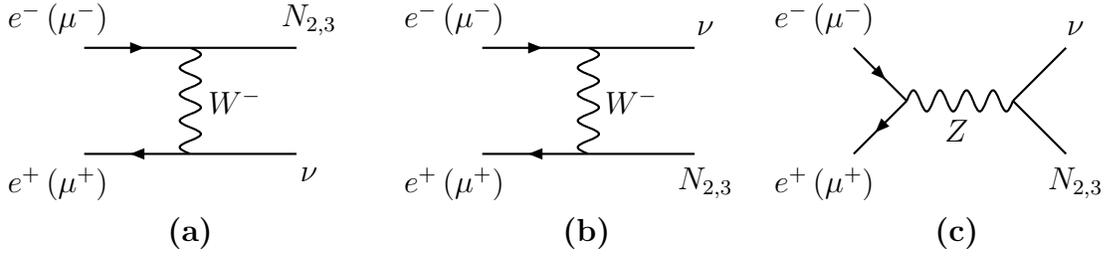

If  heavy  Majorana    neutrinos have   electroweak-scale  masses  and
appreciable  couplings  to electrons  and    muons they can  be
copiously produced  at future  $e^+e^-$~\cite{BG,AAS} and $\mu^+\mu^-$
colliders.  As  shown  in  Fig.~\ref{fig:prod}, this    is exactly the
kinematic  situation   for  the heavy   Majorana neutrinos   $N_{2,3}$
described  by the R$\tau$L models. The  heavy  Majorana neutrino $N_1$
has a very small coupling to leptons and it would be very difficult to
produce this state directly.

For   collider  c.m.s.~energies  $\sqrt{s}  \gg  m_N$, the $t$-channel
$W^-$-boson exchange graphs will  dominate over the $Z$-boson exchange
graph,   which     is    $s$-channel  propagator     suppressed   (see
Fig.~\ref{fig:prod}). In this  high-energy limit, the production cross
section for heavy  Majorana neutrinos  approaches a constant~\cite{BG},
i.e.
\begin{equation}
  \label{cross} 
\sigma \Big[\,e^+e^-\,(\mu^+\mu^-) \to N_{2,3}\,\nu\,\Big] \ \approx\
\frac{\pi\,  \alpha^2_w}{4\, M^2_W}\: (s^{\nu_{e  (\mu)}}_L)^2\ \approx\
10~{\rm fb}\,\times\, \Bigg(\,
\frac{s^{\nu_{e (\mu)}}_L}{10^{-2}}\,\Bigg)^2\ .
\end{equation}
Since  $s^{\nu_\tau}_L  \approx 0$  in  R$\tau$L  models, the produced
heavy Majorana   neutrinos  $N_{2,3}$   will have the   characteristic
signature that they will predominantly decay into electrons and muons,
but      {\em   not}    into    $\tau$      leptons.    Assuming  that
$m_N\stackrel{>}{{}_\sim}  M_H$, the branching  fraction of  $N_{2,3}$
decays into charged   leptons    and into $W^\pm$   bosons    decaying
hadronically is
\begin{equation}
  \label{BRNs}  
B \Big( N_{2,3} \to e^\mp,\,\mu^\mp\,W^\pm (\to {\rm jets})\Big)\ \approx\
\frac{1}{2}\,\times\, \frac{2}{3}\ =\ \frac{1}{3}\; .
\end{equation}
Given~(\ref{cross}), (\ref{BRNs})  and   an integrated  luminosity  of
100~fb$^{-1}$, we expect to be able to analyze about 100 signal events
for $(s^{\nu_{e (\mu)}}_L)^2 = 10^{-2}$ and $m_N \stackrel{<}{{}_\sim}
300$~GeV,  at  future   $e^+e^-$  and    $\mu^+\mu^-$  colliders  with
c.m.s.~energy $\sqrt{s} = 0.5$--$1~{\rm TeV}$.

These  simple estimates  are  supported by  a  recent analysis,  where
competitive   background   reactions   to   the   signal   have   been
considered~\cite{AAS}.   This analysis  showed that  the  inclusion of
background processes reduces  the number of signal events  by a factor
of  10.   The  authors  in~\cite{AAS}  find that  an  $e^+e^-$  linear
collider with c.m.s.~energy $\sqrt{s}  = 0.5$~TeV will be sensitive to
values of $s^{\nu_{e}}_L  = |a| v/m_N \sim 0.7  \times 10^{-2}$.  This
amounts to the  same level of sensitivity to  the parameter $|a|$, for
R$\tau$L   scenarios   with   $m_N   =  250$~GeV.    The   sensitivity
to~$s^{\nu_{e}}_L$   could   be   improved   by   a   factor   of   3,
i.e.~$s^{\nu_{e}}_L  \sim 0.2  \times 10^{-2}$,  in  proposed upgraded
$e^+e^-$ accelerators such as CLIC.

A similar analysis should be envisaged to hold for future $\mu^+\mu^-$
colliders,  leading to  similar  findings for  $s^{\nu_{\mu}}_L =  |b|
v/m_N$.  In  general, we expect that  the ratio of  the two production
cross  sections of  $N_{2,3}$  at the  two  colliders under  identical
conditions of c.m.s.~energy and  luminosity will give a direct measure
of the  ratio of $|a|^2/|b|^2$.  This information, together  with that
obtained  from   low-energy  LFV  observables,  $0\nu\beta\beta$-decay
experiments,  and  neutrino  data,  will significantly  constrain  the
parameters of the R$\tau$L  models.  Finally, since the heavy Majorana
neutrinos $N_{2,3}$  play an  important synergetic role  in resonantly
enhancing  $\delta^\tau_{N_1}$, potentially  large  CP asymmetries  in
their decays will determine the theoretical parameters of these models
further.  Evidently, more detailed studies are needed before one could
reach a  definite conclusion concerning the  exciting possibility that
electroweak-scale   R$\tau$L  models   may   naturally  constitute   a
laboratory testable solution to the cosmological problem of the BAU.

\setcounter{equation}{0}
\section{Conclusions}

We have  studied a novel  variant of RL,  which may take place  at the
electroweak phase transition.  This RL  variant gives rise to a number
of phenomenologically  testable signatures for  low-energy experiments
and  future high-energy colliders.   The new  RL scenario  under study
makes  use of  the  property that,  in  addition to  $B  - L$  number,
sphalerons  preserve the  individual quantum  numbers  $\frac{1}{3}B -
L_{e,\mu,\tau}$~\cite{HT}.   The  observed  BAU  can  be  produced  by
lepton-to-baryon conversion  of an individual lepton  number.  For the
case  of  the $\tau$-lepton  number  this  mechanism  has been  called
resonant $\tau$-leptogenesis~\cite{APtau}.

In studying  leptogenesis, we have  extended previous analyses  of the
relevant network  of BEs. More explicitly, we  have consistently taken
into account  SM chemical potential  effects, as well as  effects from
out  of  equilibrium  sphalerons   and  single  lepton  flavours.   In
particular, we  have found that  single lepton flavour  effects become
very  important in  R$\tau$L  models.  In  this  case, the  difference
between our improved formalism of BEs and the usual formalism followed
in  the literature  could be  dramatic. The  predictions of  the usual
formalism could  lead to  an erroneous result  which is  suppressed by
many orders of magnitude.  The suppression factor could be enormous of
order   $10^{-6}$   for   the   R$\tau$L  scenarios   considered   in
Section~\ref{sec:num}.   Even within leptogenesis  models with  a mild
hierarchy between the heavy neutrino masses, the usual formalism turns
out to be inadequate to  properly treat single lepton flavour effects;
its  predictions may differ  even up  to one  order of  magnitude with
respect to those obtained with our improved formalism.

One generic  feature of R$\tau$L  models is that their predictions for
the final baryon asymmetry   are  almost independent of  the   initial
values for the primordial   $B$-number, $L$-number and heavy  Majorana
neutrino   abundances.    Specifically,   we   have  investigated  the
dependence of the BAU on the initial  conditions, as a function of the
heavy  neutrino  mass  scale  $m_N$.   We have   found  that for  $m_N
\stackrel{>}{{}_\sim}  250$~GeV, the dependence of  the  BAU is always
less than 7\%,  even if the initial baryon  asymmetry is as large  as
$\eta^{\rm in}_B = 10^{-2}$ at $z = m_N/T  = 0.1$.  For smaller values
of $m_N$,  this dependence starts   getting larger.  Thus, for $m_N  =
100$~GeV, the dependence of the final  baryon asymmetry on the initial
conditions is stronger,   unless  the primordial baryon asymmetry   is
smaller than $\sim 10^{-6}$ at $z=0.1$.

In order to have successful  leptogenesis in the R$\tau$L models under
study,  the  heavy  Majorana  neutrinos  are  required  to  be  nearly
degenerate.  This  nearly degenerate heavy neutrino  mass spectrum may
be obtained by enforcing an SO(3) symmetry, which is explicitly broken
by the Yukawa interactions  to a particular SO(2) sub-group isomorphic
to  a   lepton-type  group~U(1)$_l$.   The   approximate  breaking  of
U(1)$_l$, which  could result from a  FN mechanism, leads  to a Yukawa
texture  that accounts  for  the existing  neutrino oscillation  data,
except those from the  LSND experiment~\cite{LSND}.  Our choice of the
breaking parameters was motivated by  the naturalness of the light and
heavy neutrino  sectors.  To obtain  natural R$\tau$L models,  we have
followed the principle that there should be no excessive cancellations
between tree-level and radiative or  thermal effects.  In this way, we
have found that R$\tau$L models  strongly favour a light neutrino mass
spectrum  with  an  inverted   hierarchy.   Moreover,  when  the  same
naturalness  condition is  applied  to the  heavy  neutrino sector,  a
particular hierarchy  for the mass  differences of the  heavy Majorana
neutrinos is obtained.  In particular, the mass difference of one pair
of  heavy  Majorana neutrinos  is  much  smaller  than the  other  two
possible pairs.

R$\tau$L models offer a number of testable phenomenological signatures
for  low-energy experiments and  future high-energy  colliders.  These
models  contain   electroweak-scale  heavy  Majorana   neutrinos  with
appreciable   couplings  to   electrons  and   muons,  e.g.~$N_{1,2}$.
Specifically, the  (normalized to  the SM) $W^\pm$-boson  couplings of
electrons and muons to the heavy Majorana neutrinos $N_{1,2}$ could be
as large as 0.01, for $m_{N_{1,2}} = 100$--300~GeV.  As a consequence,
these heavy Majorana particles can  be produced at future $e^+e^-$ and
$\mu^+\mu^-$  colliders, operating  with a  c.m.s.~energy  $\sqrt{s} =
0.5$--1~TeV.  Another feature of R$\tau$L models is that thanks to the
inverted  hierarchic structure  of the  light neutrino  mass spectrum,
they can  account for sizeable $0\nu\beta\beta$  decay.  The predicted
effective neutrino  mass $|({\bf m}^\nu )_{ee}  |$ can be  as large as
0.02~eV, which is within the sensitivity of the proposed next round of
$0\nu\beta\beta$    decay     experiments.     The    most    striking
phenomenological feature of 3-generation (non-supersymmetric) R$\tau$L
models  is  that  they  can predict  $e$-  and  $\mu$-number-violating
processes,  such  as the  decay  $\mu \to  e\gamma$  and  $\mu \to  e$
conversion in nuclei, with observable rates.  In particular, these LFV
effects could be as large as $10^{-12}$ for $B (\mu \to e\gamma )$ and
as large  as $5\times 10^{-13}$ for  a $\mu \to e$  conversion rate in
${}^{48}_{22}$Ti,  normalized to  the $\mu$  capture rate.   The above
predicted values are  within reach of the experiments  proposed by the
MEG and MECO collaborations.

Although  the present study    improves previous analyses of  the  BEs
related to   leptogenesis  models,  there are  still   some additional
smaller but relevant  effects that  would require special   treatment.
The  first   obvious improvement would  be   to  calculate the thermal
effects on  the collision terms,  beyond the HTL approximation.  These
corrections would eliminate some of the uncertainties pertinent to the
actual  choice of  the IR regulator   in some of the  collision terms.
These effects  limit the accuracy of our  predictions and introduce an
estimated  theoretical uncertainty   of 30\% for  leptogenesis  models
operating well above the electroweak phase transition, with relatively
large $K$ factors, i.e.~$K_{lN_\alpha}  \stackrel{>}{{}_\sim} 5$.  For
models at the  electroweak  phase transition,  the IR problem  is less
serious, but   larger uncertainties may  enter due  to the   lack of a
satisfactorily accurate quantitative framework for sphaleron dynamics.
Although the implementation of  the sphaleron dynamics  in our BEs for
RL  models   was  based   on  the  calculations  of~\cite{CLMW,KS,LS},
particular treatment would  be    needed, if the   electroweak   phase
transition was a  strong first-order one.   In this case, the dynamics
of the  expanding  bubbles during   the  electroweak phase  transition
becomes  relevant~\cite{EWBAU}.    This  possibility may   emerge   in
supersymmetric versions of  RL models.  Nevertheless, the inclusion of
the  aforementioned additional effects is  expected  not to modify the
main  results  of the present analysis    drastically and will  be the
subject of a future communication.

\bigskip

\subsection*{Acknowledgements}
We thank Mikko Laine,  Costas Panagiotakopoulos, Graham Ross, Kiriakos
Tamvakis and  Carlos Wagner for useful discussions  and comments.  The
work of  AP has been supported  in part by the  PPARC research grants:
PPA/G/O/2002/00471 and PP/C504286/1. The work of TU has been funded by
the PPARC studentship PPA/S/S/2002/03469.

\newpage

\def\theequation{\Alph{section}.\arabic{equation}}
\begin{appendix}

\setcounter{equation}{0}
\section{Collision Terms}
\subsection{Useful Notation and Definitions}

The following notation and  definitions are used  in the derivation of
the BEs. The  number density, $n_a$, of a  particle species, $a$, with
$g_a$ internal degrees of freedom is given by \cite{KW}
\begin{eqnarray}
  \label{na}
n_a (T) &=& g_a\, \int \frac{d^3{\bf p}}{(2\pi)^3}\  \exp\Big[ -
\Big(\sqrt{{\bf p}^2 + m^2_a} -  \mu_a (T)\Big)/T\,\Big]\nonumber\\
&=&  \frac{g_a\, m^2_a\,T\ e^{\mu_a (T)/T}}{2\pi^2}\
K_2\bigg(\frac{m_a}{T}\bigg)\; ,
\end{eqnarray}
where $\mu_a$ is the  $T$-dependent chemical potential and $K_n(x)$ is
the $n$th-order modified   Bessel function~\cite{AS}.  In our  minimal
leptogenesis model, the factors $g_a$ are: $g_{W^a} = 3 g_{B} = 6$ and
$g_\Phi = g_{\Phi^\dagger} = 2$, and  for the $i$th family: $g_{N_\alpha} =
2$, $g_{L_i}   =  g_{L^C_i} = 4$,  $g_{Q_i}  =  g_{Q^C_i}  =  12$, and
$g_{u_i} = g_{u_i^C} =  6$. Using the same  formalism as \cite{PU} the
CP-conserving  collision term for a generic  process $X \to Y$ and its
CP-conjugate $\overline{X} \to \overline{Y}$ is defined as
\begin{equation}
  \label{CT}
\gamma^X_Y\ \equiv \ \gamma ( X\to Y)\: +\: \gamma ( \overline{X}
\to \overline{Y} )\; ,
\end{equation}
with
\begin{equation}
  \label{gamma}
\gamma ( X\to Y)\ =\ \int\! d\pi_X\, d\pi_Y\, (2\pi )^4\,
\delta^{(4)} ( p_X - p_Y )\ e^{-p^0_X/T}\, |{\cal M}( X \to Y )|^2\; .
\end{equation}
In the above, $|{\cal M}( X \to Y  )|^2$ is the squared matrix element
which  is summed but  {\em not} averaged  over the internal degrees of
freedom of the initial and   final multiparticle states $X$ and   $Y$.
Moreover,   $d\pi_X$    represents the    phase  space   factor  of  a
multiparticle state $X$,
\begin{equation}
  \label{dpiX}
d\pi_X\ =\ \frac{1}{S_X}\, \prod\limits_{i=1}^{n_X}\,
\frac{d^4 p_i}{(2\pi )^3}\ \delta ( p^2_i - m^2_i )\; \theta (p^0_i)\; ,
\end{equation}
where $S_X = n_{\rm id}!$ is a symmetry factor depending on the number
of identical particles, $n_{\rm id}$, contained in $X$.

As  CPT is preserved,   the CP-conserving collision  term $\gamma^X_Y$
obeys the relation
\begin{equation}
\gamma^X_Y\ =\ \gamma^Y_X\,.
\end{equation}
Analogously, it is possible to define a CP-violating collision term
$\delta \gamma^X_Y$ as
\begin{equation}
  \label{dgamma}
\delta \gamma^X_Y\ \equiv\ \gamma (X\to Y)\ -\ \gamma (\overline{X}
\to \overline{Y})\ 
=\ -\,\delta \gamma^Y_X\; ,
\end{equation}
where the last equality follows from CPT invariance.

\subsection{CP-Conserving Collision Terms}

In numerically solving   the   BEs, we introduce  the    dimensionless
parameters:
\begin{equation}
z\  =\  \frac{m_{N_1}}{T}\,, \quad x\  =\ \frac{s}{m_{N_1}^2}\,, \quad
a_\alpha\ =\ \left(\frac{m_{N_\alpha}}{m_{N_1}}\right)^2,\quad a_r\ =\
\left(\frac{m_{\rm   IR}}{m_{N_1}}\right)^2,\quad  c^{\,l}_\alpha\  =\
\left(\,\frac{\Gamma^{\,l}_{N_\alpha}}{m_{N_1}}\,\right)^2,
\end{equation}
where $\alpha = 1,2,3$ labels the heavy  Majorana neutrino states, $s$
is the usual Mandelstam variable and $m_{\rm IR}$ is an infra-red (IR)
mass regulator which is discussed below.

In    terms    of   the   resummed   effective      Yukawa   couplings
$(\bar{h}^{\nu}_\pm)_{l\alpha}$ introduced in      \cite{PU},      the
radiatively corrected decay width $\Gamma_{N_\alpha}^{\,l}$ of a heavy
Majorana neutrino $N_\alpha$ into a lepton flavour $l$ is given by
\begin{equation}
\Gamma_{N_\alpha}^{\,l}\    =\   \frac{m_{N_\alpha}}{16\pi}\   
\Big[\, (\bar{h}^{\nu}_+)^*_{l\alpha}\,(\bar{h}^\nu_+)_{l\alpha}\:
+\: 
(\bar{h}^{\nu}_-)^*_{l\alpha}\,(\bar{h}^\nu_-)_{l\alpha}\,\Big]\; .
\label{Nwidth}
\end{equation}
By means  of (\ref{gamma}), the  $1\to2$  CP-conserving collision term
$\gamma^{N_\alpha}_{L_l \Phi}$ is found to be
\begin{eqnarray}
\gamma^{N_\alpha}_{L_l \Phi}\ =\ \gamma (N_\alpha \to L_l \Phi)\: +\:
\gamma (N_\alpha \to L^C_l \Phi^\dagger) \!&=&\!  
\Gamma^{\,l}_{N_\alpha}\, g_{N_\alpha}\, \int \frac{d^3{\bf
p}_{N_\alpha}}{(2\pi)^3}\,\frac{m_{N_\alpha}}{E_{N_\alpha}({\bf p})}\,
e^{-E_{N_\alpha}({\bf p})/T} \nonumber\\ &=&\!  \frac{m^4_{N_1}
a_i\,\sqrt{c^{\,l}_i}}{\pi^2\, z}\ K_1(z \sqrt{a_i})\,,
\end{eqnarray}
where $E_{N_\alpha}({\bf p})  = \sqrt{{\bf p}^2 + m^2_{N_\alpha}}$ and
$g_{N_\alpha}   = 2$  is the number    of internal  degrees of freedom
of~$N_\alpha$.     Upon summation  over    lepton  flavours~$l$,  this
collision term  reduces  to  the corresponding   one   given in  (B.4)
of~\cite{PU}.

For $2\to2$ processes,  one can make use  of the reduced cross section
$\widehat{\sigma}(s)$ defined as
\begin{equation}
  \label{reducedxs}
\widehat{\sigma}(s)\ \equiv\ 8\pi\,\Phi (s)\int\! d\pi_Y\: (2\pi)^4
\,\delta^{(4)} (q-p_Y)\: \left|{\cal M}(X\rightarrow Y)\right|^2\; ,
\end{equation}
where $s=q^2$ and the initial phase space integral is given by
\begin{equation}
  \label{Phi}
\Phi (s)\ \equiv\ \int\! d\pi_X\: (2\pi)^4\,\delta^{(4)} (p_X-q)\,.
\end{equation}
These expressions simplify to give
\begin{equation}
  \label{sigmat}
\widehat{\sigma}(s)\ =\ \frac{1}{8\pi s}\
\int\limits_{t_-}^{t_+}\! dt\ \left|{\cal M}(X\rightarrow Y)\right|^2\; ,
\end{equation}
where   $t$ is the usual    Mandelstam variable,  and the  phase-space
integration limits $t_\pm$ will be specified below.

In  processes,  such as   $N_{\alpha}   V_{\mu} \to L_{l}  \Phi$,  the
exchanged particles (e.g. $L$ and $\Phi$) occurring in the $t$ and $u$
channels are massless. These collision terms possess IR divergences at
the phase-space   integration    limits  $t_{\pm}$  in~(\ref{sigmat}).
Within a more appropriate  framework, such as finite temperature field
theory,  these IR  singularities  would  have been  regulated  by  the
thermal masses of the  exchanged particles. In  our $T=0$ field theory
calculation, we have  regulated the IR divergences  by cutting off the
phase-space  integration limits  $t_{\pm}$  using  a universal thermal
regulator $m_{\rm IR}$  related to the  expected thermal masses of the
exchanged  particles.  This  procedure  preserves chirality  and gauge
invariance, as  would be expected  within  the framework of  a  finite
temperature field theory~\cite{MBellac}.

Thermal masses for the  Higgs and leptons are predominantly  generated
by gauge and top-quark Yukawa interactions.  In the HTL approximation,
they are given by \cite{Weldon}
\begin{eqnarray}
\frac{m^2_L (T)}{T^2} & = & \frac{1}{32}\,\Big(3\,g^2 +
g^{\prime\,2}\Big)\,,\nonumber\\ 
\frac{m_{\Phi}^2 (T)}{T^2} & = & 2 d\,\Bigg(1-\frac{T_c^2}{T^2}\Bigg)\,,
\end{eqnarray}
where   $d = (8M_W^2   + M_Z^2  + 2  m_t^2   + M_H^2)/(8v^2)$. In  our
numerical estimates,  we choose  the  regulator  $m_{\rm IR}$ to  vary
between the lepton and  Higgs  thermal masses, evaluated  at $T\approx
m_N$. The resulting variation in the predicted baryon asymmetry can be
taken as a contribution to  the theoretical uncertainties in our  zero
temperature calculation.

For reduced  cross-sections with an apparent  singularity at the upper
limit $t_+$, the following upper and lower limits are used:
\begin{equation}
t_{+} \ =\ -m_{\rm IR}^2\,, \qquad t_{-} \ =\ m_{N_{\alpha}}^2 - s\,.
\end{equation}
Likewise,  for  reduced cross-sections  with apparent singularities at
both the   upper and lower limits $t_\pm$,   the following  limits are
employed:
\begin{equation}
t_{+} \ =\ -m_{\rm IR}^2\,, \qquad t_{-} \ =\ m_{N_{\alpha}}^2 +
m_{\rm IR}^2 - s\;.
\end{equation}
It is important to remark here that  the collision terms do not suffer
from   IR singularities at   $T\stackrel{<}{{}_\sim} T_c$, because the
leptons, $W$ and $Z$ bosons receive $v(T)$-dependent masses during the
electroweak phase transition.  The full implementation of such effects
will be given elsewhere.

Substituting (\ref{reducedxs}) and (\ref{Phi}) into (\ref{gamma}), one
obtains
\begin{equation}
  \label{22CT}
\gamma^X_Y\ =\ \frac{m^4_{N_1}}{64\,\pi^4 z}\
\int\limits_{x_{\rm thr} }^\infty\! dx\ 
\sqrt{x}\;K_1(z\sqrt{x})\;\widehat{\sigma}^X_Y (x)\; ,
\end{equation}
where $x_{\rm  thr}$ is the kinematic  threshold for a  given $2\to 2$
process.

For $2\to2$ $\Delta  L=1$ processes, one  can repeat  the procedure in
\cite{PU} (Appendix~B), without  summing over  lepton flavours.   Each
$\Delta L=1$  process      has an  identical    factor  dependent   on
$\bar{h}^{\nu}_{\pm}$. To produce the $\Delta L=1$ collision terms for
each   lepton flavour,  this  factor  needs  to be   replaced with its
un-summed equivalent,
\begin{equation}
(\bar{h}^{\nu}_+)^*_{l\alpha}\,(\bar{h}^\nu_+)_{l\alpha}\: +\: 
(\bar{h}^{\nu}_-)^*_{l\alpha}\,(\bar{h}^\nu_-)_{l\alpha}\,,
\end{equation}
exactly as  was done in~(\ref{Nwidth}).  The remainder of the analytic
expression for each of these terms is presented in \cite{PU}.

In addition to the Higgs and gauge mediated  $\Delta L=1$ terms, there
are also $2\to2$ $\Delta   L=2$ processes. As before,  these processes
are   $L_k  \Phi \leftrightarrow  L_l^C    \Phi^\dagger$ and  $L_k L_l
\leftrightarrow \Phi^\dagger \Phi^\dagger$  where  the former has  its
real intermediate  states  subtracted.  The  analytic  forms of  these
collision terms  are identical  to the  total lepton  number  case but
lepton flavour is   not summed over.  The reduced  cross sections  are
given by
\begin{eqnarray}
   \label{LHtoLH}
\widehat{\sigma}^{\,\prime\, L_k \Phi}_{L_l^C\Phi^\dagger}
\!\!&=&\!\!  \sum_{\alpha,\beta=1}^{3}\ {\rm Re}\, \Bigg\{\,
\bigg[\,(\bar{h}^{\nu}_+)^*_{k\alpha}\,(\bar{h}^\nu_+)_{k\beta}\,
(\bar{h}^{\nu}_+)^*_{l\alpha}\,(\bar{h}^\nu_+)_{l\beta}\: +\:
(\bar{h}^{\nu}_-)^*_{k\alpha}\,(\bar{h}^\nu_-)_{k\beta}\,
(\bar{h}^{\nu}_-)^*_{l\alpha}\,(\bar{h}^\nu_-)_{l\beta}\,\bigg]
\:\mathcal{A}^{(ss)}_{\alpha\beta}\nonumber\\*
\!\!&&\qquad\qquad +\:2\,\bigg[\,(\bar{h}^{\nu}_+)^*_{l\alpha}\,
h^{\nu}_{l\beta}\, 
(\bar{h}^{\nu}_+)^*_{k\alpha}\,h^{\nu}_{k\beta}\: +\: 
(\bar{h}^{\nu}_-)^*_{l\alpha}\,h^{\nu *}_{l\beta}\,
(\bar{h}^{\nu}_-)^*_{k\alpha}\,h^{\nu *}_{k\beta}\,\bigg]
\mathcal{A}^{(st)*}_{\alpha\beta}\nonumber\\*
\!\!&&\qquad\qquad + \: 2\, 
\Big( h^{\nu *}_{k\alpha}\,h^\nu_{k\beta}\,h^{\nu *}_{l\alpha}\,
h^\nu_{l\beta} \Big)\; \mathcal{A}^{(tt)}_{\alpha\beta}\,\Bigg\}\,,
\end{eqnarray}
and
\begin{equation}
   \label{LLtoHH}
\widehat{\sigma}^{L_k L_l}_{\Phi^\dagger\Phi^\dagger} \ = \
\sum_{\alpha,\beta=1}^{3}\; {\rm Re}\,
\Big(h^{\nu *}_{k\alpha}\,h^\nu_{k\beta}\,h^{\nu *}_{l\alpha}\, 
h^\nu_{l\beta} \Big)
\; \mathcal{B}_{\alpha\beta}\;,
\end{equation}
where  the $\mathcal{A}$  and $\mathcal{B}$  factors  are presented in
\cite{PU}.

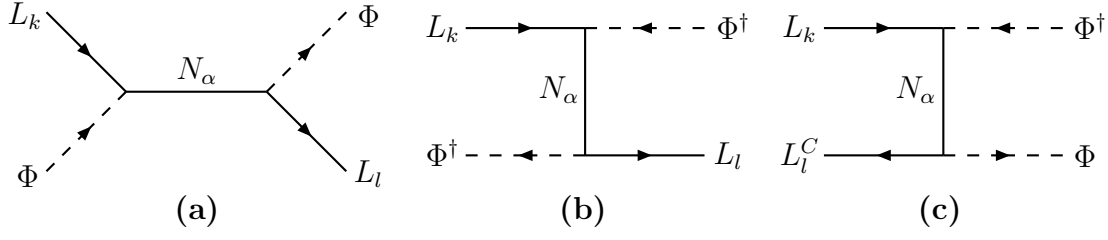
\begin{figure}
\begin{picture}(433,90) (14,-8)
\SetWidth{0.8}
\Text(94,53)[]{\normalsize{$N_\alpha$}}
\Text(30,13)[]{\normalsize{$\Phi$}}
\Text(29,74)[]{\normalsize{$L_k$}}
\Text(159,74)[]{\normalsize{$\Phi$}}
\Text(159,14)[]{\normalsize{$L_l$}}
\Text(322,69)[]{\normalsize{$L_k$}}
\Text(322,21)[]{\normalsize{$L_l^C$}}
\Text(432,70)[]{\normalsize{$\Phi^\dagger$}}
\Text(430,21)[]{\normalsize{$\Phi$}}
\Text(366,46)[]{\normalsize{$N_\alpha$}}
\Text(297,70)[]{\normalsize{$\Phi^\dagger$}}
\Text(187,69)[]{\normalsize{$L_k$}}
\Text(187,22)[]{\normalsize{$\Phi^\dagger$}}
\Text(295,21)[]{\normalsize{$L_l$}}
\Text(231,46)[]{\normalsize{$N_\alpha$}}
\SetWidth{0.8}
\ArrowLine(37,75)(67,45)
\DashArrowLine(37,15)(67,45){5}
\Line(67,45)(120,45)
\ArrowLine(120,45)(150,15)
\DashArrowLine(120,45)(150,75){5}
\ArrowLine(375,21)(330,21)
\Line(375,69)(375,21)
\DashArrowLine(420,69)(375,69){5}
\DashArrowLine(375,21)(420,21){5}
\ArrowLine(330,69)(375,69)
\ArrowLine(195,69)(240,69)
\DashArrowLine(240,21)(195,21){5}
\Line(240,69)(240,21)
\DashArrowLine(285,69)(240,69){5}
\ArrowLine(240,21)(285,21)
\Text(94,0)[]{\normalsize{\bf (a)}}
\Text(240,0)[]{\normalsize{\bf (b)}}
\Text(375,0)[]{\normalsize{\bf (c)}}
\end{picture}
\caption{\sl $\Delta L=0$ interactions between leptons of different
flavours mediated by heavy Majorana neutrinos. (a) corresponds to the
process $L_k \Phi \leftrightarrow L_l \Phi$, (b) corresponds to the
process $L_k \Phi^\dagger \leftrightarrow L_l \Phi^\dagger$ and (c)
corresponds to the process $L_k L^C_l \leftrightarrow \Phi^\dagger
\Phi$.}
\label{DeltaL0Fig}
\end{figure}

As we  now  consider lepton flavours   separately, it is  necessary to
include $\Delta L=0$, but lepton flavour  violating interactions.  The
three   lowest  order    $2\leftrightarrow  2$   processes  are  shown
diagrammatically in Figure~\ref{DeltaL0Fig}: $L_k \Phi \leftrightarrow
L_l  \Phi$, $L_k \Phi^\dagger   \leftrightarrow L_l \Phi^\dagger$  and
$L_k  L^C_l  \leftrightarrow  \Phi^\dagger  \Phi$  (note  that $k  \ne
l$). The first of these reactions contains heavy Majorana neutrinos as
RISs. These    need   be removed using     the  procedure outlined  in
\cite{PU}. The reduced cross section for each of these processes is
\begin{equation}
\widehat{\sigma}^{\,\prime\,L_k \Phi}_{\,L_l \Phi} = 
\sum_{\alpha,\beta=1}^3
\Big[(\bar{h}^\nu_+)^*_{l\alpha}\,(\bar{h}^\nu_-)^*_{k\alpha}\,
(\bar{h}^\nu_+)_{l\beta}\,   
(\bar{h}^\nu_-)_{k\beta} + 
(\bar{h}^\nu_-)^*_{l\alpha}\,(\bar{h}^\nu_+)^*_{k\alpha}\,
(\bar{h}^\nu_-)_{l\beta}\,
(\bar{h}^\nu_+)_{k\beta}\Big]\,\mathcal{C}_{\alpha\beta}
\end{equation}
with
\begin{equation}
  \label{Cab}
\mathcal{C}_{\alpha\beta} \ =\ \left\{
\begin{array}{lc}
\frac{\displaystyle x a_\alpha}{\displaystyle 4\pi |D^2_\alpha|}\ \to\ 0
                                        &\quad (\alpha = \beta)\\[5mm]
\frac{\displaystyle x\sqrt{a_\alpha\,a_\beta}}{\displaystyle 
4\pi P^*_\alpha P_\beta}\ &\quad (\alpha\neq \beta)
\end{array} \right.
\end{equation}
In~(\ref{Cab}),  $P^{-1}_\alpha (x)$ is  the Breit--Wigner $s$-channel
propagator
\begin{equation}
P_\alpha^{-1} (x) \ =\ \frac{1}{x -a_\alpha + 
i\sqrt{a_\alpha c_\alpha}}\ .
\end{equation}
Therefore,  following  the procedure in \cite{PU},  the RIS-subtracted
propagator is determined by
\begin{equation}
|D_\alpha^{-1} (x)|^2 \ =\ |P_\alpha^{-1} (x)|^2 - \frac{\pi}{\sqrt{a_\alpha
 c_\alpha}}\,\delta(x-a_\alpha) \ \to\ 0. 
\end{equation}

Processes (b) and (c) in Fig.~\ref{DeltaL0Fig} do not contain RISs and
have the following reduced cross sections:
\begin{eqnarray}
\widehat{\sigma}^{\,L_k \Phi^\dagger}_{L_l \Phi^\dagger} \!& = &\!
\sum_{\alpha,\beta=1}^3 \mathrm{Re} \Big( h^{\nu
*}_{l\alpha}\,h^{\nu}_{k\alpha}\, 
h^{\nu}_{l\beta}\,h^{\nu *}_{k\beta} \Big)\,\mathcal{D}_{\alpha\beta}\,,\\[3mm]
\widehat{\sigma}^{\,L_k L^C_l}_{\Phi^\dagger \Phi} \!& = &\!
\sum_{\alpha,\beta=1}^3 \mathrm{Re} 
\Big( h^{\nu *}_{l\alpha}\,h^{\nu}_{k\alpha}\, 
h^{\nu}_{l\beta}\,h^{\nu *}_{k\beta} \Big)\,\mathcal{E}_{\alpha\beta}\,,
\end{eqnarray}
where for $\alpha\ne \beta$,
\begin{eqnarray}
\mathcal{D}_{\alpha\beta} \!& = &\! \frac{\sqrt{a_\alpha a_\beta}}{\pi
x (a_\alpha - a_\beta)}\: 
\Bigg[\, (x+a_\beta)\ln\Bigg(\frac{x+a_\beta}{a_\beta}\Bigg)-(x+a_\alpha)
\ln\Bigg(\frac{x+a_\alpha}{a_\alpha}\Bigg)\, \Bigg]\,,\\[3mm]
\mathcal{E}_{\alpha\beta} \!& = &\! \frac{\sqrt{a_\alpha a_\beta}}{\pi
(a_\alpha - a_\beta)}\:
\ln\Bigg(\frac{a_\alpha (x+a_\beta)}{a_\beta (x+a_\alpha)}\Bigg)\,,
\end{eqnarray}
and for $\alpha=\beta$,
\begin{eqnarray}
\mathcal{D}_{\alpha\alpha} \!& = &\! \frac{a_\alpha}{\pi x}\:
\Bigg[\,\frac{x}{a_\alpha} - \ln\Bigg(\frac{x+a_\alpha}{a_\alpha}\Bigg)\,
\Bigg]\,,\\[3mm]
\mathcal{E}_{\alpha\alpha} \!& = &\! \frac{x}{\pi (x+a_\alpha)}\,.
\end{eqnarray}
\end{appendix}


\newpage
\begin{thebibliography}{99}

\bibitem{WMAP} D.N. Spergel {\em et al.}, Astrophys.\ J.\ Suppl.\ {\bf
148} (2003) 175.

\bibitem{reviews}    For  recent    reviews,   see,\\ 
W.~Buchm\"uller, R.~D.~Peccei    and     T.~Yanagida, hep-ph/0502169;\\
M.~Dine and A.~Kusenko,  Rev.\ Mod.\  Phys.\ {\bf  76}  (2004) 1;\\ 
K.~Enqvist and A.~Mazumdar, Phys.\ Rept.\ {\bf 380} (2003) 99.

\bibitem{FY} M. Fukugita and T. Yanagida, Phys.\ Lett.\ B {\bf 174}
(1986) 45.

\bibitem{seesaw}  P.~Minkowski, Phys.\ Lett.\ B {\bf 67} (1977) 421;\\
M. Gell-Mann, P.  Ramond and R. Slansky, in  {\em Supergravity}, 
eds.~D.Z.  Freedman  and  P.~van Nieuwenhuizen  (North-Holland,  Amsterdam,
1979);\\ 
T.  Yanagida, in  Proc.\ of  the {\em  Workshop on  the Unified
Theory and the  Baryon Number in the Universe},  Tsukuba, Japan, 1979,
eds.\ O.~Sawada and  A.~Sugamoto;\\ 
R.~N.~Mohapatra and G.~Senjanovi\'c, Phys.\ Rev.\ Lett.\ {\bf 44} (1980) 912.

\bibitem{KRS} V.~A.~Kuzmin, V.~A.~Rubakov and M.~E.~Shaposhnikov,
  Phys.\ Lett.\ B {\bf 155} (1985) 36.

\bibitem{AHJMP} For an alternative suggestion, see,\\ 
R.~Allahverdi, S.~Hannestad, A.~Jokinen, A.~Mazumdar and S.~Pascoli,\\
hep-ph/0504102.

\bibitem{DI} S.~Davidson and A.~Ibarra, Phys.\ Lett.\ B {\bf 535}
 (2002) 25.

\bibitem{BBP} W. Buchm\"uller, P. Di Bari and M. Pl\"umacher, Nucl.\
 Phys.\ B {\bf 643} (2002) 367.

\bibitem{GCBetal} G.C.~Branco, R.~Gonzalez Felipe, F.R.~Joaquim,
  I.~Masina, M.N.~Rebelo and C.A.~Savoy, Phys.\ Rev.\ D {\bf 67} 
  (2003) 073025.

\bibitem{CT} P.H.~Chankowski and K.~Turzynski, Phys.\ Lett.\ B 
  {\bf 570} (2003) 198;\\ 
  T.~Hambye, Y.~Lin, A.~Notari, M.~Papucci and A.~Strumia, Nucl.\
Phys.\ B {\bf 695} (2004) 169. 

\bibitem{PDG} Particle Data Group  (S. Eidelman et al.), Phys.\ Lett.\
  B {\bf 592} (2004) 1.

\bibitem{APRD} A.~Pilaftsis, Phys.\ Rev.\ D {\bf 56} (1997) 5431; 
  Nucl.\ Phys.\ B {\bf 504} (1997) 61.

\bibitem{APreview} A.~Pilaftsis, Int.\ J. Mod.\ Phys.\ A {\bf 14} (1999) 1811.

\bibitem{LB}T.~Hambye, Nucl.\ Phys.\ B {\bf 633} (2002) 171;\\
L. Boubekeur, hep-ph/0208003;\\
L.~Boubekeur, T.~Hambye and G.~Senjanovic, Phys.\ Rev.\ Lett.\ {\bf 93} (2004) 111601;\\
A.~Abada, H.~Aissaoui and M.~Losada, hep-ph/0409343;\\
L.~J.~Hall, H.~Murayama and G.~Perez, hep-ph/0504248.

\bibitem{LiuSegre} J.~Liu and G.~Segr\'e, Phys.\ Rev.\ D {\bf 48} (1993)
  4609.

\bibitem{Paschos} M.~Flanz, E.A.~Paschos and U.~Sarkar, Phys.\ Lett.\
  B~{\bf 345} (1995) 248;\\ 
  L.~Covi, E.~Roulet and  F.~Vissani, Phys.\ Lett.\
  B~{\bf 384} (1996) 169.

\bibitem{PU} A.~Pilaftsis and T.E.J. Underwood, Nucl.\ Phys.\ B {\bf 692}
    (2004) 303.

\bibitem{MV} R.N. Mohapatra and J.W.F. Valle, Phys.\ Rev.\ D {\bf 34}
  (1986) 1642;\\ 
  S. Nandi and U. Sarkar, Phys.\ Rev.\ Lett.\ {\bf 56}
  (1986) 564.

\bibitem{FN} C.D. Froggatt and H.B. Nielsen, Nucl.\ Phys.\ B {\bf 147}
  (1979) 277.

\bibitem{HMW} T.~Hambye, J.~March-Russell and S.~M.~West, JHEP {\bf
  0407} (2004) 070;\\  
  S.~M.~West, Phys.\ Rev.\ D {\bf 71} (2005) 013004.

\bibitem{softL} J.~R.~Ellis, M.~Raidal and T.~Yanagida, Phys.\ Lett.\
B {\bf 546} (2002) 228;\\ 
Y.~Grossman, T.~Kashti, Y.~Nir and E.~Roulet,
  Phys.\ Rev.\ Lett.\  {\bf 91} (2003) 251801;\\
  G.~D'Ambrosio, G.~F.~Giudice and M.~Raidal, Phys.\ Lett.\ B
  {\bf 575} (2003) 75;\\
  E.~J.~Chun, Phys.\ Rev.\ D {\bf 69} (2004) 117303;\\
  R.~Allahverdi and M.~Drees,
  Phys.\ Rev.\ D {\bf 69} (2004) 103522;\\
  Y.~Grossman, R.~Kitano and H.~Murayama, hep-ph/0504160.

\bibitem{DLR} T.~Dent, G.~Lazarides and R.~Ruiz de Austri, Phys.\
  Rev.\ D {\bf 69} (2004) 075012;\\ 
  S.~Dar, S.~Huber, V.N.~Senoguz and Q.~Shafi, 
  Phys.\ Rev.\ D {\bf 69} (2004) 077701;\\ 
  T.~Dent, G.~Lazarides and R.~R.~de Austri, hep-ph/0503235.

\bibitem{GJN} R.~Gonzalez Felipe, F.~R.~Joaquim and B.~M.~Nobre,
Phys.\ Rev.\ D {\bf 70} (2004) 085009. 

\bibitem{AFS} E.~K.~Akhmedov, M.~Frigerio and A.~Y.~Smirnov, JHEP {\bf
0309} (2003) 021. 

\bibitem{AB} C.H.~Albright and S.M.~Barr, Phys.\ Rev.\ D {\bf 69}
  (2004) 073010.

\bibitem{APtau} A. Pilaftsis, Phys.\ Rev.\ Lett.\ {\bf 95} (2005)
  081602.

\bibitem{KS} S.~Y.~Khlebnikov and M.~E.~Shaposhnikov, Nucl.\ Phys.\ B
  {\bf 308} (1988) 885.

\bibitem{HT} J.A.~Harvey and M.S.~Turner, Phys.\ Rev.\ D {\bf 42}
  (1990) 3344.

\bibitem{DR} H. Dreiner and G.G. Ross, Nucl.\ Phys.\ B {\bf 410} (1993)
  188.

\bibitem{LS} M.~Laine and M.~E.~Shaposhnikov, Phys.\ Rev.\ D {\bf 61}
  (2000) 117302.

\bibitem{Borzumati}  For alternative  suggestions, albeit  in extended
  supersymmetric settings, see,\\ 
F.~Borzumati and Y.~Nomura,
%``Low-scale see-saw mechanisms for light neutrinos,''
Phys.\ Rev.\ D {\bf 64} (2001) 053005;\\
N.~Arkani-Hamed, L.~J.~Hall, H.~Murayama, D.~R.~Smith and N.~Weiner,
%``Small neutrino masses from supersymmetry breaking,''
Phys.\ Rev.\ D {\bf 64} (2001) 115011.

\bibitem{Anupam}  Under  rather   generic  conditions,  the  reheating
temperature in supersymmetric theories could  be very low, even as low
as TeV,  because of the  presence of quasi-flat directions  with large
VEV's,  which  slow  down  the  thermalization process  in  the  early
Universe  [R.~Allahverdi and  A.~Mazumdar,  arXiv:hep-ph/0505050].  In
such a scenario, electroweak-scale RL may be the only viable mechanism
for successful baryogenesis.

\bibitem{CLMW} L.~Carson, X.~Li, L.~D.~McLerran and R.~T.~Wang, Phys.\
  Rev.\ D {\bf 42} (1990) 2127.

\bibitem{MEG}   See   proposal    by   MEG   collaboration   at   {\tt
  http://meg.web.psi.ch/docs/index.html}.

\bibitem{MECO} MECO collaboration, {\tt http://meco.ps.uci.edu/};\\
M.~Hebert  (MECO Collaboration), Nucl.\ Phys.\ A {\bf 721} (2003) 461.

\bibitem{CA} C. Aalseth et al., hep-ph/0412300;\\
  S. Pascoli, S.T. Petcov and T. Schwetz, hep-ph/0505226.

\bibitem{GGR} We thank Graham Ross for useful discussions on this point.

\bibitem{BGL} G.~C.~Branco, W.~Grimus and L.~Lavoura, Nucl.\ Phys.\
 B {\bf 312} (1989) 492.

\bibitem{AZPC} A. Pilaftsis, Z.\ Phys.\ C {\bf 55} (1992) 275.

\bibitem{APmix} A. Pilaftsis, Phys.\ Rev.\ D {\bf 65} (2002) 115013. 

\bibitem{PV} G. Pasarino and M. Veltman, Nucl.\ Phys.\ B {\bf 160}
  (1979) 151.

\bibitem{Antusch} 
P.~H.~Chankowski and Z.~Pluciennik,
  Phys.\ Lett.\ B {\bf 316} (1993) 312;\\
K.~S.~Babu, C.~N.~Leung and J.~T.~Pantaleone,
  Phys.\ Lett.\ B {\bf 319} (1993) 191;\\
For recent studies, see,\\
S.~Antusch, M.~Drees, J.~Kersten, M.~Lindner and M.~Ratz,
  Phys.\ Lett.\ B {\bf 519} (2001) 238;\\ 
  S.~Antusch, J.~Kersten, M.~Lindner, M.~Ratz and M.~A.~Schmidt, JHEP
{\bf 0503} (2005) 024. 

\bibitem{MBellac} M. Le Bellac, {\em Thermal Field Theory}, (Cambridge
  University Press, Cambridge, England, 1996);\\ 
  J.I. Kapusta, {\em Finite-Temperature Field Theory}, 
  (Cambridge University Press,
  Cambridge, England, 1989).

\bibitem{Weldon} H.A. Weldon, Phys.\ Rev.\ D {\bf 26} (1982) 2789.

\bibitem{KW} E.~W.~Kolb and S.~Wolfram, Nucl.\ Phys.\ B {\bf 172}
  (1980) 224 [Erratum-ibid.\ B {\bf 195} (1982) 542].

\bibitem{MAL} M.~A.~Luty, Phys.\ Rev.\ D {\bf 45} (1992) 455.
  
\bibitem{Keldysh/Schwinger} J.~S.~Schwinger,
 J.\ Math.\ Phys.\  {\bf 2} (1961) 407;\\
  L.~V.~Keldysh, Zh.\ Eksp.\ Teor.\ Fiz.\ {\bf 47} (1964) 1515
  [Sov.\ Phys.\ JETP {\bf 20} (1965) 1018];\\
  G.~Raffelt, G.~Sigl and L.~Stodolsky, Phys.\ Rev.\ Lett.\ {\bf 70} (1993)
  2363;\\
  G.~Sigl and G.~Raffelt, Nucl.\ Phys.\ B {\bf 406} (1993) 423.

\bibitem{MPspect} W.~Buchmuller and M.~Plumacher, Phys.\ Lett.\ B {\bf
  511} (2001) 74.

\bibitem{thooft} G.~'t Hooft, Phys.\ Rev.\ Lett.\  {\bf 37} (1976) 8.

\bibitem{AM} P.~Arnold and L.~D.~McLerran, Phys.\ Rev.\ D {\bf 36} (1987) 581.

\bibitem{CKO} J.M. Cline, K. Kainulainen and K.A. Olive, Phys.\ 
  Rev.\ D {\bf 49} (1994) 6394.

\bibitem{KM} F.~R.~Klinkhamer and N.~S.~Manton, Phys.\ Rev.\ D {\bf 30}
  (1984) 2212.

\bibitem{JVdata} For a recent analysis, see,
 M.~Maltoni, T.~Schwetz, M.~A.~Tortola and J.~W.~Valle, New J. Phys.\
{\bf 6} (2004) 122.

\bibitem{BCST} R.~Barbieri, P.~Creminelli, A.~Strumia and N.~Tetradis,
  Nucl.\ Phys.\ B {\bf 575} (2000) 61.

\bibitem{GNRRS}  G.F.~Giudice, A.~Notari, M.~Raidal, A.~Riotto and A.~Strumia,
  Nucl.\ Phys.\ B {\bf 685} (2004) 89.

\bibitem{doi85} M.~Doi, T.~Kotani and E.~Takasugi, Prog.\ Theor.\
  Phys.\ Suppl.\ {\bf 83} (1985) 1.

\bibitem{Klapdor} For example, see the textbook by K. Grotz and
  H.V. Klapdor, ``The Weak Interaction in Nuclear, Particle und
  Astrophysics,'' (Adam Hilger, Bristol, 1989).

\bibitem{HKK} H.V. Klapdor-Kleingrothaus, ``Sixty Years of Double Beta
  Decay,'' (World Scientific, Singapore, 2001);\\
  H.V. Klapdor-Kleingrothaus, A. Dietz, H.L. Harney, I.V. Krivosheina,
  Mod.\ Phys.\ Lett.\ A {\bf 16} (2001) 2409;\\ 
  H.V. Klapdor-Kleingrothaus, A. Dietz and I. Krivosheina, 
  Particles and Nuclei {\bf 110} (2002) 57; 
  Foundations of Physics {\bf 32} (2002) 1181.

\bibitem{KKS} H.V.  Klapdor-Kleingrothaus and U. Sarkar,  Mod.\ Phys.\
  Lett.\ A {\bf 16} (2001) 2469;\\ 
  H.V.~Klapdor-Kleingrothaus, H. P\"as and A.  Yu. Smirnov, 
  Phys.\  Rev.\ D~{\bf 63} (2001) 073005;\\ 
  W.~Rodejohann, Nucl.\ Phys.\ B {\bf 597} (2001) 110; J.\ Phys.\ G {\bf
  28} (2002) 1477;\\ 
  H. Minakata and  H. Sugiyama, Phys.\ Lett.\ B {\bf 532} (2002)  275;\\
  S. Pascoli and S.T.  Petcov,  Phys.\ Lett.\ B {\bf 544} (2002) 239;\\ 
  H.~Nunokawa, W.~J.~C.~Teves and R.~Zukanovich Funchal, Phys.\ Rev.\
  D {\bf 66} (2002) 093010. 

\bibitem{HMexp} H.V.~Klapdor-Kleingrothaus, A.~Dietz, I.V.~Krivosheina
  and O.~Chkvorets, Nucl.\ Instrum.\ Meth.\ {\bf A522} (2004) 371.

\bibitem{CL} T.P.~Cheng and L.F.~Li, Phys.\ Rev.\ Lett.\ {\bf 45}
  (1980) 1908.

\bibitem{IP} A. Ilakovac and A. Pilaftsis, Nucl.\ Phys.\ B {\bf 437}
  (1995) 491.

\bibitem{Ara} A. Ioannisian and A. Pilaftsis, Phys.\ Rev.\ D {\bf 62}
  (2000) 066001. 

\bibitem{KPS} J.G. K\"orner, A. Pilaftsis and K. Schilcher, Phys.\
  Lett.\ B {\bf 300} (1993) 381.

\bibitem{BKPS} J. Bernab\'eu, J.G. K\"orner, A. Pilaftsis and K.\
  Schilcher, Phys.\ Rev.\ Lett.\ {\bf 71} (1993) 2695.

\bibitem{MPLAP} A. Pilaftsis, Mod.\ Phys.\ Lett.\ A~{\bf 9} (1994) 3595;\\ 
  M.C. Gonzalez-Garcia and J.W.F. Valle, Mod.\ Phys.\ Lett.\ A~{\bf
  7} (1992) 477; Erratum~{\bf 9} (1994) 2569.

\bibitem{LFVrev} L.N. Chang, D.  Ng and J.N. Ng, Phys.\ Rev.\ D {\bf
  50} (1994) 4589;\\ 
  G.  Bhattacharya, P. Kalyniak and I. Mello, Phys.\
  Rev.\ D {\bf 51} (1995) 3569;\\ 
  A.  Pilaftsis, Phys.\ Rev.\ D {\bf 52} (1995) 459;\\ 
  A. Ilakovac, B.A.  Kniehl, and A. Pilaftsis, Phys.\ Rev.\ 
  D {\bf 52} (1995) 3993;\\ 
  A. Ilakovac, Phys.\ Rev.\ D {\bf 54} (1996) 5653;\\ 
  M. Frank and H.  Hamidian, Phys.\ Rev.\ D {\bf 54} (1996) 6790;\\
  P.  Kalyniak and I. Mello, Phys.\ Rev.\ D {\bf 55} (1997) 1453;\\ 
  G. Barenboim and M.  Raidal, Nucl.\ Phys.\ B {\bf 484} (1997) 63;\\
  Z. Gagyi-Palffy, A.  Pilaftsis and K.  Schilcher, Nucl.\ Phys.\ B {\bf
  513} (1998) 517;\\ 
  S.  Fajfer and A.  Ilakovac, Phys.\ Rev.\ D {\bf 57} (1998) 4219;\\ 
  M.  Raidal and A.  Santamaria, Phys.\ Lett.\ B {\bf 421} (1998) 250;\\ 
  M.~Czakon, M.~Zralek and J.~Gluza, Nucl.\ Phys.\ B {\bf 573} (2000) 57;\\ 
  J.~I.~Illana and T.~Riemann, Phys.\ Rev.\ D {\bf 63} (2001) 053004;\\ 
  G.~Cvetic, C.~Dib, C.~S.~Kim and J.~D.~Kim, Phys.\ Rev.\ D {\bf 66} (2002)
  034008; hep-ph/0504126;\\ 
  A.~Masiero, S.~K.~Vempati and O.~Vives, New J.\ Phys.\  {\bf 6} (2004) 202.
  
\bibitem{LL} P. Langacker and D. London, Phys.\ Rev.\ D {\bf 38} (1988) 886.

\bibitem{LLfit} C.~P.~Burgess, S.~Godfrey, H.~Konig, D.~London and I.~Maksymyk, Phys.\ Rev.\ D {\bf 49} (1994) 6115;\\
E.~Nardi, E.~Roulet and D.~Tommasini, Phys.\ Lett.\ B {\bf 327} (1994) 319;\\
D.~Tommasini, G.~Barenboim, J.~Bernabeu and C.~Jarlskog, Nucl.\ Phys.\
B {\bf 444} (1995) 451;\\ 
S.~Bergmann and A.~Kagan, Nucl.\ Phys.\ B {\bf 538} (1999) 368.

\bibitem{FW} G. Feinberg and S. Weinberg, Phys.\ Rev.\ Lett.\ {\bf 3}
  (1959) 111; 244 (Erratum);\\ 
  W.J. Marciano and A.I. Sanda, Phys.\ Rev.\ Lett.\ {\bf 78} (1977) 1512;\\ 
  O. Shanker, Phys.\ Rev.\ D {\bf 20} (1979) 1608;\\ 
  J. Bernab\'eu, E. Nardi and D. Tommasini, Nucl.\ Phys.\ B {\bf409} (1993) 69.

\bibitem{JDV}  For reviews, see, J.D.  Vergados, Phys.\ Rep.\ {\bf 133}
  (1986) 1;\\ 
  T.S.  Kosmas, G.K. Leontaris and J.D. Vergados, 
  Prog.\ Part.\ Nucl.\ Phys.\ {\bf 33} (1994) 397.

\bibitem{Zeff} J.C. Sens, Phys.\ Rev.\ {\bf 113} (1959) 679;\\ 
  K.W. Ford and J.G. Wills, Nucl.\ Phys.\ {\bf 35} (1962) 295;\\ 
  R. Pla and J. Bernab\'eu, An.\ F\'\i s.\ {\bf 67} (1971) 455;\\ 
  H.C. Chiang, E. Oset, T.S. Kosmas, A.  Faessler and J.D. Vergados, 
  Nucl.\ Phys.\ A {\bf 559} (1993) 526.

\bibitem{SMR} T. Suzuki, D.F. Measday and J.P. Roalsvig, Phys.\ Rev.\
C {\bf 35} (1987) 2212.

\bibitem{FPap} For instance, see,   B. Frois and C.N. Papanicolas,  Ann.\
 Rev.\ Nucl.\ Sci.\ {\bf 37} (1987) 133, and references therein.

\bibitem{SINDRUM}  C. Dohmen et  al.\  (SINDRUM  II Collaboration),
 Phys.\ Lett.\ {\bf B317} (1993) 631.

\bibitem{NN} D. Ng and J.N. Ng, Mod.\ Phys.\ Lett.\ A {\bf 11} (1996) 211;\\ 
  J.P. Archaumbault, A. Czarnecki and M. Pospelov, hep-ph/0406089;\\
  W.-F. Chang and J.N. Ng, hep-ph/0411201.

\bibitem{BG} W.~Buchm\"uller and C.~Greub, Nucl.\ Phys.\ B {\bf 363} 
  (1991) 345;\\
  G.~Cvetic, C.S.~Kim and C.W.~Kim, 
  Phys.\ Rev.\ Lett.\ {\bf 82} (1999) 4761;\\
 S.~F.~King and T.~Yanagida, hep-ph/0411030.

\bibitem{AAS} F.~del Aguila, J.~A.~Aguilar-Saavedra, A.~Martinez de la Ossa
  and D.~Meloni, Phys.\ Lett.\ B {\bf 613} (2005) 170;\\
  F.~del Aguila and J.~A.~Aguilar-Saavedra, hep-ph/0503026.

\bibitem{LSND} A.~Aguilar et al. (LSND Collaboration), Phys.\ Rev.\ D
  {\bf 64} (2001) 112007.

\bibitem{EWBAU} For recent analyses, see,\\
M.~Carena, M.~Quiros, M.~Seco and C.~E.~M.~Wagner,
  %``Improved results in supersymmetric electroweak baryogenesis,''
  Nucl.\ Phys.\ B {\bf 650} (2003) 24;\\
T.~Konstandin, T.~Prokopec and M.~G.~Schmidt,
  %``Kinetic description of fermion flavor mixing and CP-violating sources  for
  %baryogenesis,''
  Nucl.\ Phys.\ B {\bf 716} (2005) 373;\\
M.~Carena, A.~Megevand, M.~Quiros and C.~E.~M.~Wagner,
  Nucl.\ Phys.\ B {\bf 716} (2005)~319;\\
T.~Konstandin, T.~Prokopec, M.~G.~Schmidt and M.~Seco,
  hep-ph/0505103.

\bibitem{AS} {\em Handbook of Mathematical Functions}, edited by
  M. Abramowitz and I.~A. Stegun (Verlag Harri Deutsch, Frankfurt,
  1984).



\end{thebibliography}
\end{document}